\documentclass{aastex62}
\usepackage{amsmath}
\usepackage{mathtools}
\usepackage{threeparttable}

\usepackage{lineno}
\graphicspath{{./}{figures/}}

\accepted{July 11, 2022}
\submitjournal{ApJ}

\shorttitle{Global modeling of nebulae. II}
\shortauthors{Estrada, Cuzzi and Umurhan}

\begin{document}

\title{Global Modeling of Nebulae With Particle Growth, Drift, and Evaporation Fronts. \\
II. The Influence of Porosity on Solids Evolution}

\correspondingauthor{Paul R. Estrada}
\email{Paul.R.Estrada@nasa.gov}

\author{Paul R. Estrada}
\affiliation{NASA Ames Research Center, MS 245-3, Moffett Field, CA 94035}

\author{Jeffrey N. Cuzzi}
\affiliation{NASA Ames Research Center, MS 245-3, Moffett Field, CA 94035}

\author{Orkan M. Umurhan}
\affiliation{NASA Ames Research Center, MS 245-3, Moffett Field, CA 94035}
\affiliation{Carl Sagan Center, SETI Institute, Mountain View, CA 94043}



\defcitealias{Est16}{Paper I}
\defcitealias{Est21}{Paper III}
\defcitealias{CarrascoGonzalezetal2019}{CG19}

\begin{abstract}

Incremental particle growth 
in turbulent protoplanetary nebulae is limited by a combination of barriers that can 
slow or 
stall growth. 
Moreover, 
particles that grow massive enough to decouple from the gas 
are subject to inward radial drift which 
could lead to the depletion of most disk solids before planetesimals can form. 
Compact particle growth 
is probably not realistic. Rather, it is more likely that grains grow as fractal aggregates 
which may 
overcome this so-called radial drift barrier because they 
remain 
more 
coupled to the gas 
than compact particles of equal mass. We model 
fractal aggregate growth and compaction in a 
viscously evolving solar-like nebula 
for a range of turbulent intensities $\alpha_{\rm{t}} = 10^{-5}-10^{-2}$. 
We do find that radial drift is less influential for 
porous aggregates 
over much of their growth phase; however, 
outside the water snowline fractal aggregates can grow to much larger masses with 
larger Stokes numbers more quickly than compact particles, leading to rapid inward radial drift. 
As a result, 
disk solids outside the snowline out to $\sim 10-20$ AU are depleted earlier than in compact growth models, but outside $\sim 20$ AU material is 
retained 
much longer 
because 
aggregate Stokes numbers there 
remain lower initially. 
Nevertheless, we conclude 
even 
fractal models will lose most disk solids
without the intervention of some leap-frog planetesimal forming mechanism 
such as the Streaming Instability (SI), 
though 
conditions for the SI are generally never satisfied, 
except for a brief period 
at the snowline for 
$\alpha_{\rm{t}}=10^{-5}$. 


 
\end{abstract}




\section{Introduction}\label{sec:intro}

Observations over the last several decades indicate that not only does planet formation appear to be extremely robust, it can lead to very diverse outcomes \citep[][and others]{Lis11,WF15,Tri17,Wit20}. Yet, despite their ubiquity, a clear picture of how these systems came to be remains lacking because we still do not possess a coherent picture of how growth proceeds from sub-micron dust grains to planetary building blocks. Primary accretion of the first 100km size planetesimals, in particular,  is not understood, but is a key initial condition for subsequent stages of planetary growth in our own solar system and beyond \citep[e.g.,][]{Mor09,KB10,Sch13,Cha14, Joh14}. The physics of secondary accretion - embryos that grow by pair-wise mergers of large bodies \citep[e.g.,][]{KL94,KI02,Cha10,Bod18}, or by pebble accretion \citep[e.g.,][]{OK10,LJ12,LJ14,Bit19} - is  more well characterized, but the $100-200$ km size pre-existing bodies adopted as the initial conditions for these models are merely assumed to be formed previously by some other means.

The single most critical factor in determining the path taken by primary accretion is the poorly known intensity of global turbulence.  Grain growth begins by low-velocity ``perfect'' sticking of sub-micron monomers that are well-coupled to the gas, but relative velocities between growing particles in turbulence increase sufficiently that they become subject to a gauntlet of barriers that by themselves, or in combination, frustrate growth beyond a certain threshold. The most elementary barrier is simply fragmentation that can effectively grind particle growth to a halt. However, even before this, grains can encounter the so-called bouncing barrier where collisional energies are insufficient to lead to any fragmentation or erosion, but large enough to prevent sticking \citep{Gut10,Zso10}. The bouncing barrier does not halt growth entirely, but it can significantly slow growth by limiting the range of sizes that a grain can grow from \citep{Win12a,Est16}. Once particles grow to sufficient size that they begin to decouple from the gas and drift radially inward due to a headwind drag from the more slowly rotating, pressure-supported gas \citep{Wei77,CW06,Joh14}, they encounter a third, so-called radial drift barrier because they drift in faster than they can grow. This is especially problematic in turbulence, which decreases the local density of the midplane particle layer \citep{Dub95}. For compact particle growth models, radial drift tends to dominate outside the snow line \citep[e.g.,][]{Bra08,Bir10,Est16,Dra16}. Even if incremental growth can somehow escape these barriers and continue to larger sizes, turbulence poses yet another barrier to growth when gas density fluctuations gravitationally excite disruptive collisions between $1-10$ km size bodies \citep{Idaetal2008, NelsonGressel2010, Gresseletal2011, OrmelOkuzumi2013}.


Indeed, after much debate over the years, it is increasingly thought that protoplanetary nebulae in the early stages of their evolution are at least weakly-to-moderately-turbulent in the regions of the disk ($\lesssim 100$ AU) in which particle growth is of the greatest interest \citep[see e.g.,][for reviews]{Tur14, LU19}. Even as Magneto-Hydrodynamic (MHD) models of turbulence have trended towards very weak turbulence even at high altitudes \citep[][and others]{BS13,Tur14,Gre15}, recent theoretical advances point to a number of purely hydrodynamical mechanisms \citep[][see \citealt{LU19}]{Nel13,Mar13,Mar15,Lyr14,Sto17,Bar18} that appear to be able to drive turbulence almost anywhere in the disk, at the level of $\alpha_{\rm{t}} \sim 10^{-5}-10^{-3}$, where $\alpha_{\rm{t}}$ is the so-called turbulence coefficient which parameterizes the magnitude of turbulence strength. 

This gauntlet of barriers to incremental growth in turbulence has led to the idea that $\gtrsim 10-100$ km bodies are assembled directly from a sufficiently dense reservoir of growth-frustrated smaller particles via some collective effect, such as the Streaming Instability \citep[SI,][]{YG05,YJ07} or Turbulent Concentration \citep[TC;][]{Cuz10, HC20} that under the right conditions can leapfrog all of these barriers. 
However, with few exceptions  \citep[e.g.,][who assumed that  meter-sized particles could exist in a fully MRI-turbulent nebula]{Joh07}, SI has only been modeled in globally nonturbulent disks and the viability of the SI under plausible turbulent conditions has not been seriously studied. Recently, \citet[][see also \citealt{CL20}]{Umu20} have established new criteria for the operation of the SI in arbitrary turbulence for arbitrary particle sizes. Comparing their results with  sizes for compact particles determined in previous versions of this work \citep{Est16}, they found that self-consistent  conditions of particle size and turbulent intensity for planetesimal formation by SI are not generally obtained. Meanwhile, it has been found that TC may be able to produce planetesimals from swarms of growth-frustrated ``pebbles" close to the realistic size range \citep{HC20}. 

It is evident that, if the nebula is even weakly turbulent, accurate estimates of the limits on size and density of particles that {\it can} grow simply by sticking are needed to provide self-consistent initial conditions for these ``leapfrog'' models of planetesimal formation. Finding ways to overcome the radial drift barrier would be especially important: if particles could grow faster than they can drift, then it might be possible for them to grow to larger sizes without being lost to the inner disk regions, or even the central star. These particles would still have to contend with possible fragmentation, but if they could grow large enough and in sufficient number, and/or achieve some enhanced degree of local solids density,  then they might more easily trigger planetesimal formation by SI or TC. 
To that end, in this paper we incorporate a model for porous particle growth and compaction in our global nebula evolution code \citep{EC08,Est16} to explore the consequences of fractal aggregates on the solids evolution within the protoplanetary gas disk. 

Similar models have been developed previously, but none have applied fractal growth and compaction to a model that incorporates all of the relevant physics and self-consistently accounts for the temporal and spatial evolution of the composition and size distribution of particles  in a dynamically evolving disk as we do. 
\citet{Oku12} modeled the formation of planetesimals outside the snowline via fractal, porous particle growth and collisional compression in a static Minimum Mass Solar Nebula \citep[MMSN,][]{Hay81}, but assumed perfect sticking and ignored fragmentation. They found that although some local compaction occurs, the porosity of a growing aggregate continues to increase as a fractal, reaching planetesimal mass with internal densities of $\sim 10^{-4}$ g cm$^{-3}$ or less. In Section \ref{sec:collcomp}, we describe how we improve on these assumptions. 
Following along these lines, \citet{Kri15} used the \citeauthor{Oku12} model in a static MMSN, but added fragmentation and erosion along with non-collisional compaction effects \citep{Kat13a}. Using a Monte Carlo model, they show that erosion and restructuring by collisions with smaller particles stall growth well below planetesimal masses. 
Most recently, \citet{HN18} have applied the \citeauthor{Oku12} model \citep[and that of][]{Kat13a,Kat13b} in an evolving nebula with infall from the molecular cloud.  Like \citeauthor{Oku12}, \citeauthor{HN18} also assume perfect sticking and ignore fragmentation and erosion, 
and only consider the growth of aggregates outside the snowline. It is found that in this hotter nebula, where the snowline evolves outwards ($\sim 7-15$ AU) due to the infall, 
growth still stalls well below planetesimal masses even with perfect sticking. 
\citet{DD18} have also recently explored a model with infall in which their snowline evolves outwards with time  during the buildup stage. They find that planetesimals mostly form later, but under certain conditions can form in the buildup stage, but these authors assume  compact particle growth and adopt different turbulent strengths for disk evolution and turbulent mixing of the dust.
 
In this paper we employ the same fractal growth and compaction recipe \citep{Suy12,Oku12,Kat13a}. However, none of these previous models consider multiple species or have a self-consistent calculation of disk opacity and temperature which depends on the evolving size-distribution \citep[][hereafter, Paper I]{Est16}. Realistic disk composition will contain a mixture of silicates, iron metal, iron sulfide, organics, and water and other ices (see Table \ref{tab:species}) each with its own associated evaporation front (EF). Our model is novel in the treatment of these, and is ideally suited to study the global implications of fractal growth over a range of (albeit  imperfectly constrained) parameters.   
In Section \ref{sec:model} we summarize the basic workings of our global nebula evolution model, and expand on changes made to the code for this paper, including how we implement fractal growth and compaction into our numerical scheme. In Section \ref{sec:results} we describe the results of our simulations, comparing models in which we include fractal growth to their  compact  particle equivalent  over a range of $\alpha_{\rm{t}}$. We also present models that vary  other  disk parameters in Appendix \ref{sec:varparm}. Generally, while lacking infall, the nebula density and temperature, and the central star properties and timescales modeled here are characteristic of the Class 0/I stage of nebula evolution ($<1$ Myr) during which time the first planetesimals form 
\citep{Kruijeretal2017}.  In Section \ref{sec:discuss} we discuss implications for planetesimal formation, as well as observations of  particle porosity. In our companion paper, we discuss the compositional evolution of our disk models in more detail \citep[][hereafter Paper III]{Est21}. In Section \ref{sec:sum} we summarize our results.   

\section{Nebula model}
\label{sec:model}
 
The simulations presented herein use our parallelized $1+1$D radial nebula code in 
which we simultaneously treat particle growth and radial migration of solids, while evolving 
the dynamical and thermal evolution of the protoplanetary gas disk. Our code includes the
self-consistent growth and radial drift of particles of all sizes, accounts for vertical
settling and diffusion of the smaller grains, radial diffusion and advection of solid and
vapor phases of multiple species of refractories and volatiles, and contains a self-consistent
calculation of the opacity and disk temperature that allows  us to track the evaporation
and condensation of all species as they are transported throughout the disk. 
In the following  sections, we only briefly summarize our code and will mostly focus on the description of additional physics or improvements that have been implemented for this work. 

One difference between our implementation here and in \citetalias{Est16} is that in our previous work we employed an asynchronous time-stepping scheme. Briefly, each radial bin had its own timestep, with the innermost radial bin of the grid defining the minimum {\it particle growth} timestep $\Delta t_{\rm{min}}$, which is a fraction of the local orbital period. This allowed for significant computational savings because only the innermost bin is called every $\Delta t_{\rm{min}}$, while other bins were called at the appropriate time interval. We utilized a much larger {\it global} timestep, typically $\Delta t_{\rm{global}} = 100 \times \Delta t_{\rm{min}}$ to ``synchronize'' all radial bins to the current absolute simulation time, and then executed spatially global calculations like solving the gas and particle evolution equations, and determining the disk temperature. In this paper, we now use $\Delta t = \Delta t_{\rm{min}}$ for all radial bins which leads to higher accuracy and faster run times, but at the expense of increased computational resources. However, we do continue to use a global timestep because as we showed previously (see Appendix C of \citetalias{Est16}) these global processes are not changing quickly over $\Delta t_{\rm{global}}$ compared to growth times, even for the highest turbulence strength we employ here. The reader is referred to \citetalias{Est16} for a more detailed description of our code.

\subsection{Gas Disk Evolution}
\label{sec:gasevol}

The initial conditions for the gas disk models used in this work are derived from the analytical expressions of \citet{LP74}, as generalized by \citet{Har98} 
in terms of an initial disk mass $M_{\rm{disk}}$ and radial scale factor $R_0$  \citepalias[see][]{Est16}. This gas surface density distribution is fairly similar to that of \citet{Des07} in the $0.3-0.5$ Myr timeframe of both models, and represents a denser inner nebula than the MMSN. The time-dependent evolution of the gas surface density $\Sigma$ and gas radial velocity $v_{\rm{g}}$ in one dimension are then obtained by solving \citep{Pri81}

\begin{equation}
    \label{equ:gasevol}
    \frac{\partial \Sigma}{\partial t} = \frac{3}{R}\frac{\partial}{\partial R}\left\{R^{1/2}\frac{\partial}{\partial R}(R^{1/2}\nu\Sigma)\right\},
\end{equation}

\begin{equation}
    \label{equ:vgas}
    v_{\rm{g}} = -\frac{3}{R^{1/2}\Sigma}\frac{\partial}{\partial R}(R^{1/2}\nu\Sigma).
\end{equation}

\noindent
The turbulent eddy kinematic viscosity $\nu = \alpha_{\rm{t}}cH$ is parametrized in terms of a turbulent intensity $\alpha_{\rm{t}}$, and depends on the evolving disk temperature $T$ both through the gas sound speed $c$, and the nebula gas density scale height $H$. The gas mass accretion rate is given by $\dot{M} = - 2\pi R\Sigma v_{\rm{g}}$ where $\dot{M} > 0$ indicates flow towards the star, and $\dot{M} < 0$ indicates mass flux outwards. The turnaround radius, at which the mass flux changes sign, is generally an increasing  function of time, and at $t=0$ is $R_{\rm{t}} \simeq R_0/2$ for our initial conditions \citep{Har98}. 
The water snowline begins around $\sim 10-15$ AU in our models, so they can perhaps best be associated with the post buildup stage due to infall \citep{DD18,HN18}.

\subsection{Evolution of Solid and Vapor Species}
\label{sec:dustvapevol}

Our model is capable of tracking multiple species in both vapor and solid phases. In Table \ref{tab:species}, we list the condensibles used in this work along with their corresponding condensation temperatures $T_i$, compact particle density $\rho_i$ and initial global mass fraction in the solid state $\bar{x}_i$ relative to H and He. 
The initial mass fractions 
$\bar{x}_i$ 
are constructed using data from Table 2 of \citet{Lod03} but differ in detail from Lodders' values. To get our values for $\bar{x}_i$, we started with the initial ``refractory organics" or ``CHONs'' fraction from \citet{Pol94} as a guide for C \citep[see also][]{Jessbergeretal1988, JessbergerKissel1991,LawlerBrownlee1992,MummaCharnley2011}, and then partitioned the remaining C in the ratio of 1:1:1 between the ``supervolatiles" CO, CH$_4$ and CO$_2$. The silicates are determined by placing all the Mg into a combination of orthopyroxene (MgSiO$_3$) and olivine (Mg$_2$SiO$_4$), while the refractory iron (metal) fraction is what remains after all of the S is placed in FeS. Some of the remaining O is used in determining the Ca-Al ``refractory" fraction which is a mixture of Al$_2$O$_3$, CaO and CaTiO$_3$, while the rest is placed in water. By contrast, \citet{Lod03} has no refractory organics, assumes no O-bearing supervolatiles, allows water to capture almost all the O that is not consumed by silicates, and assumes the C is in methane ice or in methane hydrate that travels with the water ice. 
The total metallicity for our disk models is $\bar{Z} = \sum_i \bar{x}_i \simeq 0.014$.

\begin{table}[h!]
\renewcommand{\thetable}{\arabic{table}}
\centering
\caption{List of Species} \label{tab:species}
\begin{tabular}{lccc}
\tablewidth{0pt}
\hline
\hline

Species & $T_i$ (K) & $\rho_i$ (g cm$^{-3}$) & $\bar{x}_i$ ($T<T_i$) \\
\hline
Ca-Al & 2000 & 4.0 & $2.38\times 10^{-4}$ \\
Iron & 1810 & 7.8 & $6.54\times 10^{-4}$ \\
Silicates & 1450 & 3.4 & $3.01\times 10^{-3}$ \\
FeS & 680 & 4.8 & $1.16\times 10^{-3}$ \\
Organics & 425 & 1.5 & $3.53\times 10^{-3}$ \\
Waterice & 160 & 0.9 & $3.46\times 10^{-3}$ \\
CO$_2$ & 47 & 1.56 & $8.16\times 10^{-4}$ \\
Methane & 31 & 0.43 & $2.97\times 10^{-4}$ \\
CO & 20 & 1.6 & $5.19\times 10^{-4}$ \\
 
\hline
\end{tabular}
\end{table}

We subdivide the particles into ``dust" and larger  ``migrator" phases - essentially particles smaller than or larger than the fragmentation  mass (see Sec. \ref{sec:partgrow}, and \citetalias{Est16}). These and the vapor phase fraction of each species are denoted by subscripts d, m and v  respectively.  As the disk evolves, these fractions change with time as particles grow and are transported throughout the nebula. The local  instantaneous mass fraction, or concentration $x_i$ of each phase of species $i$, which is  the ratio of the surface density of the species to the gas surface density, varies with time and is given by

\begin{equation}
    \label{equ:alphas}
    x_i^{\rm{v,d,m}} = \frac{\Sigma_i^{\rm{v,d,m}}}{\Sigma}.
\end{equation}

\noindent
The total phase fraction of all species $f_{\rm{v,d,m}}$ is just the sum over $i$ in Eq. (\ref{equ:alphas}), e.g. for the smaller ``dust" fraction, $f_{\rm{d}} = \sum_i x^{\rm{d}}_i$. Similarly, the locally varying total mass fraction of condensables is the sum over all {\it i}, $x_i^{\rm{v}} + x_i^{\rm{d}} +x_i^{\rm{m}}$, whereas the instantaneous metallicity (only species in the solid form) is $Z = \sum_i (x_i^{\rm{d}} + x_i^{\rm{m}})$. 

The evolution of the gas surface density (Sec. \ref{sec:gasevol}) includes the vapor fraction of all species, but we do not include the contribution of the vapor phase in the molecular weight of the gas which would affect the sound speed, and thus viscosity \citep[see, e.g.,][]{SO17,Cha21}. Large enhancements in the vapor phase could lead to a significant decrease in $c$ at EFs \citepalias[see][]{Est21} which then might lead to a pressure bump that can produce even stronger enhancements in solids, and expand the conditions under which, for instance, leap-frog planetesimal formation mechanisms may be satisfied (see Sec. \ref{sec:pform} for more discussion). We will revisit the influence of the gas molecular weight in a future paper. 

\subsubsection{Particle Stopping Times}
\label{sec:stoptime}

Particle growth in the protoplanetary disk begins by sticking of sub-micron grains which are 
dynamically coupled to the nebula gas. As grains grow
into larger particles and agglomerates, they begin to decouple from the gas and collide at
higher velocities. This influence of the nebula gas on the motion of particles is determined
by the Stokes number

\begin{equation}
\label{equ:St}
    {\rm{St}} \equiv \frac{t_{\rm{s}}}{t_{\rm{ed}}},
\end{equation}

\noindent
where $t_{\rm{ed}}$ is some characteristic eddy turnover time which we take to be the integral
scale, or the turnover time of the largest eddy $\Omega^{-1}$ for global turbulence 
\citep[e.g.,][]{Cuz01,Car10}. Here, $\Omega = \sqrt{GM_\star/R^3}$ is the orbital frequency at 
semi-major axis $R$. The stopping time $t_{\rm{s}}$ is the time needed for gas drag to dissipate  the momentum of a particle of mass $m$ relative to a gas with local volume density $\rho$. The drag force a particle of radius $r$ feels depends on its size relative to the molecular mean-free path $\lambda_{\rm{mfp}}$: 

\begin{equation}
    \label{equ:stoptime}
    t_{\rm{s}} = 
    \begin{dcases}
    \;\;\;\;\frac{3m}{4\rho c A} & {\rm{if}}\, r\le (9/4)\lambda_{\rm{mfp}}
    \;\;\; {\rm{(Epstein}} \,  {\rm{regime)}}; \\
    \frac{2m}{C_{\rm{d}}\rho \Delta v_{\rm{pg}}A} & {\rm{if}}\, r > (9/4)\lambda_{\rm{mfp}} \;\;\; {\rm{(Stokes}} \, {\rm{regime)}},
    \end{dcases}
\end{equation}

\noindent
where $\lambda_{\rm{mfp}} = \mu_{\rm{m}}/\rho c$, with $\mu_{\rm{m}} = 1.3\times 10^{-4}$ g cm$^{-1}$ s$^{-1}$ is the dynamic viscosity of the gas, and $A$ is the projected area of an aggregate. Equation (\ref{equ:stoptime}) describes two distinct flow regimes, the Epstein flow regime 
and the Stokes regime. 
Particles can be well or poorly coupled to the gas in either regime, in principle. In the Stokes regime, particles' stopping times are affected by a drag coefficient $C_{\rm{d}}$ which depends on the particle-to-gas relative velocity $\Delta v_{\rm{pg}}$ through the particle Reynolds number ${\rm{Re}}_{\rm{p}} = 2r\rho \Delta v_{\rm{pg}}/\mu_{\rm{m}}$. 
For the simulations in this paper, Stokes flow particles typically have ${\rm{Re}}_{\rm{p}}\lesssim 1$ so that $C_{\rm{d}} = 24/{\rm{Re}}_{\rm{p}}$ \citep{Wei77}. 

\subsubsection{Radial Diffusion-advection and Vertical Diffusion}
\label{sec:diffadv}

The radial motions of both solid and vapor species are determined for each species and phase from the advection-diffusion equation \citep{Des17}

\begin{equation}
    \label{equ:diffadv}
    \frac{\partial \Sigma_i}{\partial t} = \frac{1}{R}\frac{\partial}{\partial R}\left\{R{\mathcal{D}}\Sigma
    \frac{\partial x_i}{\partial R} - R\bar{v}\Sigma_i\right\} + {\mathcal{S}}_i,
\end{equation}

\noindent
where $\bar{v}$ is the net, inertial space velocity (advection $+$ drift), ${\mathcal{D}}$ is the diffusivity and ${\mathcal{S}}_i$ represents sources and sinks for solids 
and vapor of species $i$ which include growth, radial transport  and destruction of migrating material \citepalias[see Sec. \ref{sec:destprob}, and][]{Est16}. 
For the vapor phase, $\bar{v} = v_{\rm{g}}$ and ${\mathcal{D}}=\nu$, whereas for the dust particles, $\bar{v}$ and ${\mathcal{D}}$ (which are both functions of height $z$) are determined at any radius $R$ in the midplane layer or ``subdisk'' (see below) through a mass volume density-weighted mean over particle masses less than or equal to the fragmentation mass  



\begin{equation}
    \label{equ:dustdiff} 
    {\mathcal{D}} = \frac{\nu}{\rho_{\rm{d}}}\sum_k \frac{\rho^{\rm{d}}_k}{1+{\rm{St}}^2_k},\;\;
    \bar{v} = \frac{1}{\rho_{\rm{d}}} \sum_k \rho^{\rm{d}}_k v^{\rm{rad}}_k.
\end{equation}

\noindent
where $\rho_{\rm{d}} = \sum_k \rho^{\rm{d}}_k$, and 

\begin{equation}
    \label{equ:rhodk}
    \rho^{{\rm{d}}}_k = \sqrt{\frac{2}{\pi}}\frac{\Sigma^{\rm{d}}_k}{2h^{\rm{d}}_k}e^{-\frac{1}{2}(z/2h^{\rm{d}}_k)^2},
\end{equation}

\noindent
with $h^{\rm{d}}_k$ the particle scale height as defined in Equation \ref{eq:hdeq} below. The radial drift of particles relative to the gas due to the local pressure gradient \citep{Nak86,TL02}

\begin{equation}
    \label{equ:vrad}
    v^{\rm{rad}}_k = \frac{v_{\rm{g}}}{1 + {\rm{St}}^2_k} + \Delta u_k,
\end{equation}

\noindent
has two components. The first is imposed by the radial motion of the gas moving with advective velocity $v_{\rm{g}}$, while the second 
is the radial velocity of the particle with respect to the gas. It is important to note that the sign of the radial drift velocity depends on particle mass -- massive particles tend to drift radially inwards ($\bar{v} < 0$), while less massive ones can be radially advected outwards with the gas ($\bar{v} > 0$). We account for this in our code by determining the particle mass (where $\bar{v} \approx 0$) that separates the population of particles moving inward from that moving outward, and then using the respective fractional masses of these populations as a weighting factor in solving the advection-diffusion equation for each population \citepalias[see][]{Est16}. For particle masses beyond the fragmentation mass,  {\it i.e.}, migrators or ``lucky particles'', we explicitly follow the evolution (including growth and radial transport) of their mass distribution in our code \citepalias[see Sec. \ref{sec:partgrow}, and][]{Est16}.  


The vertical distribution of the mass volume density of particles $\rho_k^{\rm{d}}(z)$ and $\rho_l^{\rm{m}}(z)$ are accounted for through a balance between settling and diffusion that combines elements of several studies \citep{Dub95,CW06,YL07}. The scale height for particles of mass $m$ in Equation \ref{equ:rhodk} is given by \citepalias[see Appendix B,][]{Est16}

\begin{equation}
    h_{\rm{d}}(m) = \frac{H}{(1  + {\rm{St}}/\alpha_{\rm{t}})^{1/2}},
\label{eq:hdeq}
\end{equation}

\noindent
and similarly for migrators.
The particle mass whose scale height sets the subdisk height, which defines the midplane volume where particle growth is followed in our code (Sec. \ref{sec:partgrow}), is taken to be one half of the mass of the mass-dominant particle at any given $R$ \citepalias[see][]{Est16}. Finally,  the total mass volume density at $R$ of solid material in the subdisk is the sum over all particle mass bins in the dust and migrator populations indexed by $k$ and $l$, $\rho_{\rm{solids}} = \sum_k \rho^{\rm{d}}_k + \sum_l \rho^{\rm{m}}_l$.

\subsection{Disk Thermal Evolution}
\label{sec:diskevol}

Because the disk temperature depends on the evolving particle size distribution through the opacity, a  self-consistent calculation  is essential to capturing the disk's dynamical evolution. We assume that the nebula is heated through a combination of internal viscous dissipation (primarily near the midplane), and external illumination by the stellar luminosity $L_\star$. The equation we use for determining the midplane temperature is given by \citet{NN94}: 

\begin{equation}
    \label{equ:temp}
    \sigma_{\rm{SB}}T^4 = \frac{9}{8}\nu\Sigma\Omega^2\left(\frac{3\tau_{\rm{R}}}{8}+\frac{1}{2\tau_{\rm{P}}}
    \right) + \frac{L_\star \phi}{4\pi R^2},
\end{equation}

\noindent
which must be solved iteratively because the Rosseland and Planck mean optical depths $\tau_{\rm{R}}$ and $\tau_{\rm{P}}$, respectively, are temperature dependent and thus affect the evolving particle size distribution and the solids and vapor fractions of all species. In Eq. (\ref{equ:temp}), the first term on the RHS is a local, vertically integrated viscous dissipation rate, while the second term on the RHS accounts for the stellar flux on each disk face, and $\phi$ is a grazing incidence angle depending on disk geometry. In our model, we assume a flared disk geometry with a general radial variation given by \citet[][see also \citet{KH87,RP91}]{CG97}

\begin{equation}
    \label{equ:graze}
    \phi(R) \sim 0.005 R^{-1}_{\rm{AU}} + 0.05 R^{2/7}_{\rm{AU}},
\end{equation}

\noindent
where $R_{\rm{AU}}$ is radial distance in AU. As we did in \citetalias{Est16}, we employ a time variable luminosity using the model for a 1 M$_\odot$ star \citep{DM94,Sie00}. For the latter, the initial luminosity $L$ is roughly $12 \,{\rm{L}}_\odot$ at the beginning of a simulation, which drops to $\approx 3 \,{\rm{L}}_\odot$ after 0.5 Myr. 
One issue identified with these models is that they are based on oversimplified initial conditions that introduce uncertainty in the evolutionary track for stellar ages $t \lesssim 1$ Myr \citep{Bar98,Bar02}. However, the luminosity reduction factor may be at most $\sim 10$\% \citep[see][]{PS99} which we consider reasonable for the purposes of our simulations. In future work, especially when considering different metallicities as well as stellar masses, we will consider more recent  evolutionary track models \citep[e.g.,][]{DiC09,Tog11}.
 
The optical depths $\tau$ in Eq. (\ref{equ:temp}) are functions of the opacity $\kappa$ as $\tau = \kappa\Sigma/2$. We define the Rosseland and Planck mean opacities from the basic wavelength-dependent opacity $\kappa_\lambda$ weighting in the standard way

\begin{equation}
    \label{equ:kappaR}
    \kappa^{-1}_{\rm{R}} = \frac{\pi}{4\sigma_{\rm{SB}}T^3}\int \kappa^{-1}_\lambda\frac{dB_\lambda}{dT}\,d\lambda;
    \,\,\,\,\kappa_{\rm{P}} = \frac{\pi}{\sigma_{\rm{SB}}T^4}\int \kappa_\lambda B_\lambda d\lambda.
\end{equation}

\noindent
To determine the $\kappa_\lambda$, we utilize the opacity model of \citet[][see also \citealt{Sen19}]{Cuz14} which includes  realistic material refractive indices for the species listed in Table \ref{tab:species}, and particle porosity. We
note that we currently use the optical constants for (crystalline) silicates from \citet{Pol94}. However, more recent applications \citep[e.g.,][]{Bir18} tend to use (amorphous) astronomical silicates \citep{Dra03} which are more absorbing, especially at shorter wavelengths by up to an order of magnitude. We will consider these in future applications since several recently discovered hydrodynamical instabilities which can lead to sustained turbulence \citep{LU19} are operationally sensitive to the magnitude of the disk solids opacity. We have not included the Rosseland and Planck {\it gas} opacities in our models as they really only start to become important for temperatures above 2000 K which are not achieved in our simulations, though we have included them in a followup paper \citep[which also includes a model for disk winds]{Sen21} using a table lookup derived from the work of \citet{Fre14}. 

In our model, we specifically treat the evaporation fronts (EFs) - those locations in the disk where phase changes between solids and vapor can occur - of all species. The evaporation of inwardly migrating material, and the subsequent recondensation of outwardly diffused vapor onto grains, can lead to significant enhancements in solids (outside) and vapor phase (inside) an EF, as well as altering the composition of solids and vapors with implications for accretion, chemistry and mineralogy. We do not treat EFs as sharp boundaries, but allow for the ``phase'' change to occur linearly over a small temperature range $T_i - \Delta T_{\rm{EF}} \le T \le T_i + \Delta T_{\rm{EF}}$, with $\Delta T_{\rm{EF}} = 0.05$ K. Allowing for this gradual radial transition prevents unrealistic drops in opacity and temperature just inside an EF, and effectively mimics buffered temperature changes as material is evaporated or condensed over a range of nebula altitudes \citepalias[see][for more discussion]{Est16}. 
This approach works well for compact particles, but it should be acknowledged that for large particles, and especially for the large fluffy aggregates in our fractal growth simulations, it may not capture a scenario in which an aggregate’s surface layers insulate volatile material in the interior. 
An improved model would require we treat the kinetics of evaporation and condensation at EFs \citep[e.g., see][]{RJ13,SO17} which we will implement in future work. 
 
For simplicity, we have also assumed that evaporation and condensation are reversible processes, but this is almost certainly not the case for the refractory organics.  We would expect that these CHONs would likely break down irreversibly into CO, CO$_2$, NH$_3$ and ``carbon chains'' like acetylene.  These products 
would have an effect on the magnitude of enhancements in both refractory carbon solids outside, and their vapor fraction inside the organics EF \citepalias[see further discussion in][]{Est21}. 
The depletion of C in this way would be more consistent with the observation that the most primitive carbonaceous chondrites contain only $\sim 10$\% of the carbon found in comets \citep[see][]{Woodwardetal2021}. On the other hand, organics may be stickier  than what we assume in this work \citep[][though see \citealt{Bis20}]{Kou02,Hom19} which might facilitate faster growth to larger sizes, and/or also allow for retention of some organics sequestered in the interiors of the larger particles and aggregates.

\subsection{Particle Growth}
\label{sec:partgrow}

We handle particle growth using the moments method solution to the coagulation equation for the ``dust" portion $f(m)$ of the particle size  distribution  \citep{EC08,EC09} smaller than the fragmentation barrier mass $m_{\rm{f}}$ (Sec. \ref{sec:barriers}), and explicitly follow growth of ``migrators" for masses larger than $m_{\rm{f}}$. The moments method greatly speeds up calculations and allows us to focus on the more interesting evolution of the larger sizes. In the following sections, we summarize the collisional physics used in our code, and describe how we add fractal, porous growth to our model.

\subsubsection{Relative Velocities}
\label{sec:relvel}

We include both stochastic and deterministic relative velocities in our code. The former are due to either collisions between grains and gas molecules, or interaction of particles with turbulent eddies within the gas, while the latter are radial and azimuthal drift, and vertical settling, that result from the friction caused by particle-gas interactions within the mean gas flow \citep[and section \ref{sec:stoptime}] {Nak86, WC93}. 

For very small grains, thermal or Brownian motion dominates relative velocities between particles of masses $m^\prime$ and $m$. The mean relative velocity is

\begin{equation}
    \label{equ:dvB}
    \Delta v_{\rm{B}}(m^\prime,m) = \sqrt{\frac{8k_{\rm{B}}T}{\pi}{\frac{m^\prime + m}{m^\prime m}}},
\end{equation}

\noindent
where $T$ is the local temperature and $k_{\rm{B}}$ is the Boltzmann constant. Generally these stochastic motions are only important for the smallest particles in our distribution, and at the highest  temperatures found in the inner disk regions.

For the turbulence-induced relative velocities between particles of arbitrary size, we employ the closed-form expressions of \citet{OC07}\footnote{  \citet{Ish18} recently confirmed a finding by \citet{PP15}, but at much higher Reynolds number, that the Ormel \& Cuzzi model predictions for relative interparticle velocities seem to be a factor of 1.9 too large for particles with ${\rm{St}} < 1$. \citet{PP15} made some suggestions about the reasons for this, which are connected to simplifying assumptions by \citet{OC07} and earlier works on the subject, and which we have assessed in a preliminary way and found plausible. Fortunately it is easy to correct the Ormel \& Cuzzi model expressions to agree with actual numerical simulations. Moreover, \citet{PP15} also note that the actual property of interest, the {\it collisional} velocity that most people use the relative velocities to represent, is actually slightly larger, decreasing the discrepancy of the Ormel \& Cuzzi expressions to a factor of 1.3 or less. Consequently here we use the expressions in \citet{OC07}.}:

\begin{equation}
    \label{equ:dvT}
    \Delta v_{\rm{t}}(m^\prime,m) = \sqrt{\Delta v_{\rm{I}}^2 + \Delta v_{\rm{II}}^2},
\end{equation}

\noindent
in which $\Delta v_{\rm{I}}$ and $\Delta v_{\rm{II}}$ are velocity contributions from  so-called Class I and Class II eddies \citep[see][]{Vol80}. In terms of the Stokes numbers of $m$ and $m^\prime$, and the turbulent large eddy gas velocity $v_{\rm{t}} = \alpha_{\rm{t}}^{1/2}c$,

\begin{equation}
    \label{equ:dv1}
    \Delta v_{\rm{I}}^2 = v_{\rm{t}}^2\frac{{\rm{St^\prime}}-{\rm{St}}}{{\rm{St^\prime}}+{\rm{St}}}
    \left(\frac{{\rm{St^\prime}}^2}{{\rm{St}^*_{12}}+{\rm{St^\prime}}}-\frac{{\rm{St^\prime}}^2}{1+{\rm{St^\prime}}}
    -\frac{{\rm{St}}^2}{{\rm{St}^*_{12}}+{\rm{St}}}+\frac{{\rm{St}}^2}{1+{\rm{St}}}\right),
\end{equation}

\begin{equation}
    \label{equ:dv2}
    \Delta v_{\rm{II}}^2 = v_{\rm{t}}^2\left(2({\rm{St^*_{12}}} - {\rm{Re}}^{-1/2}) +\frac{{\rm{St^\prime}}^2}{{\rm{St^*_{12}}}+{\rm{St^\prime}}}-\frac{{\rm{St^\prime}}^2}{{\rm{St^\prime}}+{\rm{Re}}^{-1/2}}+\frac{{\rm{St}}^2}{{\rm{St^*_{12}}}+{\rm{St}}}-\frac{{\rm{St}}^2}
    {{\rm{St}}+{\rm{Re}}^{-1/2}}\right),
\end{equation}

\noindent
where ${\rm{Re}} = \rho\nu/\mu_{\rm{m}}$ is the gas Reynolds number. Equations (\ref{equ:dv1}) and (\ref{equ:dv2}) illustrate the different coupling that exists between particles and eddies of different sizes, and ${\rm{St^*_{12}}} = {\rm{MAX(St^{*\prime},St^*)}}$ is the larger of the two particle Stokes numbers at the boundary between these two eddy classes. We solve for ${\rm{St^*_{12}}}$ using Eq. (22) of \citet{OC07} which  takes into account eddy-crossing effects. When these effects become important, it can influence the Stokes number at which fragmentation occurs (see Sec. \ref{sec:barriers}).

We note that for very small particles, 
when ${\rm{St}} < {\rm{Re}}^{-1/2} \equiv {\rm{St}}_\eta$, $\Delta v_{\rm{t}} = \Delta v_{\rm{I}} = 0$ for same-sized particles. Here ${\rm{St}}_\eta = t_\eta/t_{\rm{ed}}$ identifies the Kolmogorov scale, where $t_{\eta}$ is the turnover time of the smallest eddy. 
However, as has been pointed out \citep[e.g.,][]{Oku11}, the dispersion in the mass-to-area ratio $m/A$ between aggregate to aggregate gives rise to a small relative velocity between them. This turns out to be particularly important in the treatment of fractal aggregates with fractal dimension $D\simeq 2$. To account for this, we 
replace the area $A$ in the differential mass-to-area ratio $\Delta(m/A)$ between aggregates with the mean projected area $\bar{A}$ of the aggregates \citep[as in][their Eq. 47]{Oku09} and the size of its standard deviation normalized by the mean mass-to-area-ratio \citep[e.g.,][]{Oku12}: 

\begin{equation}
    \label{equ:diffmA}
    \left|(m/A)\right|^2 = \left|m/\bar{A} -m^\prime/\bar{A}^\prime\right|^2 + \delta^2\left[(m/\bar{A})^2 
    +(m^\prime/\bar{A}^\prime)^2\right]],
\end{equation}

\noindent
where we adopt $\delta = 0.1$ as in \citet{Oku11}.

For larger particles, systematic pressure-induced velocities arise from drag forces (primarily azimuthal)  from the nebula gas, which rotates more slowly due to the radial gas pressure gradient. We solve for the radial and azimuthal components of the velocity for the particles and the gas  
using a set of equations generalized from \citet{Nak86} for a particle size distribution using matrix inversion \citepalias[see appendix A.4,][]{Est16}. The mean particle-to-gas deterministic relative velocity can be constructed in a similar fashion. In the case where the local solids fraction is high enough to produce a feedback effect on the gas, the relative velocities must be determined iteratively. The normalized gas pressure gradient\footnote{In \citetalias{Est16}, there is an exponent sign error in Eq. (39) where $R^{3/2}$ should be $R^{-3/2}$.}

\begin{equation}
    \label{equ:pgrad}
    \beta(R,t) = -\frac{1}{2\rho\Omega v_{\rm{K}}}\frac{\partial p}{\partial R} = -\frac{1}{2}\left(\frac{c}
    {v_{\rm{K}}}\right)^2\frac{\partial \,{\rm{ln}}\,p}{\partial \,{\rm{ln}}\, R},
\end{equation}

\noindent
where $v_{\rm{K}}= \Omega R$ is the local Kepler velocity, can be quite strong at the outer edge of the disk as $\rho$ can decrease quite sharply there. This can lead to rapid inward radial drift of even very small particles. In most cases when the local solids-to-gas mass volume  density ratio $\epsilon = \rho_{\rm{solids}}/\rho$ is not too large, the particle drift velocity can be well approximated by \citep{Wei77,CW06}

\begin{equation}
    \label{equ:uk}
    u \approx \frac{2{\rm{St}}\beta v_{\rm{K}}}{1 + {\rm{St}}^2}.
\end{equation}

\noindent
We make use of this approximation in Sec. \ref{sec:destprob}.

The rms relative systematic velocity between $m^\prime$ and $m$ due to this drag, as well as differential settling, is

\begin{equation}
    \label{equ:dvD}
    \Delta v_{\rm{D}}(m^\prime,m) = \sqrt{(u^\prime - u)^2 + (v^\prime - v)^2 + (w^\prime - w)^2},
\end{equation}

\noindent
with the vertical settling velocity $w = - \Omega^2 z t_{\rm{s}} = -\Omega z\cdot{\rm{St}}$. A detailed description of how we calculate these are given in Appendix A4 of \citetalias{Est16}.

We then obtain the traditional mean particle-to-particle collisional velocities by combining all of the relative velocity components in quadrature \citep[e.g.,][]{Tan05,Oku12}

\begin{equation}
    \label{equ:dVpp}
    \Delta v_{\rm{pp}}(m^\prime,m) = \sqrt{\left(\Delta v_{\rm{B}}\right)^2 + \left(\Delta 
    v_{\rm{D}}\right)^2 + \left(\Delta v_{\rm{t}}\right)^2}.
\end{equation}

\noindent
Our models employ a Gaussian relative velocity PDF centered on these mean values (Sec. \ref{sec:destprob}). This was acknowledged to be a simplification in \citetalias{Est16}, since it is not strictly correct to include the fluctuating turbulent relative velocity with the deterministic velocities in this way. \citet{Gar13} have further emphasized that Eq. (\ref{equ:dVpp}) is not accurate unless the systematic velocities are  dominant. In most cases, the turbulence-induced velocities are dominant for this work. For comparison, we consider a Maxwellian approach in Appendix \ref{sec:vpdf}; the differences are not large (also see Sec. \ref{sec:destprob}).

\subsubsection{Bouncing, Fragmentation, Mass Transfer and Erosion}
\label{sec:barriers}

Small grains are efficient at sticking at smaller sizes owing to their low relative velocities. But as they grow, their relative velocities increase, gradually decreasing their sticking efficiency $S$ and eventually halting further growth altogether. We account for these barriers in our collisional kernel using mass-and-velocity-dependent sticking coefficients 
for $m^\prime \le m$. One such barrier to growth derived from experimental work 
is the so-called bouncing barrier \citep{Gut10,Zso10} in which the relative velocity between two grains is too high for any sticking, but too low for fragmentation. If bouncing is included in the kernel, then  

\begin{equation}
    \label{equ:sbounce}
    S_{\rm{b}}(m^\prime,m) = 1 - \frac{m^\prime}{m^\prime + m}\frac{(\Delta v_{\rm{pp}})^2}
    {v^2_{\rm{b}}} \geq 0,
\end{equation}

\noindent
gives the first barrier encountered during growth as $S_{\rm{b}}\rightarrow 0$
where the threshold velocity $v_{\rm{b}}$ for bouncing collisions for either compact  or porous aggregates is mass-dependent. For instance, for similar-sized compact silicate particles, \citet{Gut10} found $v_{\rm{b}} \simeq (C_0/m)^{1/2}$, with $C_0 = 10^{-7}$ g cm$^2$ s$^{-2}$. For  compact grains with density $\rho_{\rm{p}}$, and when turbulence dominates their motion, the Stokes number ${\rm{St}}_{\rm{b}}$ at which bouncing of equal sized particles occurs under typical nebula conditions is approximately 

\begin{equation}
\label{equ:Stbounce}
{\rm{St}}_{\rm{b}} \simeq \left(\frac{3 \sqrt{2\pi} C_0 \rho_{\rm{p}}^2}{4 \alpha_{\rm{t}} c^2 \Sigma^3}\right)^{1/4},
\end{equation}

\noindent
where $S_{\rm{b}}=0$ defines the bouncing mass $m_{\rm{b}}$ that corresponds to ${\rm{St}}_{\rm{b}}$.
For fractal, porous grains (Sec. \ref{sec:fracgrow}), we adopt a different form for $v_{\rm{b}}$ with a knee derived from \citet{Gut10} for similar-sized aggregates:

\begin{equation}\label{equ:vbpor}
    v_{\rm{b}} \simeq 
    \begin{dcases}
    (C_1/m)^{1/2} & {\rm{if}}\, m < 3\times 10^{-7}\,{\rm{g}}; \\
    (C_2/m)^{1/4} & {\rm{if}}\, m \ge 3\times 10^{-7}\, {\rm{g}},
    \end{dcases}
\end{equation}

\noindent
with $C_1 = 1.55\times 10^{-7}$ g cm$^2$ s$^{-2}$, and $C_2 = 8\times 10^{-8}$ g cm$^4$ s$^{-4}$.The bouncing barrier does not halt growth, but merely slows it, by restricting  the size range from which a given aggregate can grow to particles smaller than $m_b$.  

Collisions become fragmenting, however, when the relative velocities between two
particles of similar size exceed a certain threshold energy $Q_{\rm{f}}$ \citep{SL09,Bei12}. In a similar way to bouncing, the fragmentation coefficient
\begin{equation}
    \label{equ:sfrag}
    S_{\rm{f}}(m^\prime,m) = 1 - \frac{m^\prime}{m^\prime + m}\frac{(\Delta v_{\rm{pp}})^2}{Q_{\rm{f}}} \geq 0,
\end{equation}

\noindent
determines the fragmentation mass $m_{\rm{f}}$ when $S_{\rm{f}}\rightarrow 0$.
Unless turbulence is very weak, the highest relative velocities tend to be between similar-sized particles, so that fragmentation will first occur there.
The fragmentation Stokes number in turbulence can be similarly derived from Eq. (\ref{equ:dv2}) when ${\rm{St}} = {\rm{St^\prime}}$ and dropping terms in ${\rm{Re}}^{-1/2}$ which are negligible in the ``fully intermediate regime'' \citep{OC07}:

\begin{equation}
    \label{equ:Stfrag}
    {\rm{St_f}} = \frac{(y^* + 1)}{y^{*2}}\frac{Q_{\rm{f}}}{\alpha_{\rm{t}}c^2} \simeq \frac{Q_{\rm{f}}}
    {\alpha_{\rm{t}}c^2}.
\end{equation}

\noindent
 We include the factor $y^* = {\rm{St^*}}/{\rm{St}}$ here explicitly to distinguish between cases where eddy-crossing effects are important. When ${\rm St} \ll 1$, 
 $y^* \approx 1.6$ so that ${\rm{St_f}}$ takes on the more familiar form of the last term on the RHS of Eq. (\ref{equ:Stfrag}). However, eddy-crossing effects can become important when ${\rm{St}}/\alpha_{\rm{t}} \gtrsim \eta^{-1}$ \citep[][also see \citealt{Jac12}]{OC07} so that $y^*$ can be greater than 1.6, and thus decrease the size at which fragmentation occurs. On the other hand, when ${\rm{St}} \sim {\rm{Re}}^{-1/2}$, Eq. (\ref{equ:Stfrag}) can underestimate the fragmentation size because terms in Re$^{-1/2}$ cannot be ignored. 
 Our models cover the full range, and both situations can and do occur.
 
It can be seen from Eq. (\ref{equ:Stfrag}) that in regions of the disk where the temperature does not vary significantly ({\it i.e.}, in the outer disk) ${\rm{St_f}}$ is nearly constant, though the value of $Q_{\rm{f}}$ may vary with composition which we allow for in our models. For silicate particles we take $Q_{\rm{f}} = 10^4$ cm$^2$ s$^{-2}$ \citep[equivalent to a threshold velocity of 1 m s$^{-1}$, e.g.,][]{BM93,SL09}, and $Q_{\rm{f}} = 10^6$ cm$^2$ s$^{-2}$ for water ice grains which are apparently stronger, and may grow larger and more rapidly than silicate grains \citep{Wad09,Wad13,Oku12}, though perhaps only over a limited range of temperatures near the evaporation point \citep{MW19}. For the more volatile ices (and for water-ice in two of our models which we refer to as ``cold H$_2$O ice'', when $T\lesssim 140$ K, see Table \ref{tab:models} and Appendix \ref{sec:varparm}), we adopt $Q_{\rm{f}}$ to be the same as for silicate particles \citep{Mus16}. Given that our aggregate particles are composed of different fractions of icy and non-icy species, we do a mass fraction-weighted mean $Q_{\rm{f}}(R) = \sum_i Q^{\rm{f}}_i x^{\rm{d}}_i(R)/f_{\rm{d}}(R)$ to determine the fragmentation energy at any disk location. 
A similar approach is taken for the bouncing threshold \citepalias{Est16}.

Finally, the outcomes of high-velocity collisions have been observed experimentally to lead to growth \citep{Wur05,Kot10} when a mutual collision between different-sized particles can erode but not fragment the larger particle, and is enough to fragment the smaller of the two. In such a circumstance, it is possible for there to be deposition of mass on the larger particle if the efficiency of accretion exceeds that of erosion. This is the so-called {\it mass transfer regime}. When one considers a PDF of velocities, as we do for growth beyond the fragmentation barrier (Sec. \ref{sec:destprob}), mass transfer can allow so-called ``lucky particles" to grow  much larger  than the mass-dominant particle size of the distribution. We use the model of \citet[][also see \citealt{Bei12}]{Win12b,Win12c} to account for mass transfer, and the description of our implementation is fully described in \citetalias{Est16}.


\subsubsection{Destruction Probabilities Beyond the Fragmentation Barrier}
\label{sec:destprob}

For particle growth beyond the fragmentation mass $m_{\rm{f}}$, we employ a statistical scheme using a PDF of relative velocities to determine the probability that a migrator particle ({\it i.e.} $m > m_{\rm{f}}$) is destroyed due to fragmenting collisions with equal-size and smaller particles in the distribution. For compact particles, the probability that a migrator of mass $m$ will be fragmented during a time $\Delta t$ by some $m^\prime \le m$ is given by

\begin{equation}
    \label{equ:probfrag}
    {\mathcal{P}}(m) = \int_t^{t+\Delta t} dt^\prime \int_{m_0}^{m} \pi
    (r^\prime + r)^2 \Delta v_{\rm{pp}}(m^\prime,m)\zeta(m^\prime,m)f(m^\prime,t^\prime)\,dm^\prime,
\end{equation}

\noindent
where $\zeta(m^\prime,m)$ is an integral over the PDF of relative velocities. In this work, as we did in \citetalias{Est16}, we primarily use 
a Gaussian PDF of relative velocities with rms velocity equal to the mean \citep{Car10,Hub12,PP13}

\begin{equation}
    \label{equ:gausspdf}
    \zeta(m^\prime,m) = \frac{1}{\sqrt{2\pi}\Delta v_{\rm{pp}}}\int_{v_c(m)}^\infty 
    e^{-\frac{(v^\prime-\Delta v_{\rm{pp}})^2}{2\Delta v^2_{\rm{pp}}}}\,dv^\prime = \frac{1}{2}
    \left[1 - {\rm{erf}}\left(\frac{v_c-\Delta v_{\rm{pp}}}{\sqrt{2}\Delta 
    v_{\rm{pp}}}\right)\right],
\end{equation}

\noindent
where the integration is over the range equal to, or greater than the critical impact velocity\footnote{There is a typo of a factor of 2 in the definition of the critical impact velocity in \citetalias{Est16}.} $v_c = \sqrt{Q_{\rm{f}}(m^\prime + m)/m^\prime}$. In Eq. (\ref{equ:gausspdf}, and also \ref{equ:garaudpdf} in the Appendix \ref{sec:vpdf}), $v^\prime$ is used as an integration variable and should not be confused with the azimuthal velocity component of a particle as defined in Sec. \ref{sec:relvel}.

Although other workers have utilized similar Gaussian PDFs, it has been argued that a Maxwellian distribution may be more appropriate, especially when relative velocities are dominated by Brownian or turbulent motion \citep{Gal11,Win12a,Gar13}. To that end, we also explore models that utilize a Maxwellian PDF \citep{Gar13} in Appendix \ref{sec:vpdf} in order to compare with our usage of a Gaussian PDF. 

\subsubsection{Fractal Growth}
\label{sec:fracgrow}

Initially, our models begin with a ``dust" powerlaw size distribution in mass such that  particle number density $f(m) \propto m^{-q}$ where $q = 11/6$, composed of compact, spherical grains with a minimum size of $r_0 = 0.1$ $\mu$m, and thus a monomer mass is $m_0 = (4/3)\pi\rho_{\rm{p}}r^3_0$ where we define $\rho_{\rm{p}}$ as the compacted density, or the density of a compact particle\footnote{That is, if the particle internal  density is constant, the number of particles $dN(m,dm)$ lying in a mass bin between $(m, m+dm)$ is $f(m)dm = N_1m^{-11/6}dm$ and expressed in terms of particle radius is $dN(r,dr)=N_2r^{-7/2}dr$.}. In this work, an initial monomer density that contains all of the solid phase species in Table \ref{tab:species} is $\rho_{\rm{p}} = 1.52$ g cm$^{-3}$ (though note $\rho_{\rm{p}}$ varies both temporally and spatially), which yields a monomer mass $m_0 = 6.4\times 10^{-15}$ g. As particles grow, they remain spherical aggregates regardless of whether growth is compact, or fractal - we cannot make a distinction between them based on shape\footnote{In our models, we assume that the lower bound of our particle-size distribution $r_0$ remains a monomer which is equivalent to assuming that fragmentation leads to redistribution of solids in the form of a power law down to this minimum size \citepalias[see][]{Est16}. However, the moments method \citep{EC08} does allow for the lower bound to be evolved as well using an additional moment. The expectation would be that this would even further restrict growth because much of the feedstock for continued particle growth comes from the smaller sizes, especially if the bouncing barrier mass is exceeded.}. 
As demonstrated in \citet{EC08} with regards to the moments method we utilize for the particle distribution with $m < m_{\rm{f}}$, fractal growth can be incorporated into the collisional kernel without any loss of generality. Porous particles of mass $m$ and radius $r$ can be defined in terms of a fractal dimension $2 \le D \le 3$ as $m = m_0(r/r_0)^D$ and $r = r_0(m/m_0)^{1/D}$.  The density, or similarly the volume, of a fractal, porous particle can also be defined in terms of a enlargement factor or inverse filling factor $\psi$  \citep[e.g.,][]{Orm07,Kri15} as

\begin{equation}
    \label{equ:psi}
    \frac{V_{\rm{p}}}{V_{\rm{comp}}} = \frac{\rho_{\rm{p}}}{\rho_{\rm{int}}} \equiv \psi,
\end{equation}

\noindent
where $V_{\rm{comp}}$ is the volume the compact ($D = 3$) particle of equivalent mass would occupy, and $\rho_{\rm{int}}$ is the internal density of a porous aggregate. A familiar description of a porous aggregate is its porosity $\varphi$ which is 1 minus the volume filling fraction: $\varphi = 1-\rho_{\rm{int}}/\rho_{\rm{p}}$. Thus, $\psi = 1/(1-\varphi)$, and the relationship between the fractal dimension and enlargement factor is 

\begin{equation}
    \label{equ:psiD}
    D = 3 \left(1 + \frac{{\rm{ln}}\,\psi}{{\rm{ln}}\,(m/m_0)}\right)^{-1}.
\end{equation}

\noindent
The stopping time in the Epstein regime (Eq. \ref{equ:stoptime})  depends on the midplane gas density $\rho = \Sigma/\sqrt{2\pi}H$ and can be rewritten in terms of the Stokes number and surface mass density \citep{EC08}\footnote{Note typo in equation 34 of \citet{EC08}, while their equation 36 is correct.} as:

\begin{equation}
    \label{equ:fracSt}
    {\rm{St}} = \frac{3\sqrt{2\pi}m_0^{2/D}}{4\pi r^2_0 \Sigma} m^{1 - 2/D}, 
\end{equation}

\noindent
It is interesting to note that  $D=2$ represents a special case that an aggregate of any size and mass has the same Stokes number as a monomer, ${\rm{St}}=\sqrt{2\pi}\rho_{\rm{p}}r_0/\Sigma$, and would remain strongly coupled to the gas. However, aggregates will eventually suffer compaction through collisions, and their ${\rm{St}}$ will grow. The main effect of porosity is that porous aggregates ($D<3$) will have smaller ${\rm{St}}$ than completely compact particles of the same mass. Thus, they will not drift as quickly, will be diffused more easily, and may have the potential to overcome the radial drift barrier.

\subsubsection{Compaction Model}
\label{sec:collcomp}


Fractal growth of particles by low-velocity sticking of compact monomers of mass $m_0$ and radius $r_0$ initially leads to very low density, fluffy aggregates, but as relative velocities increase, these aggregates will be subject to compaction due to collisions with other aggregates. Modeling of this process is generally done with Monte Carlo coagulation codes that allow individual collisions  to be followed directly in order to quantitatively assess the amount of compaction that occurs \citep[e.g.,][]{Orm07,Zso10,Kri15,Lor18} in a growing aggregate using a recipe for collisional compaction. However, the computational space is limited in these models, because adding porosity and compaction makes the already cumbersome Smoluchoswki equation multi-dimensional and thus far more numerically intensive to solve. Apart from being impractical for a global model such as ours, it is probably too elaborate given the uncertainties.

\citet[][see also \citealt{HN18}]{Oku12} solve the coagulation equation explicitly for the particle mass distribution $f(m)$, and have addressed  compaction by deriving a corresponding equation for $f(m)V(m)$ using a volume averaging approximation \citep{Oku09}, which reduces the computational burden of solving a multi-dimensional equation to solving two one-dimensional ones. In this approximation, $V(m)$ represents an average or characteristic volume for aggregates of mass $m$. However, this technique for including compaction does not lend itself readily to our method of moments approach, even if pure fractal growth does (Sec. \ref{sec:fracgrow}). 

Since we are most concerned with the rate of growth of the largest particle in the distribution, we are only interested in {\it its} average volume. Thus we apply a statistical approach where we calculate the change in volume by an estimate of the number of collisions an aggregate experiences due to other aggregates of different sizes in a time step $\Delta t$. The dust population then is characterized by a single $D(t)$ which characterizes the mass dominant particle. 
On the other hand, because we follow the growth of migrator particles explicitly beyond the fragmentation barrier, migrators can have individually different $D$. We describe our procedure below. In what follows, unprimed quantities refer to the growing aggregate, and primed to the impactor.

The primary goal of a recipe for collisional compaction is to determine the enlargement factor $\psi$ of a growing aggregate $m$ after a collision. Much like our treatment of bouncing and fragmentation (Eqns. [\ref{equ:sbounce}] and [\ref{equ:sfrag}]), at the heart of any growth with compaction scheme is consideration of the collisional impact energy between an aggregate of mass $m$ and other aggregates with $m^\prime \le m$: 

\begin{equation}
    \label{equ:Eimp} 
    E_{\rm{imp}} = \frac{1}{2}\frac{m^\prime m}{m^\prime + m}(\Delta v_{\rm{pp}})^2,
\end{equation}

\noindent
which is the total kinetic energy in the center-of-momentum frame.
The amount of compaction an aggregate suffers depends on the ratio between the impact energy and the so-called ``rolling energy'' $E_{\rm{roll}}$, which is defined as the energy required for a single monomer to roll over another by an angle of 90$^o$ \citep{DT97,BW00}. Physically, low energy collisions ($E_{\rm{imp}}\ll E_{\rm{roll}}$) lead to perfect sticking with little or no internal restructuring of an aggregate, while for increasing relative velocities ($E_{\rm{imp}} > E_{\rm{roll}}$), collisions begin to 
modify the aggregate's internal structure. 
For equal-sized monomers, \citet{DT97} and \citet{BW00} give an expression for $E_{\rm{roll}}$ which we write here as

\begin{equation}
    \label{equ:Eroll}
    E_{\rm{roll}} = \frac{1}{2}\pi r_0 (\gamma/\gamma_0)F_{\rm{roll}}
\end{equation}

\noindent
where the rolling force $F_{\rm{roll}} = (8.5 \pm 1.6)\times 10^{-5}$ g cm s$^{-2}$ and the surface energy density $\gamma_0 = 14 \pm 2$ g s$^{-2}$ are measured for uncoated SiO$_2$ spheres \citep{Hei99}. The  factor $\gamma/\gamma_0$ \citep[e.g.,][]{Orm07} allows us a simple way to scale the rolling energy for different materials. For this work, we choose  the most recent values of $\gamma \simeq 25$ and $200$ g s$^{-2}$ \citep{Gun11,Kri14} for silicates and water ice, respectively. As in the case for fragmentation and bouncing, because our aggregates are composed of both ice and non-ice, we do a weighted average $E_{\rm{roll}}(R) = \sum_i E^{\rm{roll}}_i x^{\rm{d}}_i(R)/f_{\rm{d}}(R)$ to determine the rolling energy at any disk location \citep[see also][]{Kri18}.  

In the limit $E_{\rm{imp}} \ll E_{\rm{roll}}$, \citet{Oku09} derived an empirical formula for the merging of two aggregates with volumes $V_{\rm{p}}$ and $V^\prime_{\rm{p}}$ via a hit-and-stick collision $V_{\rm{HS}} = V^\prime_{\rm{p}} + V_{\rm{p}} + V_{\rm{void}}$, where

\begin{equation}
    \label{equ:vvoid}
    V_{\rm{void}} = {\rm{MIN}}\left\{0.99 - 1.03\,{\rm{ln}}\left(\frac{2}{V_{\rm{p}}/V^\prime_{\rm{p}}+1}
    \right),\,6.94\right\}V^\prime_{\rm{p}}.
\end{equation}

\noindent
For $V_{\rm{p}} \sim V^\prime_{\rm{p}}$, $V_{\rm{HS}} \simeq 3V_{\rm{p}}$ which is basically equivalent to pure fractal growth where $D \simeq 2$, and aggregates grow without any {\it net} compaction, with a nonuniform internal structure. In the opposite limit $(E_{\rm{imp}} \gg E_{\rm{roll}})$ for a head-on collision of equal-sized aggregates, the ``merged'' volume is now explicitly a function of the rolling energy \citep{Suy08,Wad08,Oku12}:

\begin{equation}
    \label{equ:Vhigh}
    \psi(E_{\rm{imp}} \gg E_{\rm{roll}}) = \frac{1}{V_{\rm{comp}}}\left[\frac{(3/5)^5E_{\rm{imp}}}{N^5 b E_{\rm{roll}} V^{10/3}_0} + 
    \left(2V^{5/6}_{\rm{p}}\right)^{-4}\right]^{-3/10}.
\end{equation}

\noindent
Here, $V_0 = m_0/\rho_0$ is the volume of a monomer, $b=0.15$ is a non-dimensional fitting parameter and $N$ is the total number of monomers contained in the merged volume. For the intermediate case when $E_{\rm{imp}}\sim E_{\rm{roll}}$, \citet{Oku12} adopt and update the analytical fit of \citet{Suy12} to complete the porous compaction recipe for the new value of the enlargement factor:

\begin{equation}
    \label{equ:fullcomp}
    \psi(V_{\rm{p}},V^\prime_{\rm{p}}) = \frac{1}{V_{\rm{comp}}}
    \begin{dcases}
    \left[(1-\xi)V^{5/6}_{\rm{HS}} + \xi\left(V^{\prime 5/6}_{\rm{p}}+V^{5/6}_{\rm{p}}\right)\right]^{6/5} &
    {\rm{if}}\,V^{5/6}_{\rm{HS}} > V^{\prime 5/6}_{\rm{p}}+V^{5/6}_{\rm{p}}\,{\rm{and}}\,\xi < 1, \\
    \left[\frac{3^6(\xi - 1)}{(5N)^5 V^{10/3}_0} + \left(V^{\prime 5/6}_{\rm{p}}+
    V^{5/6}_{\rm{p}}\right)^{-4}\right]^{-3/10} & {\rm{if}}\,V^{5/6}_{\rm{HS}} > V^{\prime 5/6}_{\rm{p}}
    +V^{5/6}_{\rm{p}}\,{\rm{and}}\,\xi > 1, \\    
    \left[\frac{3^6\xi}{(5N)^5 V^{10/3}_0} + V^{-10/3}_{\rm{HS}}\right]^{-3/10} &
    {\rm{if}}\,V^{5/6}_{\rm{HS}} < V^{\prime 5/6}_{\rm{p}} +V^{5/6}_{\rm{p}},
    \end{dcases}
\end{equation}

\noindent
where $\xi = E_{\rm{imp}}/3bE_{\rm{roll}}$.

In order to implement this scheme in our code, we approximate the compaction of a particle statistically, due to the number of collisions it suffers with other aggregates in a given $\Delta t$. The number of collisions that a particle of mass $m \ge m^\prime$ experiences due to collisions with particles of mass $m^\prime$ in a time $\Delta t$ is estimated using

\begin{equation}
    \label{equ:tcoll}
    N_{\rm{coll}} = \frac{\Delta t}{t_{\rm{coll}}} = \sqrt{2}\pi(r^\prime +r)^2n(m^\prime)
    \Delta v_{\rm{pp}}(m^\prime,m)\Delta t,
\end{equation}

\noindent
where $n(m^\prime)$ is the number density of particles of $m^\prime$ \citepalias[e.g., see Eq. 75 of][]{Est16}. We then sum over all collisions with aggregates $m^\prime$ using  Eq. (\ref{equ:fullcomp}) modifying $V_{\rm{p}}$ after each collision. We do not include all particle masses as this would  quickly become cumbersome for a very large difference in mass between the aggregates (see below). \citet[][also see \citealt{Kri15}]{Oku12} note several caveats in using their prescription, in particular, that their formulation is only tested for similar-sized aggregates with mass ratio $m^\prime/m \gtrsim 1/16$. For these authors, as in the case here, this is not a major concern since the majority contribution to growth in our models occurs between similar-sized particles. 
For $m < m_{\rm{f}}$, the end result of tallying all collisions using Eq. (\ref{equ:fullcomp}) is a new $D$ derived for the largest particle\footnote{In the moments method, the largest particle in the particle mass distribution is, by definition, \begin{equation*} m_{\rm{L}} \equiv \frac{\int m^2 f(m) dm}{\int m f(m) dm}, \end{equation*} and thus is equivalent to the mass-weighted mean of the distribution \citepalias{Est16}.} $m=m_{\rm{L}} \le m_{\rm{f}}$ after each growth step, which is then used in the next growth step for the entire population with $m < m_{\rm{f}}$. In this way the particle distribution ($m<m_f$) has a single power-law dependence for the enlargement factor $\psi(m) = \psi_{\rm{d}} = (m/m_0)^{3/D-1}$ consistent with our assumption of a power-law size distribution for the particle $f(m)$ for masses less than the fragmentation mass. On the other hand, the $f(m)$ for migrators ($m>m_{\rm{f}}$) does not generally follow a power law, so that migrators can have both different $D$ and $\psi$ since the growth of the migrator branch of the particle distribution is done explicitly \citepalias{Est16}.

\begin{figure}[ht!]
\centering
\includegraphics[width=0.5\textwidth]{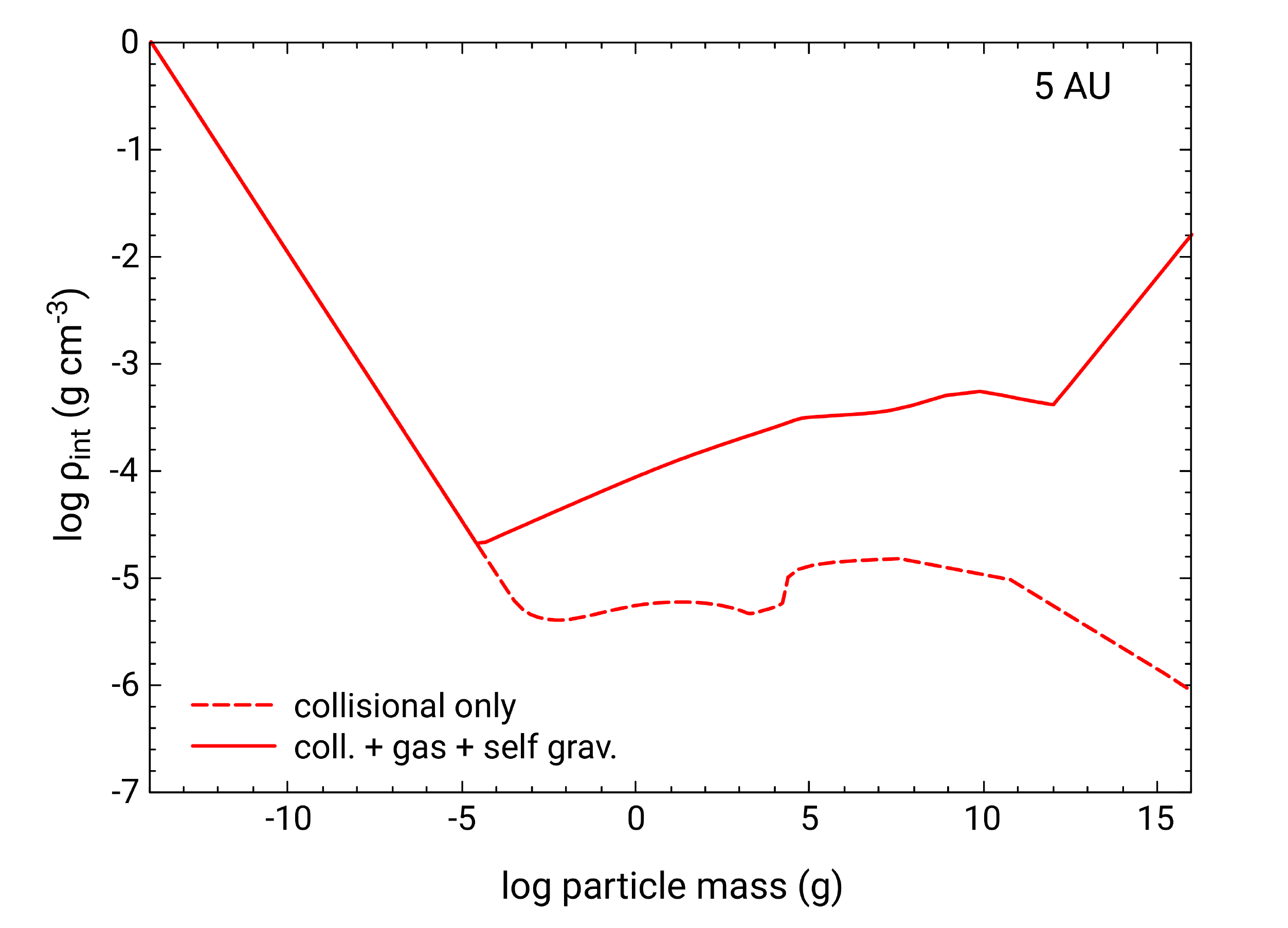}
\caption{
Evolution of the internal density $\rho_{\rm{int}} = \rho_{\rm{p}}/\psi$ of the largest (mass-dominant) aggregate for a static MMSN model at 5 AU.  Perfect sticking is assumed. Models are shown in which only collisional compaction is included (dashed curve) employing the compaction formulae from Eq. (\ref{equ:fullcomp}), and with both collisional and non-collisional compaction effects (solid curve) using Eqns. (\ref{equ:Pc}-\ref{equ:Pgasgrav}), and asssuming $\alpha_{\rm{t}}=10^{-3}$. In both cases, aggregates begin as monomers which grow fractally ($D\sim 2$) with decreasing internal density until either $E_{\rm{imp}} \gtrsim E_{\rm{roll}}$ and/or gas ram pressure begins to compact the aggregate. When non-collisional compaction effects are included, aggregates will generally continue to compact as they grow, whereas if only collisional compaction is considered, aggregates can remain very underdense even at planetesimal masses.
\label{fig:psi}}
\end{figure}

As aggregates grow to larger sizes, collisions between these and similar-sized aggregates can occur on a timescale longer than a simulation timestep. 
This is taken into account naturally when solving the full coagulation equation(s), but 
when the timestep $\Delta t < t_{\rm{coll}}$, important collisions would be 
omitted 
that otherwise might occur at later times. Likewise, even in the case where there are many collisions between different-sized aggregates, $N_{\rm{coll}}$ most likely will not be an integer. We account for this by rounding $N_{\rm{coll}}$ down to the next integer value $n_{\rm{coll}}$, and we count the final collision (or in the case when $N_{\rm{coll}} < 1$, $n_{\rm{coll}}=0$, the only collision), as a single collision, calculate what the contribution $\Delta \psi$ would be to $\psi$,  but then do a simple weighting in which
$\psi(n_{\rm{coll}}+1) = \psi(n_{\rm{coll}}) + \Delta \psi (N_{\rm{coll}}-n_{\rm{coll}})$. This assures that there are contributions to the volume change of an aggregate of mass $m$ from the full range of particles with mass $m^\prime \le m$ we consider during each time step.

In this work, we also consider the compaction for porous aggregates due to gas ram pressure, or via their own self-gravity. In general, our particles do not get large enough for self-gravity to be a factor, because they don't grow sufficiently massive {\it or}  porous, but we include it for completeness. The external pressure that an aggregate of low internal density can withstand is \citep{Kat13a,Kri15}: 

\begin{equation}
    \label{equ:Pc}
    P_{\rm{c}} = \frac{E_{\rm{roll}}}{r^3_0}\psi^{-3},
\end{equation}

\noindent
which is compared to the pressures associated with the gas and self-gravity \citep{Kat13b}:

\begin{equation}
    \label{equ:Pgasgrav}
    P_{\rm{gas}} = \frac{m \Delta v_{\rm{pg}}}{\pi r^2 t_{\rm{s}}}; \,\,\,\, 
    P_{\rm{grav}} = \frac{G m^2}{\pi r^4}.
\end{equation} 

\noindent
Implementation of these conditions is quite straightforward in our code. As in the case with collisional compaction, we conduct a check on the largest particle in the particle distribution which may be as large as the fragmentation mass, and for all migrator particles in the distribution beyond the fragmentation barrier. If $ P_{\rm{gas}} > P_{\rm{c}}$, we adjust $\psi$ until these quantities are equal, and likewise in the case that $P_{\rm{grav}}$ is larger.

As a demonstration that the above approach works sufficiently well, in Figure \ref{fig:psi} we plot the evolution of the internal density $\rho_{\rm{int}}$ of the largest aggregate under collisional compaction (dashed curve), and compaction due to collisions, ram pressure and gravity (solid curve). For simplicity and comparison, we have chosen a static, \citet{Hay81} MMSN disk model at 5 AU with $\bar{Z} = 0.01$, a monomer particle density of $\rho_{\rm{p}}=1.4$ g cm$^{-3}$, and for the purpose of collisional velocities we adopt $\alpha_{\rm{t}}=10^{-3}$ \citep[cf.,][]{Oku12,Kri15,HN18}. In those models, bouncing, fragmentation and erosion are not included, so that growth to larger sizes for this test is done only using our moments method and thus strictly remains a power law where the largest particle size in the dust distribution is equivalent to the mass-weighted mean \citepalias{Est16}. We find that a mass ratio of $m^\prime/m \gtrsim 6.25\times 10^{-6}$ works fairly well to reproduce the qualitative behavior in the evolution of the internal density $\rho_{\rm{int}} = \rho_{\rm{p}}/\psi$ as seen by other workers. 

\begin{table}[h!]
\renewcommand{\thetable}{\arabic{table}}
\centering
\caption{List of Simulations} \label{tab:models}
\begin{threeparttable}

\begin{tabular}{lcccccc}
\tablewidth{0pt}
\hline
\hline

\tnote{a} Model & Type & $\alpha_{\rm{t}}$ & $M_{\rm{disk}}$ (M$_\odot$) & $R_0$ (AU) & $Q^{\rm{ice}}_{\rm{f}}$ (cm$^2$ s$^{-2}$) \\
\hline
fa2g & fractal & $10^{-2}$ & 0.2 & 20 & $10^6$ \\
\tnote{c} fa2m & fractal & $10^{-2}$ & 0.2 & 20 & $10^6$ \\
fa3g & fractal & $10^{-3}$ & 0.2 & 20 & $10^6$  \\
\tnote{c} fa3m & fractal & $10^{-3}$ & 0.2 & 20 & $10^6$  \\
\tnote{b} fa3Qg & fractal & $10^{-3}$ & 0.2 & 20 & $10^4$  \\
\tnote{b} fa3R060g & fractal & $10^{-3}$ & 0.2 & 60 & $10^6$  \\
\tnote{b} fa3M01g & fractal & $10^{-3}$ & 0.1 & 20 & $10^6$ \\
\tnote{b} fa3M002g & fractal & $10^{-3}$ & 0.02 & 20 & $10^6$ \\
fa4g & fractal & $10^{-4}$ & 0.2 & 20 & $10^6$  \\
fa5g & fractal & $10^{-5}$ & 0.2 & 20 & $10^6$  \\
sa2g & compact & $10^{-2}$ & $0.2$ & 20 & $10^6$  \\
sa3g & compact & $10^{-3}$ & $0.2$ & 20 & $10^6$  \\
\tnote{c} sa3m & compact & $10^{-3}$ & 0.2 & 20 & $10^6$  \\
sa4g & compact & $10^{-4}$ & 0.2 & 20 & $10^6$  \\
sa5g & compact & $10^{-5}$ & 0.2 & 20 & $10^6$  \\

\hline
\end{tabular}
\begin{tablenotes}
 \item[a] g = Gaussian, m = Maxwellian \citep{Gar13}.
 \item[b] These models are discussed in Appendix \ref{sec:varparm}. 
 \item[c] These models are discussed in Appendix \ref{sec:vpdf}. 
\end{tablenotes}
\end{threeparttable}
\end{table}

Growth proceeds fractally ($D\sim 2$) up until the point where $E_{\rm{imp}} \approx E_{\rm{roll}}$ which occurs roughly at $m \simeq 10^{-5}$ g, whence compaction begins. However, we find that there is a slight delay as to when the ``turnover'' occurs where the aggregate's internal density remains roughly constant in the collisional-compaction-only case. The slight increase in internal density occurring near $m\sim 1$ g is also where $r \simeq \lambda_{\rm{mfp}}$, and is likely due to the bridging function 
we use between Epstein and Stokes regimes \citep[][see \citetalias{Est16}]{Pod88}, while a decrease in $\rho_{\rm{int}}$ as $r \gtrsim (9/4)\lambda_{\rm{mfp}}$ ($m \sim 10^3$ g) is because the aggregate has transitioned to the Stokes regime. The sharp increase in $\rho_{\rm{int}}$ seen at $m \sim 10^4-10^5$ g is the point where ${\rm{St}} \simeq {\rm{St}}_\eta$. 
Here the aggregate  first begins to experience type II eddies (section \ref{sec:relvel}), which cause a sharp increase in the turbulent relative velocity to occur, leading to further compaction \citep{Zso10}. 
At around $m \simeq 10^{10}$ g, ${\rm{St}} \gtrsim 1$ and the radial drift velocity begins to decrease, causing the internal density to decrease again.

On the other hand, for the case that also includes non-collisional compaction effects (Fig. \ref{fig:psi}, solid curve), we note that the point at which $E_{\rm{imp}} = E_{\rm{roll}}$ roughly coincides with the point where gas ram-pressure compaction occurs. This is likely because $\Delta v_{\rm{pg}}$ is not too different from $\Delta v_{\rm{pp}}$ in the intermediate regime. 
From then, the evolution is dictated by the flow regime the aggregate is in. Initially in the Epstein regime, the internal density increases systematically as it becomes more and more decoupled from the gas with $\rho_{\rm{int}} \propto (\Delta v_{\rm{pg}})^{1/3}$. The point at which the particle enters the Stokes regime ($r \sim 9/4\lambda_{\rm{mfp}}$) occurs at a higher mass than the collisional-compaction-only case. 
Once in the Stokes regime, the three different sub regimes are apparent - the internal density increases more slowly when ${\rm{Re}}_{\rm{p}} < 1$, $\rho_{\rm{int}} \propto (\Delta v_{\rm{pg}})^{1/3}\,r^{-1/3}$, but begins to increase more sharply in the intermediate (${\rm{Re}}_{\rm{p}} \gtrsim 1$) sub-regime where $\rho_{\rm{int}} \propto (\Delta v_{\rm{pg}})^{7/15}\,r^{-1/5}$, and finally $\rho_{\rm{int}} \propto (\Delta v_{\rm{pg}})^{2/3}$ when ${\rm{Re}}_{\rm{p}} \gg 1$. The internal density decreases once ${\rm{St}} \gtrsim 1$, as the particle-to-gas relative velocity decreases ($m \sim 10^{10}$ g) until self-gravity compaction kicks in around $m \sim 10^{12}$ g.

\section{Results} 
\label{sec:results}

As we did in \citetalias{Est16}, we choose as our baseline model a nebula with a stellar mass of $M_\star = 1$ M$_\odot$, and an initial disk mass of $M_{\rm{disk}} = 0.2$ M$_\odot$. We adopt a fiducial turbulent parameter of $\alpha_{\rm{t}} = 10^{-3}$, but we explore a range of values from $10^{-5}-10^{-2}$. The value of the scaling parameter $R_0$ (Sec. \ref{sec:gasevol}) is a free parameter which sets the initial surface density profile in terms of the initial disk mass. The rationale for the chosen value of $R_0$ in the literature has not always been clear, however. A value of $R_0 = 10$ AU was preferred by \citet{Har98} based on young star statistics \citepalias[a value used by][]{Est16}, but it has been set to even larger values of $20-60$ AU \citep{CC06,Gar07,Bra08,HA12,YC12}. Using a different functional form, \citet{Cuz03} chose the equivalent parameter to match the specific angular momentum of the solar system. In this paper we nominally choose $R_0 = 20$ AU, giving general agreement with the radially compact solar nebula of \citet{Des07}, but explore a model with $R_0 = 60$ AU for comparison, and  we also consider models with different initial disk masses to show how these choices influence our model simulations. The latter models are compared with our fiducial model in Appendix \ref{sec:varparm}. All of our models include the growth barriers, mass transfer and erosion discussed in Sec. \ref{sec:barriers}. In total, we have conducted 15 simulations for both fractal and compact particle growth which are summarized in Table \ref{tab:models}. We focus in this paper on the evolution of the particle mass and porosity distribution, and the ambient properties of the disk. We discuss the evolution of disk bulk composition through the redistribution of refractory and volatile species for these same models in our companion \citetalias{Est21}.

When discussing the evolving particle mass distribution over time, we will in some plots overlay approximations of the different barrier masses which we derive from their Stokes number expressions \citepalias[][and also Sec. \ref{sec:barriers}]{Est16}. The first is the bouncing barrier, which we estimate by equating the threshold bouncing velocity to the turbulent velocity:  $v_{\rm{b}} \sim c\sqrt{2\alpha_{\rm{t}}{\rm{St}}}$. These particles are small and always in the Epstein regime, so a general expression is

\begin{equation}
    \label{equ:mb}
    m_{\rm{b}} \approx \left[\left(\frac{4\pi}{3\rho^2_{\rm{p}}}\right)^{1/3}\frac{\Sigma C_{\rm{b}}}{2\sqrt{2\pi}c^2\alpha_{\rm{t}}} m_0^{2/3-2/D}\right]^{n_{\rm{b}}},
\end{equation}

\noindent
where for compact particles ($D=3$) $C_{\rm{b}} = C_0$ and $n_{\rm{b}} = 3/4$. For fractal aggregates, $C_{\rm{b}} = C_1$ and $n_{\rm{b}} = D/(2D-2)$, or $C_{\rm{b}} = \sqrt{C_2}$ and $n_{\rm{b}} = 2D/(3D-4)$ for the limits on $m$ defined in Eq. (\ref{equ:vbpor}). Similarly for the fragmentation barrier, we equate $Q_{\rm{f}} \sim c^2\alpha_{\rm{t}}{\rm{St}}$ and find an expression for $m_{\rm{f}}$ in the Epstein regime 

\begin{equation}
    \label{equ:mf}
    m_{\rm{f}} \approx \left[\left(\frac{4\pi}{3\rho^2_{\rm{p}}}\right)^{1/3}\frac{\Sigma Q_{\rm{f}}}{\sqrt{2\pi}c^2\alpha_{\rm{t}}} m_0^{2/3-2/D}\right]^{D/(D-2)}.
\end{equation}

\noindent
In some instances, the fragmentation may occur in the Stokes regime. In such cases, using the appropriate expression for St \citepalias[for instance, see Eq. 58 of][]{Est16} will lead to a better approximation. In terms of mass, the particle where the transition to the Stokes regime occurs is $m_\lambda = m_0(9\lambda_{\rm{mfp}}/4r_0)^D$. We note that these barriers correspond to collisions by same-sized aggregates.
Finally, the radial drift barrier can crudely be approximated by equating the radial drift time $t_{\rm{dr}} = R/u \sim (1 + {\rm{St}}^2)/2\beta\Omega {\rm{St}}$ of an aggregate to its growth time $t_{\rm{gr}} = m/\dot{m}$, where $\dot{m} \sim \pi r^2 \rho_{\rm{solids}} \Delta v_{\rm{pp}} \bar{S}$ with $\bar{S} < 1$ some average sticking coefficient. We nominally choose $\bar{S} = 0.5$, though with bouncing included, a smaller value may be more appropriate which would increase growth times. Determining the appropriate drift mass $m_{\rm{d}}$ over the various drag regimes can be complicated because $t_{\rm{dr}} = t_{\rm{gr}}$ leads to a transcendental equation in St which we solve iteratively up to a maximum Stokes number of unity where $t_{\rm{dr}}$ has a minimum value. When the growth and drift times intersect, they do so first at a value of ${\rm{St}} < 1$, beyond which the growth time becomes longer. In order to reintersect the drift curve and once again have shorter growth times than drift, a particle would then have to grow sufficiently to reach $\rm{St} > 1$ which is generally not possible because drift limits their growth. Hence the radial drift barrier is not a single, well-defined mass, but an entire range of particle growth space in which the particle masses exceed $m_{\rm{d}}$, but remain smaller than $m_{\rm{f}}$ (which is a well-defined, local barrier). More discussion of $m_{\rm{d}}$ follows in the next section.



\subsection{Fiducial Model}
\label{sec:fiducial}

\begin{figure}
\includegraphics[width=1.0\textwidth]{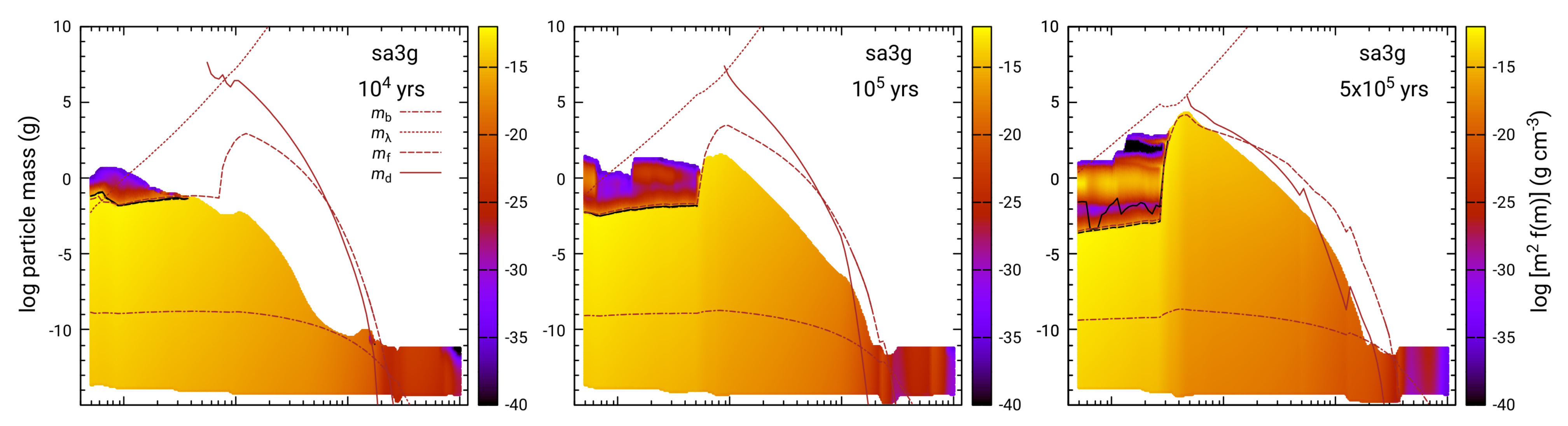}
\includegraphics[width=1.0\textwidth]{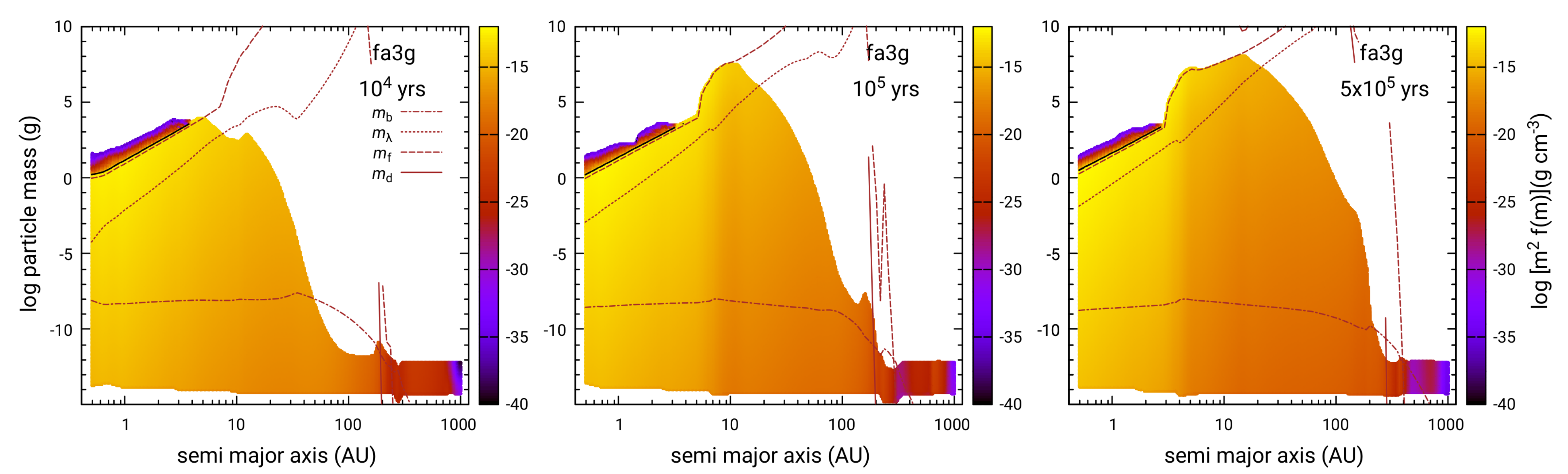}
\caption{Fiducial model with $\alpha_{\rm{t}}=10^{-3}$ for both compact (sa3g) and fractal (fa3g) growth at, from left-to-right, $10^4$, $10^5$ and $5\times 10^5$ years. Plotted in each panel is a snapshot of the particle mass distribution as a function of semi-major axis. The color scale on the right indicates the particle solids mass volume density per bin of width $\Delta m$, $m^2 f(m)$. 
The brown curves overlaying the distributions are the mass that corresponds to the mean free path $m_\lambda$ (dotted), and estimates for the bouncing $m_{\rm{b}}$, fragmentation $m_{\rm{f}}$, and radial drift $m_{\rm{d}}$ curve masses (above which growth times are longer than drift times, see text and Sec. \ref{sec:barriers}) as indicated in the leftmost panels. When and where present, the dashed black curve corresponds to locations where the actual fragmentation barrier has been reached. Growth beyond the fragmentation barrier occurs to varying degrees, up to ``lucky particles'' which are few in number and contain little mass. When the mass distribution extends beyond the fragmentation barrier, the mass dominant particle (particle size containing most of the mass) is shown by the black solid curve. 
We note that in the fractal models, the $m_{\rm{d}}$ increases very steeply to masses well beyond the plotted range. We have chosen to have a break in the curves ($m_{\rm{f}}$ as well) so that they do not cut through the upper right label and legend. 
The particle size distributions for models sa3g and fa3g after 0.5 Myr can be seen in Appendix \ref{sec:sizes}.
\label{fig:a3g}}
\end{figure}

In this section we compare our fiducial model with $\alpha_{\rm{t}} = 10^{-3}$ for both compact (sa3g) and fractal (fa3g) growth assuming the Gaussian velocity PDF (in Appendix \ref{sec:vpdf}, we compare with a Maxwellian PDF). In Figure \ref{fig:a3g}, we show the evolution of the particle mass distribution as a function of semi-major axis at $10^4$, $10^5$ and $5\times 10^5$ years. We chose to cut off the evolutions at $5\times 10^5$ years because (a) the particle properties were not evolving quickly (in great part due to the bouncing barrier, e.g., see \citealt{Zso10}), and (b) it appears that planetesimal formation (which is not taken into account in our current model) had proceeded to a significant extent in the inner nebula, and probably at the snowline, by that time \citep{Kruijeretal2017}. The corresponding color scale shows the solids mass volume density $m^2 f(m)$ per bin of width $\Delta m$ 
as a function of particle mass and disk radius. Plotted in each panel are the estimates of the bouncing barrier mass (brown dot-dashed), fragmentation mass (brown dashed), mass of a particle with the same size as the gas mean free path (brown dotted), and the radial drift mass (brown solid curve) above which particles drift in faster than they can grow ({\it i.e.}, $t_{\rm{gr}} > t_{\rm{dr}}$). When present, we also plot the actual fragmentation mass (black dashed), and the mass dominant particle, if it is beyond the fragmentation barrier (black solid curve). When the fragmentation barrier has not been reached, the mass dominant particle is simply the most massive particle in the distribution. The minimum mass in the distribution corresponds to the monomer mass. These curves are most easily studied on-screen with some magnification. 

\begin{figure*}

\gridline{\fig{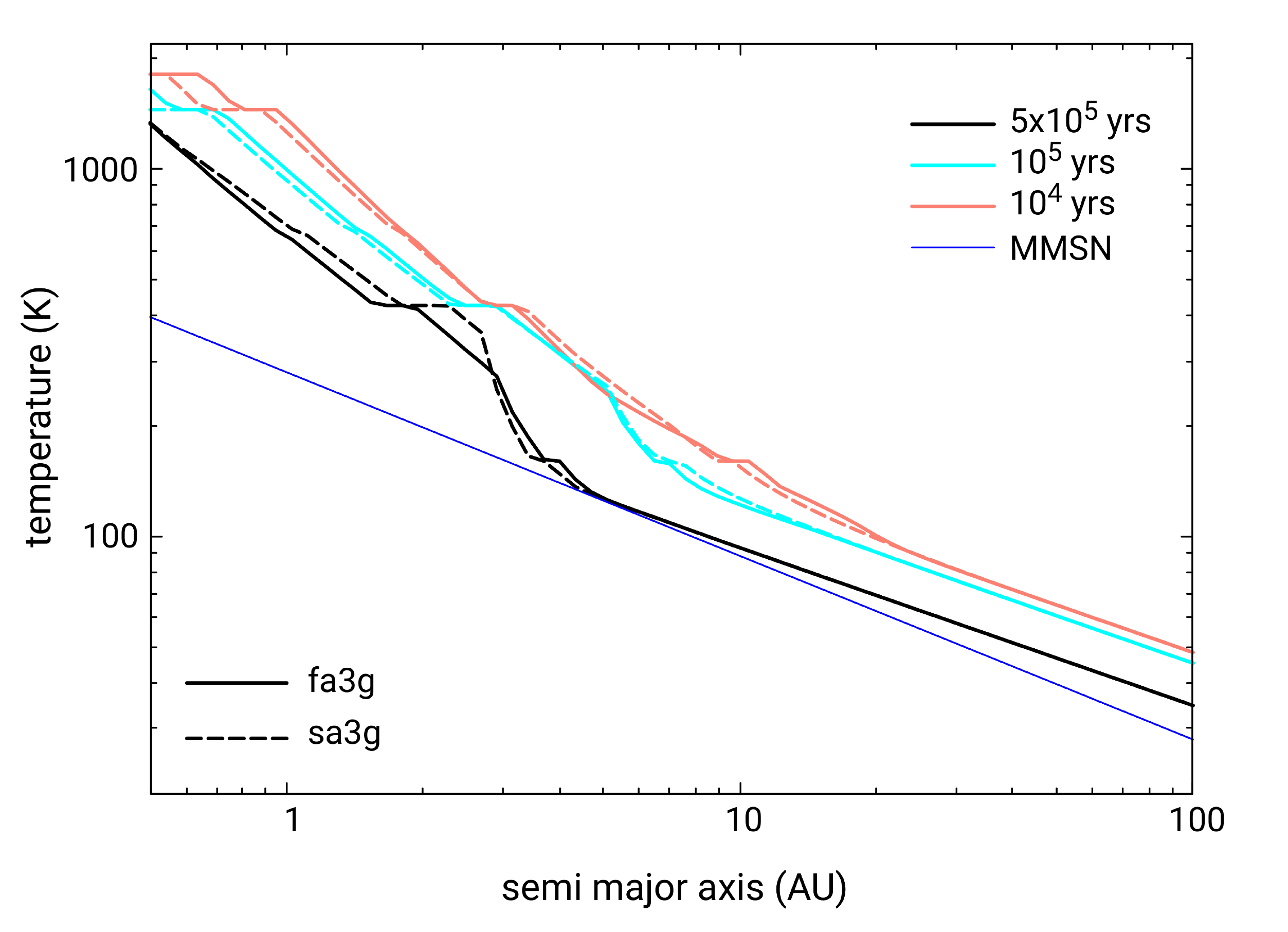}{0.45\textwidth}{}
          \fig{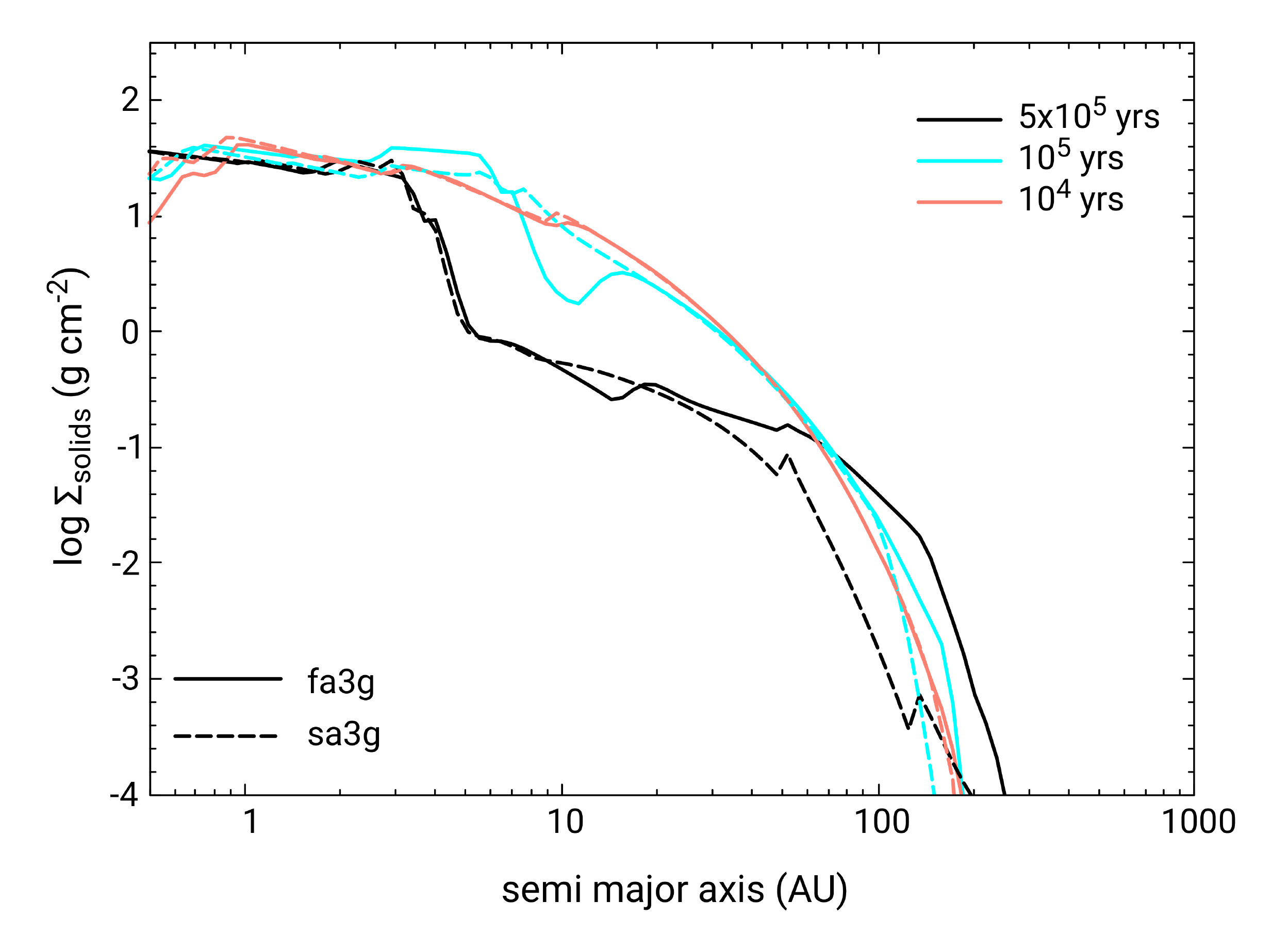}{0.45\textwidth}{}}

\caption{\uline{Left panel}: Disk midplane temperature for the compact model sa3g (dashed curves)  and  fractal model fa3g  (solid curves) at the same times given in Fig. \ref{fig:a3g}. ``Flat'' regions seen in some parts of the profile are EFs, which migrate inwards considerably over the course of the simulation. The blue curve is a standard MMSN temperature profile \citep{Hay81} for comparison, which beyond 10 AU is steeper than our profile that has a $\sim R^{-3/7}$ dependence due to our flared disk geometry. Inside the water snowline, the mean slope of $T$ has roughly a $\sim R^{-1}$ dependence. \uline{Right panel}: Total solids surface density at the same times. Both compact and fractal growth models are characterized by a sharp decrease in surface density outside the snowline and beyond, as growth proceeds quickly to large particles there which drift in. This is visible by the shift from yellow to orange color in Fig. \ref{fig:a3g}, for fractals at $10^5$ yr, and for both by $5\times 10^5$ yr.  However, the fractal case retains much more material outside starting outside of $\sim 20$ AU because aggregates are increasingly more underdense. Note that in the fiducial model, the water snowline migrates inwards from outside 10 AU to $\sim 4$ AU after 0.5 Myr. Kinks in $\Sigma_{\rm{solids}}$ seen at $\sim 50$ and $\sim 125$ AU at 0.5 Myr most notable in the compact  growth case correspond to enhancements at the CO$_2$ and CH$_4$ EFs.
\label{fig:a3gtemp}}
\end{figure*}

Following the particle mass distributions in time from left to right, it can be seen that the fragmentation mass (black dashed curve) in the inner disk\footnote{In this work, ``inner disk'' refers to inside the water snowline, while ``outer disk'' refers to regions outside the snowline. The outer disk can also be separated into two sub regions which generally display distinct behavior, from the snowline out to $\sim 20-30$ AU, and the other beyond $\gtrsim 20-30$ AU.} 
 regions is reached relatively quickly for both the compact and fractal cases extending out to $\sim 4-5$ AU after only $10^4$ years. After the fragmentation barrier is reached, $m_{\rm{f}}$ evolves slowly   due both to evolving nebula conditions and the inward loss of solids via slow radial drift. The fragmentation barrier is much more difficult to achieve further out in the disk, even after 0.5 Myr, due to a much higher $Q_{\rm{f}}$ outside the water snowline (which migrates inwards from outside $10$ AU to $\sim 4$ AU after 0.5 Myr as the disk cools; see Fig. \ref{fig:a3gtemp}, left panel). 
With a higher $Q_{\rm{f}}$, particles can grow to much larger sizes and Stokes numbers (see Fig. \ref{fig:a3gst}). This  increases their radial drift velocities, allowing even fractal aggregates to drift inwards more rapidly from the snowline outwards to $\sim 20$ AU. This leads to a significant drop in the solids surface density in this region (see Fig. \ref{fig:a3gtemp}, right panel). The growth rate in this region begins to significantly slow as the fragmentation barrier (dashed brown curve) is approached, not only due to the decrease in the solids volume density at all sizes (seen as a discernible vertical inflection from yellow to orange), but also because of the bouncing barrier which largely limits particle or aggregate growth from masses $m < m_{\rm{b}}$ (though in the fractal case, bouncing can still lead to compaction). As a consequence, once the fragmentation barrier is reached, no further growth occurs. 

On the other hand, in the inner disk, growth does proceed beyond the fragmentation barrier leading to a population of ``lucky particles'' or migrators 
the largest of which generally contain negligible mass \citepalias{Est16}. 
At $10^4$ years, both compact and fractal growth cases are characterized by a steep dropoff in solids density with masses larger than the fragmentation mass, which are gradually destroyed due to our use of a velocity PDF. As time goes on, however, continued incremental growth by {\it compact} lucky particles eventually leads to a secondary peak in the particle mass density at around 1 g ($\sim 0.3-0.4$ cm in radius), as seen at 0.5 Myr (more detail revealing additional growth peaks can be seen in the size distribution, Appendix \ref{sec:sizes}). 
This is due to mass transfer combined with  the velocity PDF \citepalias[see, e.g.,][]{Est16}. No such secondary peak is seen in the fractal growth case as time goes on. At first, it may seem as if this behavior is related to whether aggregates are in the Epstein or Stokes drag regime. From comparison with the $m_{\lambda}$ curves, most of the distribution in the compact growth case is in the Epstein regime, while the fractal aggregates are in the Stokes regime. However, we believe it is simply due to the fractal particles having a much larger $m_{\rm{f}}$ than the corresponding compact growth case. By entrapping more mass, fractal aggregates limit the abundance of small particles available for ``lucky'' particles to grow beyond $m_{\rm f}$. As was seen outside the snowline, growth is mostly restricted by the bouncing barrier, which slows growth enough that a secondary peak cannot be achieved (cf. Fig. \ref{fig:a2g}). It is interesting to note that the largest mass achieved in the inner disk is roughly the same in both compact and fractal cases, suggesting that the growth limit imposed by the combination of bouncing and radial drift is independent of particle porosity. 

What we believe the drag regime does determine is the slope of the fragmentation limit curve as seen in these simulations. In the compact growth case (Epstein regime), the fragmentation curve is relatively flat in the inner disk. 
In the Epstein regime (see Eq. [\ref{equ:mf}]), and for constant $Q_{\rm{f}}$, the compact growth $m_f \propto (\Sigma/T)^3$ suggesting the outward decrease in $\Sigma$ and $T$ (Fig. \ref{fig:a3gtemp}) are similar (though EFs and changes in particle density can affect it to a lesser extent). In the Stokes regime for ${\rm{Re}_{p}} < 1$, $m_f \propto R^3/T^{3/2}$ which should always increase outwards. The same similarly holds true for fractal growth.

The curves for the drift mass $m_{\rm{d}}$ require some explanation. Quite generally, the ``drift barrier'' has been associated with ${\rm{St}}=1$ particles because it is at Stokes unity where a particle's inward drift velocity is maximum \citep[though see, e.g.,][]{Bir12}. However, the drift barrier is more well defined as $t_{\rm{gr}}=t_{\rm{dr}}$, a function of ambient conditions in the nebula (e.g., the local metallicity), showing that the drift barrier is not a sharp transition. For instance, particles can still drift inwards even if their growth times are shorter ($t_{\rm{gr}} < t_{\rm{dr}}$); conversely, particles can still continue to grow even if their drift times are shorter ($t_{\rm{gr}} > t_{\rm{dr}}$). Moreover, there are ambient conditions under which 
the ``barrier'' can be undefined ($t_{\rm{gr}}$ and $t_{\rm{dr}}$ never intersect): the growth time can always be shorter than the drift time for all particle or aggregate masses in the distribution. One can also find a situation in which the growth times are always longer (e.g., for very low $\rho_{\rm{solids}}$). 
These conditions are reflected in regions where the $m_{\rm{d}}$ curve is not plotted.   
Naturally, because the drift barrier is a function of ambient conditions (in particular, changes in the local $Z$), the $m_{\rm{d}}$ curve evolves with time. Generally, the drift barrier is only a factor in the icy outer nebula and there, mostly for compact  particles. 
The fractal model cases are a bit more complicated because the local maximum mass fractal dimension $D$ is used in calculating $t_{\rm{gr}}$, and thus $t_{\rm{gr}}$ depends on how compacted the aggregates are. As a result $m_{\rm{d}}$ can increase very steeply to extremely large masses over a short radial range (this also applies to $m_{\rm{f}}$). 

The above conditions are demonstrated by the brown solid curves of Fig. \ref{fig:a3g}. In the compact particle growth model, the drift curve $m_{\rm{d}}$ is present outside the snowline over the course of the simulation. Growth times are initially shorter than drift times out to $\sim 150$ AU and particles grow, but as the simulation progresses, the drift mass curve begins to flatten somewhat and by 0.5 Myr, has evolved {\it below} the pre-existing largest particles from $\sim 100-150$ AU. Much of the material outside the water ice EF, despite being mostly below the radial drift curve, has drifted inwards (see Fig. \ref{fig:a3gZ}) inside the snowline by this time. For porous aggregates the $m_{\rm d}$ curve seen near $\sim 200$ AU increases sharply to masses well beyond the plotted range, so growth times for all aggregates out to this distance are much shorter than drift times. Initially, aggregates are not very compacted, but as they are compacting the $m_{\rm{d}}$ curve should also decrease. In this $\alpha_{\rm{t}}=10^{-3}$ model, that decrease is very modest after 0.5 Myr (but will be much more evident in our other $\alpha_{\rm{t}}$ models, see Sec. \ref{sec:strong} and \ref{sec:weak}). 
We see in the lower panels of Fig. \ref{fig:a3g} that all aggregates from the water snowline (roughly 4 AU by 0.5 Myr) outwards to $\sim 15-20$ AU have masses that are well below $m_{\rm{d}}$ (which is off the top of the plot), so they are still growing rapidly, but through most of this region their growth is limited by fragmentation. Because of the large masses and St of these aggregates, much of the mass in solids outwards of the snowline has drifted inwards (indicated by the density step decrease to darker orange colors as one moves outwards from the snowline).
In fact, $\Sigma_{\rm{solids}}$ for the fractal model is even lower than the compact case between $\sim 10-15$ AU; however, the reverse is true outside $\sim 20$ AU (see below).   
Conversely, outside of $\sim 150-300$ AU where $\rho_{\rm{solids}}$ is very low, the growth times are much longer than the drift times even for monomers, regardless of whether growth is fractal or compact. 
Inside the water snowline 
the $\Sigma_{\rm{solids}}$ and $Z$ after 0.5 Myr remain similar. 
In this region no curve for $m_{\rm{d}}$ is shown for either model because the growth times are always shorter than the drift times for these nebula conditions. 

Figure \ref{fig:a3gtemp} (left panel) shows the midplane disk temperature for our fiducial models at the same evolution times as in Fig. \ref{fig:a3g}. The blue curve gives the temperature profile for a standard MMSN \citep{Hay81} for comparison. Locations in the disk where the radial temperature profile flattens correspond to EFs. Early on at $10^4$ years, EFs for water (160 K) to Fe (1810 K) are present. The EFs for the supervolatiles initially lie much further out in the disk, e.g., CO$_2$ is located at $\sim 105$ AU. As the disk cools, these EFs are seen to evolve inward significantly over time. For example, the water snowline moves inwards from $\sim 9-10$ AU to inside $4$ AU by 0.5 Myr, in which time the silicate EF (1450 K) has evolved to within the inner boundary of our computational grid. Though not noticeable here, the CO$_2$ EF has evolved inwards to $\sim 50$ AU, but kinks associated with enhancements in solids at this EF can be seen in the  $\Sigma_{\rm{solids}}$ profile of Fig. \ref{fig:a3gtemp} (right panel), which includes both dust and migrators. More detailed analysis of the compositional evolution is discussed in \citetalias{Est21}.

\begin{figure*}
\gridline{\fig{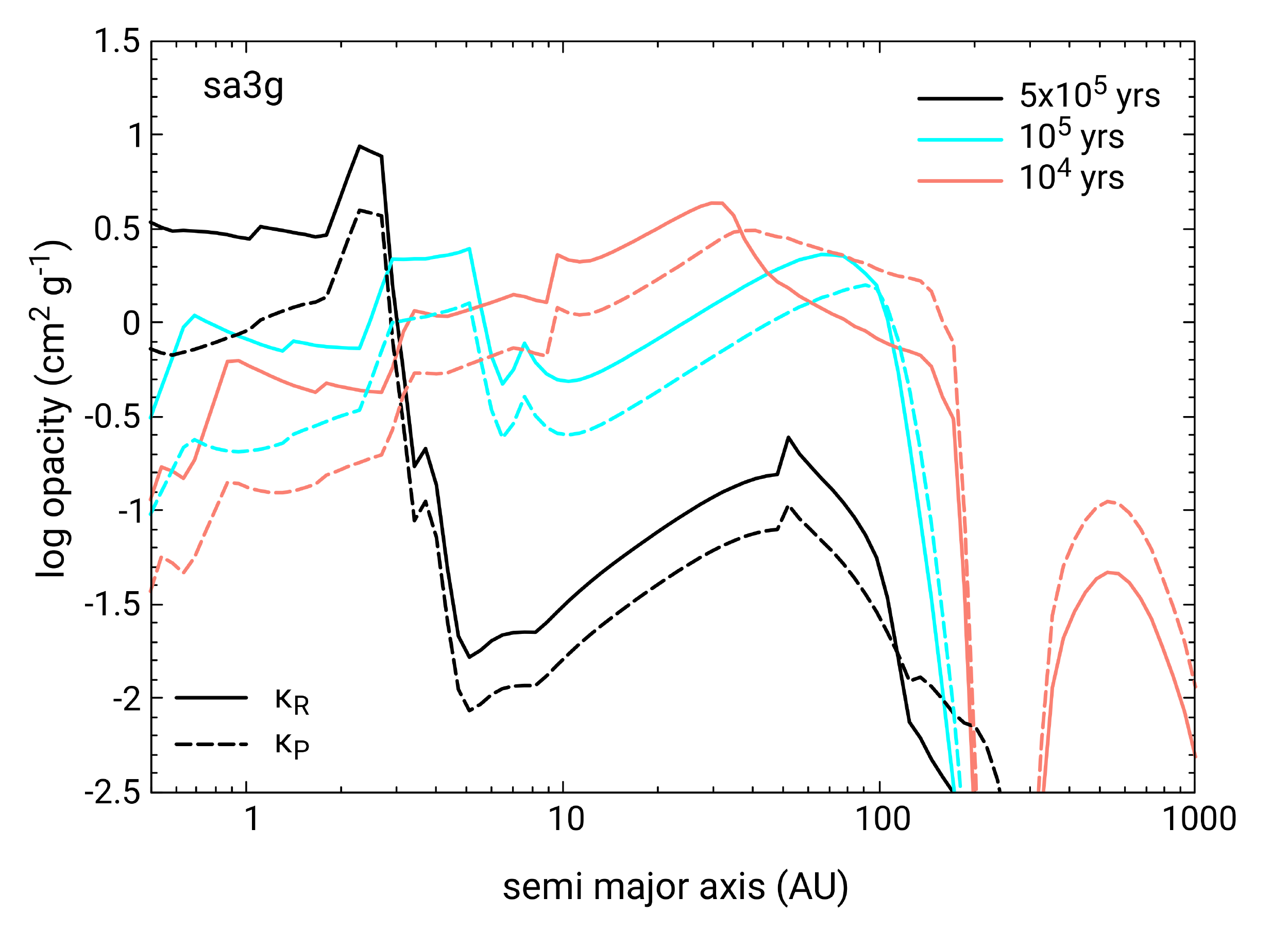}{0.45\textwidth}{}
          \fig{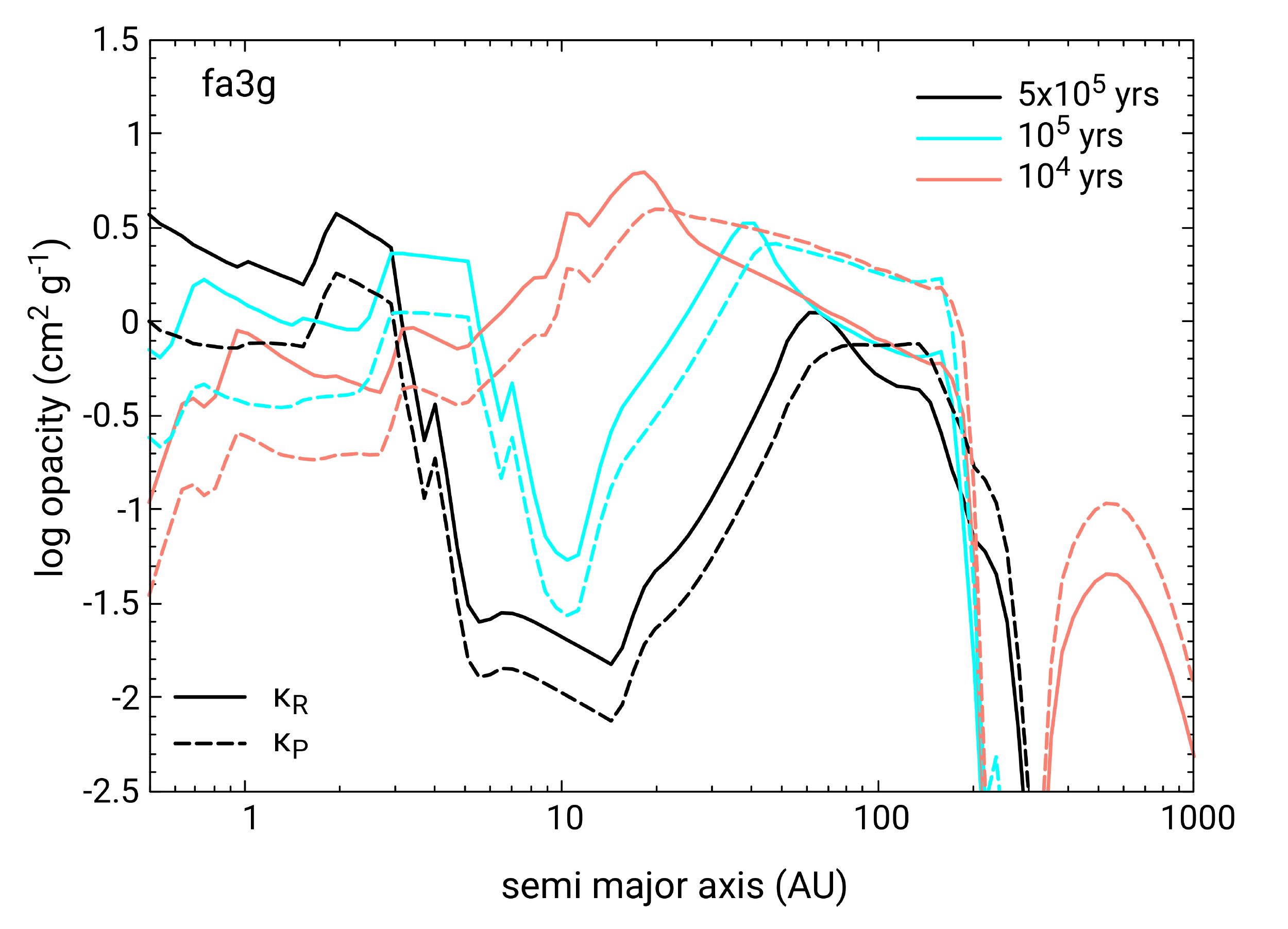}{0.45\textwidth}{}}
 
\caption{Rosseland (solid curves) and Planck (dashed curves) mean opacities as a function of semi-major axis and at the same times as the  compact  (left panel) and fractal (right panel) growth simulations shown in Fig. \ref{fig:a3g}. In both cases the decrease in opacity outside the snowline (4-5AU) is due to larger particles and material drifting inwards and  lowering the local effective $Z$. However, the fractal case retains much more mass in the outer disk from $\gtrsim 40$ to $100$ AU. \label{fig:a3gopac}} 
\end{figure*}

\begin{figure*}
\centering
\includegraphics[width=0.5\textwidth]{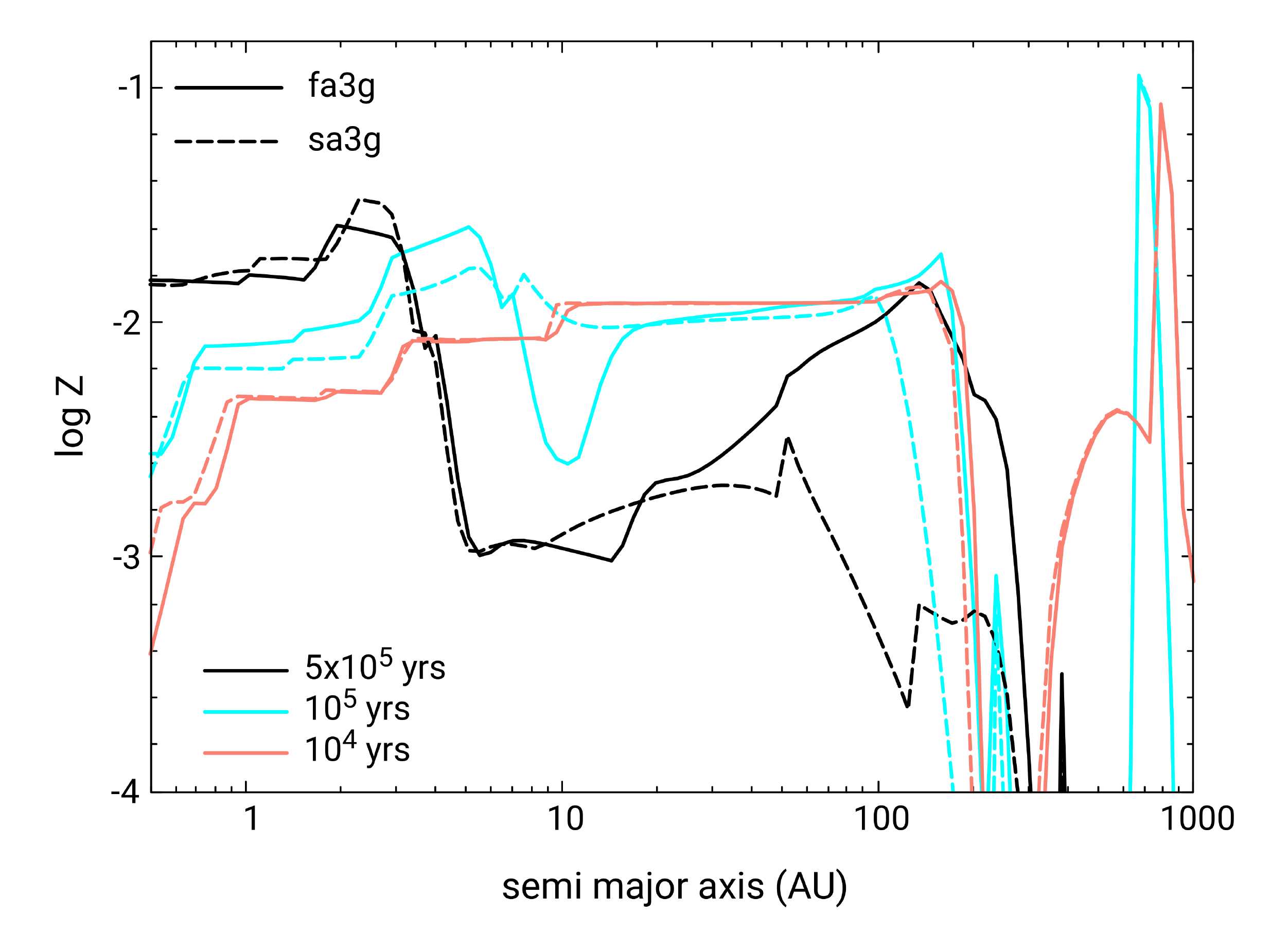}
\caption{Metallicity $Z$ for the compact model sa3g (dashed curves)  and fractal model  fa3g (solid curves)  at the same times given in Fig. \ref{fig:a3g} as a function of semi-major axis. 
The sharp decrease in the total solids surface density outside the snowline (located at $\sim 4$ AU after 0.5 Myr) and beyond seen in Fig. \ref{fig:a3gtemp} (right panel) is reflected more strongly here as an order of magnitude drop in $Z$ compared to the initial disk metallicity of $\bar{Z}\simeq 0.014$.
\label{fig:a3gZ}}
\end{figure*}

\begin{figure}
\includegraphics[width=1.0\textwidth]{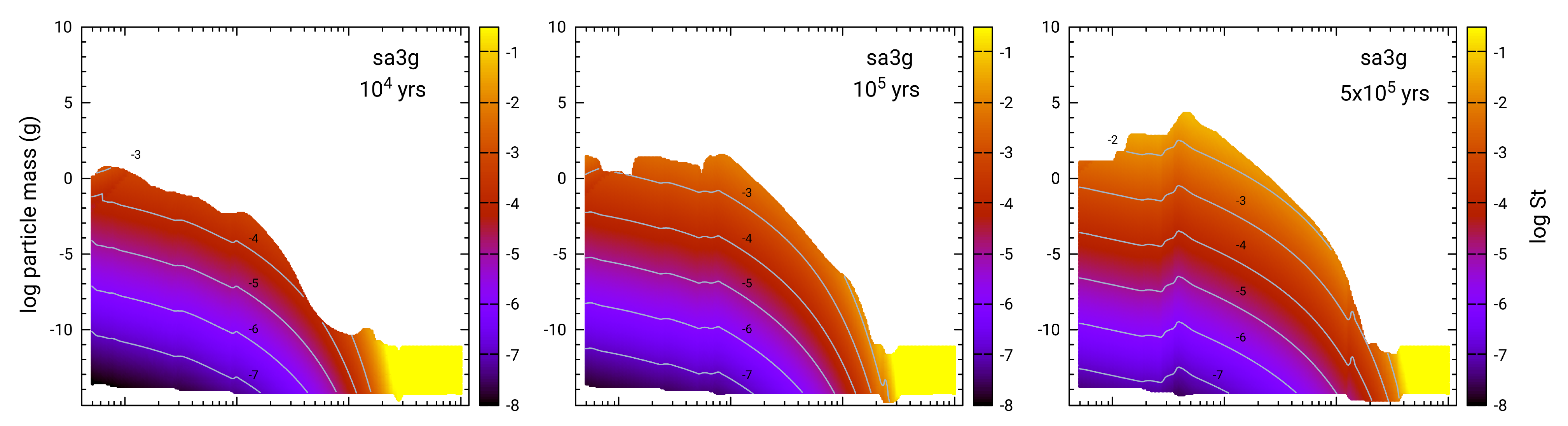}
\includegraphics[width=1.0\textwidth]{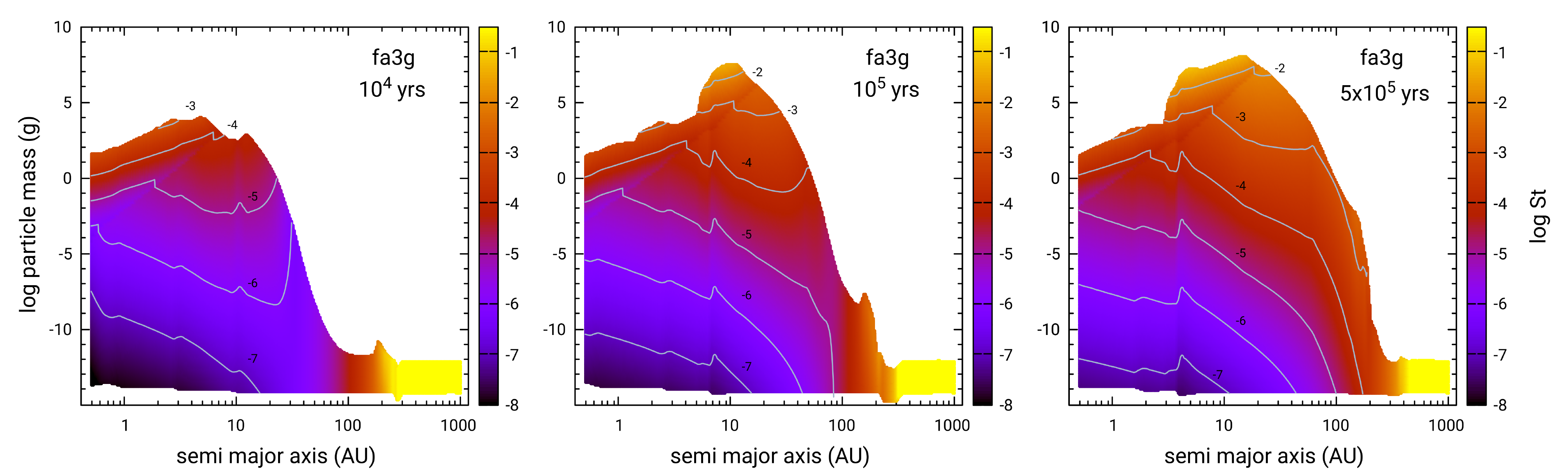}

\caption{Distribution of Stokes numbers associated with the evolution of the particle mass distributions for the compact (sa3g) and fractal (fa3g) fiducial growth models shown in Fig. \ref{fig:a3g}. Stokes numbers of the distribution are indicated by the color map at right, and labeled (grey) curves of constant St are plotted to aid in visualization. As noted in the text, the more rapid growth of fractal particles beyond the snowline leads to more massive, and higher St particles by $10^5$ years compared to compact growth (middle panels). Note that outside of $200-400$ AU, even small grains have very large ${\rm{St}} \gg 1$ making them essentially unaffected by radial drift. The total mass fraction is very low so growth times are exceedingly long. With careful inspection, one can see the outline of the water ice EF (e.g., $\sim 4$ AU at 0.5 Myr), as well as the outline of $m_\lambda$ in the fractal growth case (by reference, the dotted brown curves in Fig. \ref{fig:a3g}; the reader may find it easier to view this onscreen and magnified). 
} 
\label{fig:a3gst}
\end{figure}

\begin{figure}
\includegraphics[width=1.0\textwidth]{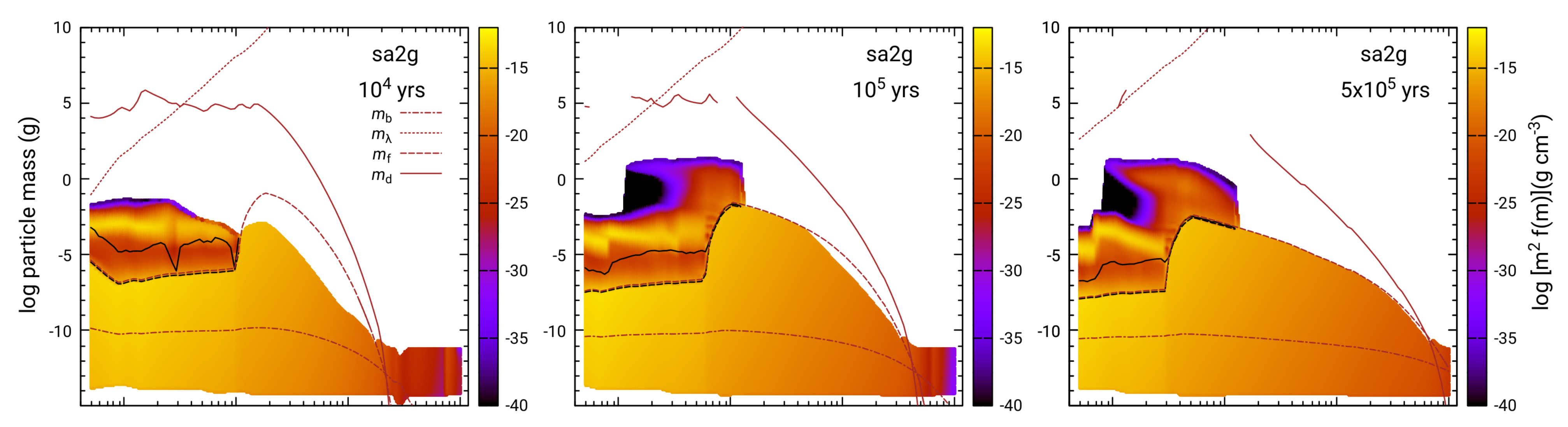}
\includegraphics[width=1.0\textwidth]{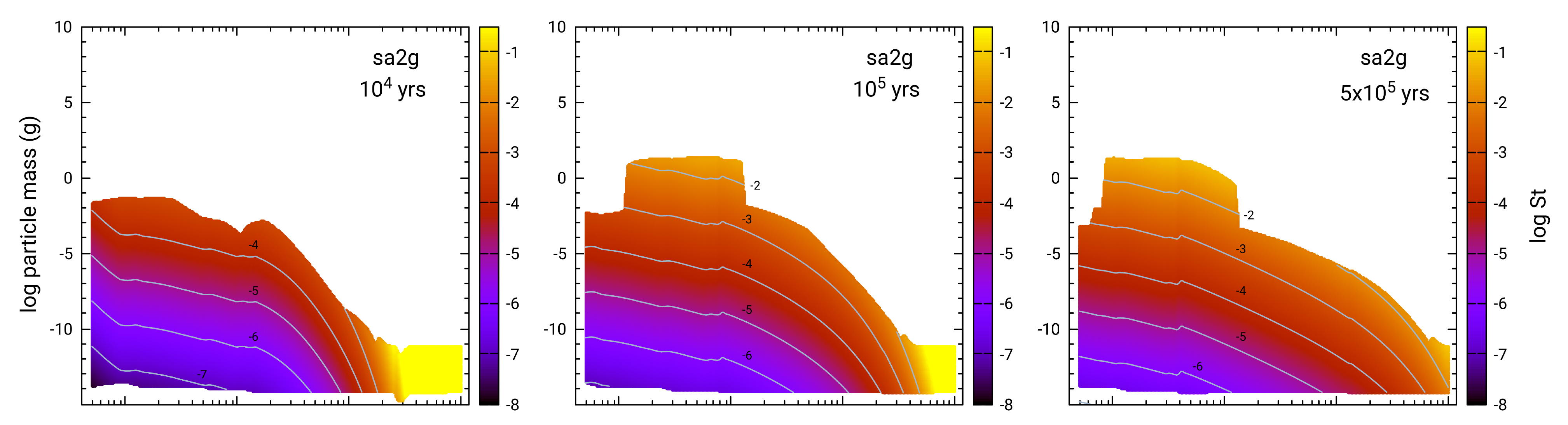}
\includegraphics[width=1.0\textwidth]{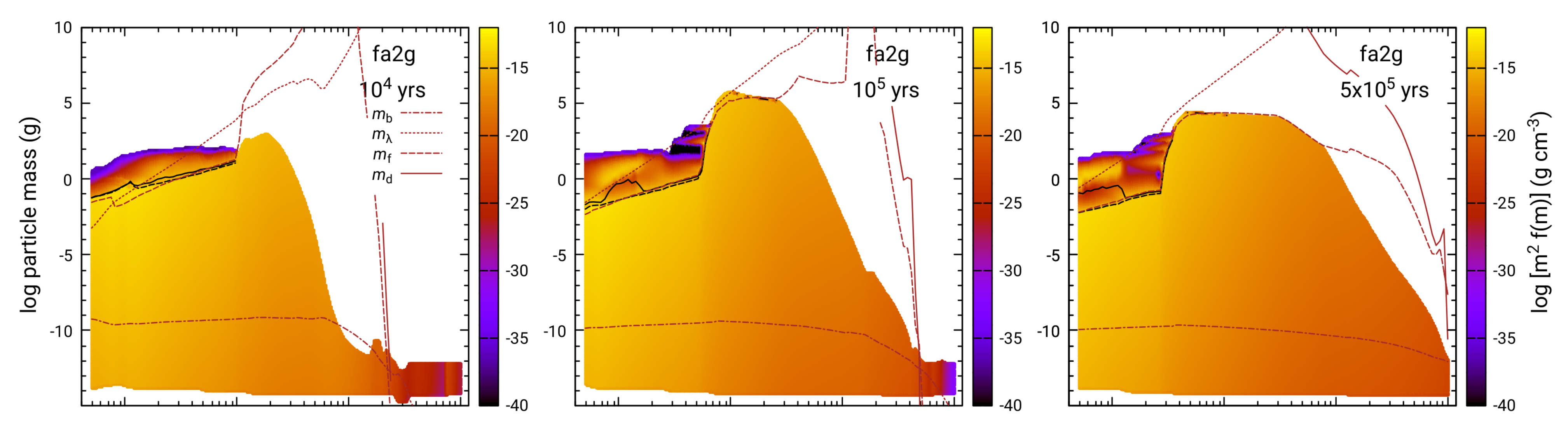}
\includegraphics[width=1.0\textwidth]{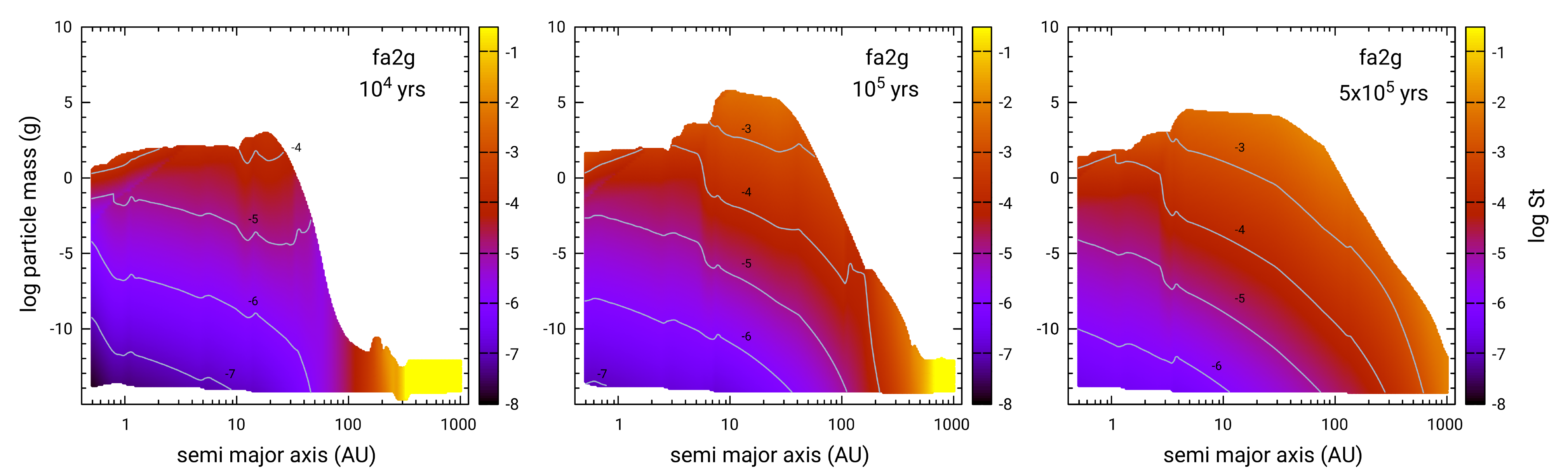}
\caption{Comparison of models with $\alpha_{\rm{t}}=10^{-2}$ for both compact  (top panels) and fractal (bottom panels) growth at, from left-to-right, $10^4$, $10^5$ and $5\times 10^5$ years. Plotted in each panel is a snapshot of the particle mass distribution as a function of semi-major axis. \underline{1st and 3rd row}: Particle solids mass volume density per bin of width $\Delta m$. \underline{2nd and 4th row}: Stokes numbers. The particle size distributions for models sa2g and fa2g after 0.5 Myr can be seen in Appendix \ref{sec:sizes}.
\label{fig:a2g}}
\end{figure}

The relatively small differences in the temperature profiles between the two models can be intimately tied to the evolution of the particle mass distributions  through their Rosseland and Planck mean opacities, which are plotted in Figure \ref{fig:a3gopac}. Early in the simulation, there is not much difference between the compact and fractal models. This changes though, and by $10^5$ years the opacity in the fractal growth model outside the water snowline  drops drastically (this region can be associated with the humps in the particle mass distributions seen in Fig. \ref{fig:a3g}). This large dip reflects that in the fractal case, growth to large sizes is occurring much more quickly 
as a result of their enhanced cross sections. The larger (and higher St) fractal aggregates have a more rapid inward drift than the compact  particles just outside the snowline, so the
local $Z$ is significantly lower at $10^5$ years (solid curves, Fig. \ref{fig:a3gZ}; and also Fig. \ref{fig:a3gtemp}, right panel). 
The sharp opacity drop also leads to a slight, but notably lower 
temperature 
(Fig. \ref{fig:a3gtemp}, left panel). By 0.5 Myr, the compact and fractal particles outside the snowline and inside 10 AU have similar opacities, but outside $\sim 20$ AU the opacity increases in the fractal case as there is more material retained in the outer disk. 
Just interior to the snow line (between $\sim 2-3$ AU), despite the surface densities being similar (Fig. \ref{fig:a3gtemp}, right panel), the opacity is lower in the fractal case (Fig. \ref{fig:a3gopac}). This is because, while thermal opacities are generally higher for porous particles relative to compact  grains of the {\it  same size} 
\citep[there are more of them;][]{Cuz14}, here the aggregates in the fractal case are much {\it larger} than in the compact case, and the local $Z$ is also somewhat lower, so the overall effect is a {\it  lower} opacity. In Section \ref{sec:obs}, we briefly discuss scattering properties of porous grains as it relates to the observations. 


In Figure \ref{fig:a3gst} we explore further the distribution of the masses and Stokes numbers in these fiducial models. The fractal aggregates always have larger masses. At $10^5$ years, the largest Stokes numbers in the inner disk are comparable between the compact and fractal models, but outside the snowline the most massive particles in the fractal case have achieved higher St than in the compact growth case even as early as $10^5$ years. The resulting higher drift velocities allow the region just outside the snowline 
to be depleted sooner in the fractal case than for the compact  growth case (the dip seen in Fig. \ref{fig:a3gZ}, in the blue curve at 10 AU). By 0.5 Myr outside the snowline, the compact case has essentially caught up
to the fractal case with maximum ${\rm{St}} \simeq 0.03-0.04$. 
The compact  particle case also shows that the very outer disk (outside of $R \gtrsim 20$ AU) is becoming more and more depleted in material as mass dominant particles have achieved higher St, and therefore radial drift velocities, relative to the fractal case. Beginning outside of $\sim 20$ AU, the model sa3g shows that mass dominant particles have ${\rm{St}} \gtrsim {\rm{few}} \times 10^{-3}$, whereas the Stokes numbers for fa3g drop sharply to values of $\sim 10^{-4}$ and lower further out,
explaining why $Z$ and  $\Sigma_{\rm{solids}}$ there remain high.

Fractal growth in mass is in general faster than compact particle growth over all sizes owing to the  larger cross section of aggregates; however, the Stokes numbers of fractal aggregates grow more slowly than compact particles in the earlier growth stages, and in fact remain similar to monomers until compaction begins to set in (see Sec. \ref{sec:fracgrow}, and Fig. 
\ref{fig:psi}). Once significant compaction occurs, the fractal aggregate St can grow more quickly than compact particles. This is evident in the region outside the snowline to $\sim 20$ AU. The St of fractal aggregates and compact particles are similar at 0.5 Myr, and also at $10^4$ years, but clearly at 0.1 Myr the fractal aggregates have already achieved much higher St values. 
This is reflected in the much lower opacity in model fa3g (see cyan curves at 0.1 Myr, Fig. \ref{fig:a3gopac}). By 0.5 Myr, the opacities (and surface densities, black curves Fig. \ref{fig:a3gtemp}, right panel) are similar for both models. Outside of $\sim 20$ AU, the difference in Stokes numbers in Fig. \ref{fig:a3gst} between sa3g and fa3g then is largely due to slower dynamical times and decreasing $\rho_{\rm{solids}}$ with distance (both which influence growth rate) such that fractal aggregates are not yet compacted enough to be comparable to the compact growth model  particles.
Thus a characteristic of fractal growth is that aggregates can be retained further out in the nebula for longer periods of time, to provide source material for potential planetesimal formation there (see black solid curve in Fig. \ref{fig:a3gZ}).


Figure \ref{fig:a3gst} also helps explain an effect seen in previous plots, of a transition that occurs around $\sim 200-400$ AU in these models where a stark depletion (though relatively subtle in total mass) in material is evident, both in the opacity (Fig. \ref{fig:a3gopac}) and metallicity (Fig. \ref{fig:a3gZ}) plots. This transition is due to the steep decrease in the gas surface density. On the one hand, the steep ``edge'' of the gas disk leads to a strong pressure gradient facilitating very rapid inward drift. On the other hand, even small aggregates quickly achieve large Stokes numbers at these low gas densities and can quickly become immobile, remaining so until the gas density increases as the disk spreads outwards at later times. All panels in Fig. \ref{fig:a3gst} demonstrate this evolution. Inside the transition, one can see a spike of increased growth produced by the increased solids fraction having St $\sim 10^{-3}-10^{-2}$. Outside the transition, despite particles being immobile with ${\rm{St}} > 0.1$, there is very little material, and thus growth times are exceedingly long regardless of whether growth is  compact or fractal.


Overall, a general result of these fiducial $\alpha = 10^{-3}$ simulations  is that fractal, porous growth is characterized by mass-dominant aggregates that are as much as $\sim 2-4$ orders of magnitude more massive than their compact counterparts, even as their internal densities are orders of magnitude smaller (Sec. \ref{sec:dens}). The masses of ``lucky" particles, however, are not too different between cases. We find that regardless of whether growth is fractal or compact, the mass-dominant aggregates are in the fragmentation regime in the inner disk regions, but for the larger $Q_{\rm{f}}$ used for icy particles beyond the snowline, aggregate or particle masses and St can be much larger and much more prone to rapid radial drift. Slowed growth as a result of the higher fragmentation mass threshold (which limits the amount of fine dust), and steady loss of material inwards, conspires to keep aggregates in the drift-dominated regime. Both compact and fractal particles quickly evacuate the region between the snowline and $\sim 40$ AU, but porous aggregates survive longer, and in greater abundances, outside of $\sim 40$ AU. 
 

\subsection{Variation with Turbulent Strength}
\label{sec:varturb}


\subsubsection{Strong Turbulence $\alpha_{\rm{t}}=10^{-2}$}
\label{sec:strong}
 
In Figure \ref{fig:a2g} we present snapshots of simulations comparing compact (sa2g) and fractal (fa2g) growth models using a higher $\alpha_{\rm{t}} = 10^{-2}$ at $10^4$, $10^5$ and $5\times 10^5$ years. Due to the higher relative velocities and collisional frequencies  induced by the stronger turbulence, growth times are more rapid and both the bouncing and fragmentation barriers are reached at much smaller sizes, with the latter generally lying within the Epstein regime in both models. As a result of lower $m_{\rm{b}}$ and $m_{\rm{f}}$, and similar to what was found in \citetalias{Est16}, lucky particles are even more ``lucky'' under stronger turbulence relative to our fiducial model for three reasons: first, the fragmentation barrier occurs at smaller $m_{\rm{f}}$ and small particles are constantly replenished from which lucky particles mostly grow; second, many decades of possible growth are left before the barriers imposed by the combination of bouncing and radial drift; and third, higher relative velocities between the largest and smallest particles increases the efficiency of mass transfer \citep{Win12a}. Much like our fiducial case $\alpha_{\rm{t}} = 10^{-3}$, the masses of the largest lucky particles are not too different between the compact and fractal models.

For this $\alpha_{\rm{t}}$, the compact growth case (top row) shows that a strong secondary peak appears inside the water ice EF relatively early on (after $10^4$ yrs), which persists throughout the simulation (another tertiary growth peak is more easily seen in the size distribution, see Appendix \ref{sec:sizes}). This is due to mass transfer (see Sec. \ref{sec:barriers}) and our velocity PDF, significantly increasing the mass of the mass-dominant particles beyond the fragmentation mass (black solid curve). This is consistent with the behavior seen in the compact growth simulations with $\alpha_{\rm{t}}=10^{-3}$, where significant growth beyond the fragmentation barrier occurs with a well-defined secondary peak appearing by $10^5$ years (Fig. \ref{fig:a3g}). 
We see a slightly smaller secondary peak in the fractal case as well (third row, Fig. \ref{fig:a2g}; also see Appendix \ref{sec:sizes}), 
whereas no distinct secondary peak was seen in the fiducial model (the largest achieved particle masses were not far beyond $m_{\rm{f}}$). We note that the fragmentation curves are flatter at this $\alpha_{\rm{t}}$, (also see below) consistent with being for the most part in the Epstein regime where $m_{\rm{f}}$ depends on the behavior of both $T$ and $\Sigma$ (cf. Fig. \ref{fig:a3g}, and Sec. \ref{sec:fiducial})\footnote{We note that in the fractal case, the fragmentation curve estimate (brown dashed curve) does not quite match the actual fragmentation curve in the vicinity of $m_\lambda$. This is due to our use of a bridging function \citep{Pod88} that transitions from the Epstein to the Stokes regime \citepalias{Est16} which would predict a higher fragmentation threshold. The apparent ``overflow'' of masses beyond $m_{\rm{f}}$ seen around $\sim 8-9$ AU at $10^5$ years is also due to this effect.}. 

Beyond the water snowline, 
the compact particle growth case is fragmentation limited out to 50AU by 0.1 Myr, and over almost the entire extent of the disk by 0.5 Myr (though lucky particles can be found as far out as $\sim 15$ AU), as the disk gas rapidly evolves outwards. 
In contrast, the fractal case is fragmentation limited to about $\sim 100$ AU or so; outside this location, growth to larger sizes is still proceeding slowly, and may still reach the fragmentation limit at later times. The strong diffusion and advection associated with $\alpha_{\rm{t}}=10^{-2}$ coupled with lower mass particles prevents the sharp drop in solids surface density seen outside the snowline in the fiducial case, and indeed $\Sigma_{\rm{solids}}$ (see cyan curves, Fig, \ref{fig:agsurf}) decreases in a similar manner to the gas over much of the disk.

The mass-dominant particles for both compact and fractal particles 
mostly lie below the radial drift mass curve where it is defined (as in the fiducial model). By $10^4$ years the $m_{\rm d}$ curve  drops especially sharply near $\sim 200$ AU, crossing to below the mass-dominant particle mass. Unlike the fiducial cases, however, this location migrates to much larger $R$ with time as the nebula gas advects outwards. Thus particles outside $200$ AU, which are initially growing slowly or not at all, eventually find themselves below $m_{\rm{d}}$ and the increasing gas density allows for both significant growth and inward drift, even very far out in the disk. Also by $10^4$ years in the compact  growth case, $m_{\rm{d}}$ is defined everywhere inside the snowline, but as time progresses, the growth times become faster than the radial drift times due primarily to the increase in $Z$ in the inner disk, causing the $m_{\rm{d}}$ curve to vanish. In the fractal case, the $m_{\rm{d}}$ curve is only mildly less restrictive than the fiducial case, at least initially, with growth times always faster than drift inside $\sim 200$ AU (after $10^4$ yrs). By 0.5 Myr, the nebula gas density has increased so much even at 1000 AU that ${\rm{St}} < 1$ everywhere. The radial drift mass curve has lowered and flattened significantly as well in both models, but $m_{\rm{d}}$ remains still far larger than the largest particles or aggregates, and mostly well above the fragmentation curve. 
 
\begin{figure}
\includegraphics[width=1.0\textwidth]{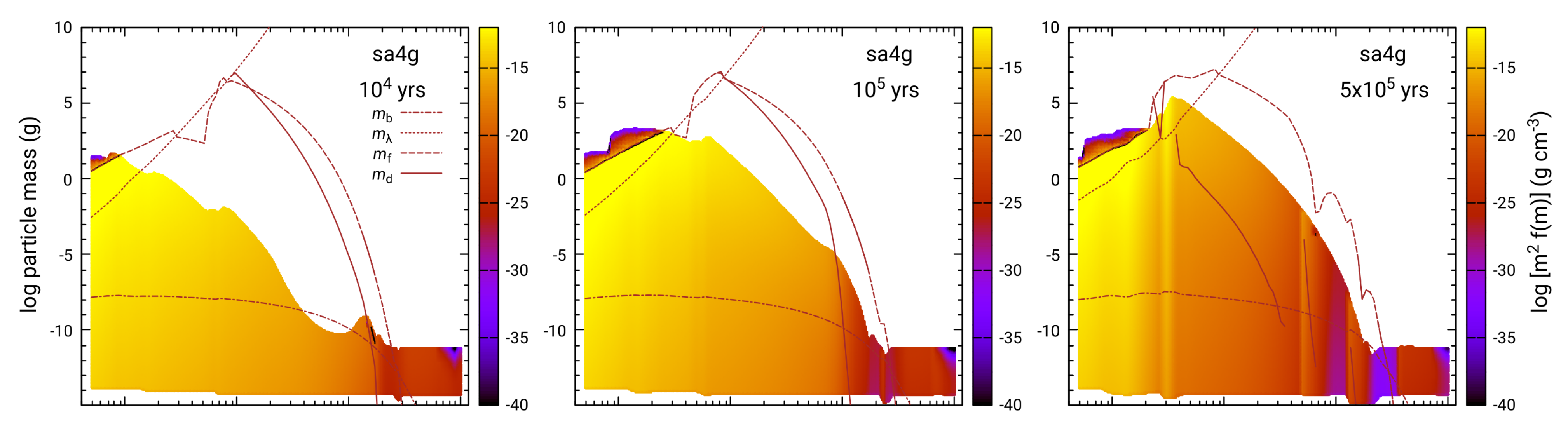}
\includegraphics[width=1.0\textwidth]{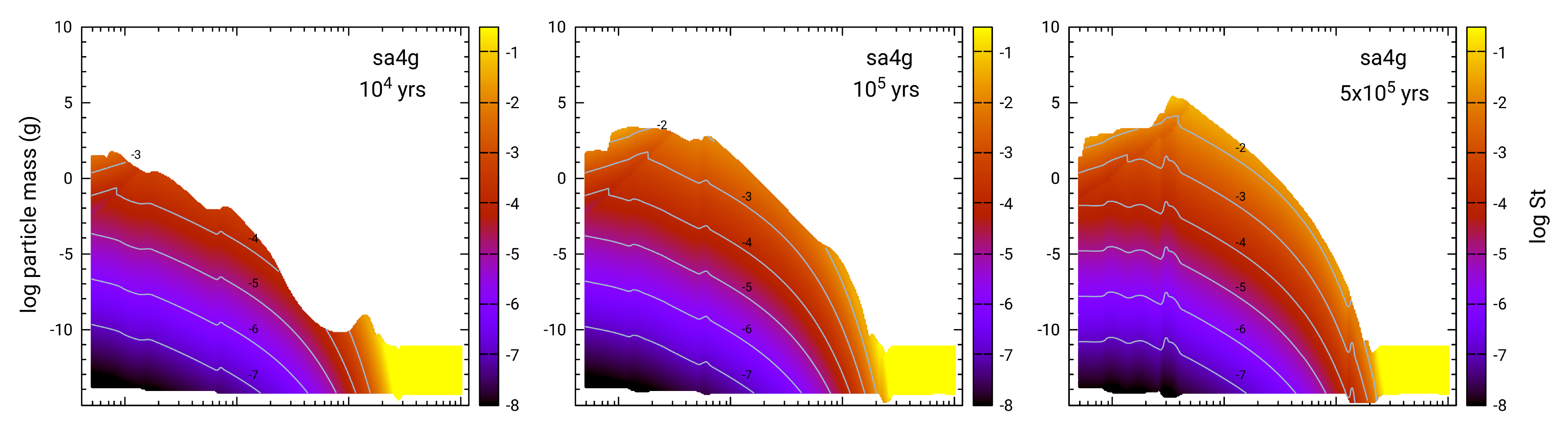}
\includegraphics[width=1.0\textwidth]{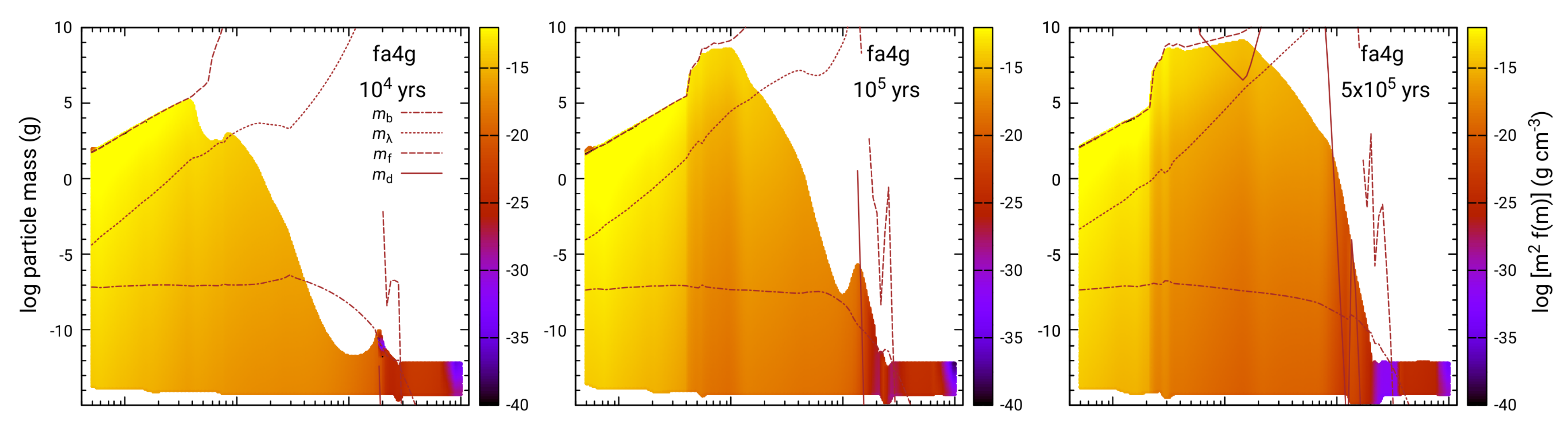}
\includegraphics[width=1.0\textwidth]{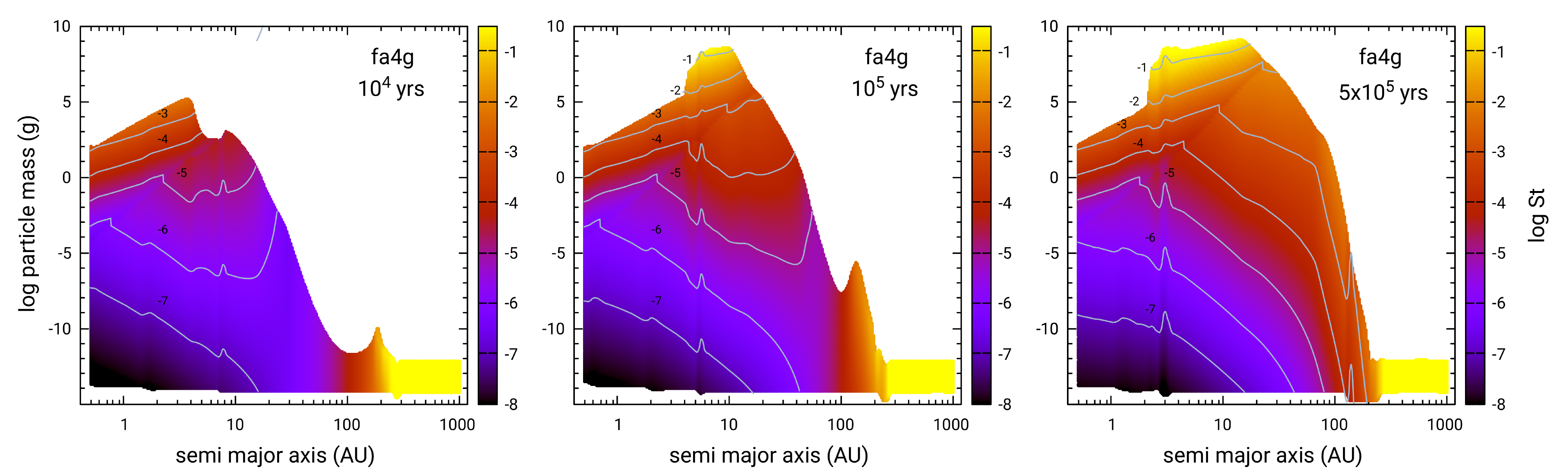}
\caption{Comparison of models with $\alpha_{\rm{t}}=10^{-4}$ for both compact  (top panels) and fractal (bottom panels) growth at, from left-to-right, $10^4$, $10^5$ and $5\times 10^5$ years. Plotted in each panel is a snapshot of the particle mass distribution as a function of semi-major axis. \underline{1st and 3rd row}: Particle solids mass volume density per bin of width $\Delta m$. \underline{2nd and 4th row}: Stokes numbers. Though it is more apparent in the compact particle models, the St of particles outside the snowline tend to be relatively constant over a large radial extent \citepalias{Est16}. For these models, the water snowline is located near $\sim 3$ AU after 0.5 Myr. The particle size distributions for models sa4g and fa4g after 0.5 Myr can be seen in Appendix \ref{sec:sizes}.
\label{fig:a4g} 
}
\end{figure}

With regards to growth, the $\alpha_{\rm{t}} =10^{-2}$ cases indicate (more clearly in the fractal growth model) that the peak in the particle mass distribution (which occurs outside the water ice EF) has been reached sometime earlier in the simulation than for $\alpha_{\rm{t}}=10^{-3}$. The fiducial models have the largest peak masses after 0.5 Myr, while the peak masses in fa2g (and sa2g) occur between $0.1-0.2$ Myr.
The peak in the mass distribution for the $\alpha_{\rm{t}}=10^{-2}$ models (which outside the snowline for the mass dominant particle is $m_{\rm{f}}$) notably begins to decrease (in fact, by over an order of magnitude) due to the rapid outward expansion of the disk gas. 
At the same time (though not discernible from Fig. \ref{fig:a4g}), the fragmentation ${\rm{St}}_{\rm{f}}$ has grown larger from 0.1 to 0.5 Myr with  
the increase in ${\rm{St}}_{\rm{f}}\propto T^{-1}$ (Eq. \ref{equ:Stfrag}) almost all due to the disk's lower temperature later in the simulation. 
The decrease in the fragmentation mass $m_{\rm{f}}\propto (\Sigma/T)^{D/(D-2)}$ \citep[Eq. \ref{equ:mf}, see also][]{Bir12} is because $\Sigma$ has decreased much faster relative to $T$ over the same timescale.
This behavior is different from the fiducial model where the peak in the particle mass distribution has steadily increased, because the gas disk has not spread out nearly as much and $\Sigma$ remains relatively large. However, it suggests that at later times the peak masses for the $\alpha_{\rm{t}}=10^{-3}$ models 
will eventually become smaller as the disk spreads and the gas density locally decreases. 
The $\alpha_{\rm{t}}=10^{-2}$ case is also interesting in that the stopping times for both the compact and fractal particles are closer to the Kolmogorov eddy time - that is,  ${\rm{St}} \sim {\rm{St}}_\eta = {\rm{Re}}^{-1/2}$ (Sec. \ref{sec:relvel}). 
As discussed in Sec. \ref{sec:barriers}, this means that the approximation for the fragmentation Stokes number 
given by Eq. (\ref{equ:Stfrag}) is inaccurate. Although $y_* = 1.6$ \citep{OC07}, one can no longer ignore terms in ${\rm{Re}}^{-1/2}$ in Eq. (\ref{equ:dvT}), 
which increases the estimate of the fragmentation mass. This has been taken into account in the curves shown in Fig. \ref{fig:a2g}. 

Finally, the difference in peak masses of the particle mass distributions between the compact and fractal growth cases (here we mean $m_{\rm{f}}$ in the inner disk and the largest particles in the outer disk)  is more pronounced at $\alpha_{\rm t} = 10^{-2}$ than for the fiducial $\alpha_{\rm t} = 10^{-3}$ case - now $\sim 4-8$ orders of magnitude by 0.5 Myr. This is due to a combination of more underdense aggregates (Sec. \ref{sec:dens}) and overall smaller 
$m_{\rm f}$, which leads to smaller radial drifts. This is also evident in the Stokes numbers (2nd and 4th rows of Fig. \ref{fig:a2g}). The Stokes numbers achieved in the compact and fractal growth models are quite similar to each other, but about an order of magnitude smaller than those for the fiducial case in Fig. \ref{fig:a3gst}. 


\subsubsection{Weak Turbulence $\alpha_{\rm{t}}=10^{-4}$ and $\alpha_{\rm{t}}=10^{-5}$}
\label{sec:weak}

The situation is quite different when choosing a smaller $\alpha_{\rm{t}}=10^{-4}$, the compact (sa4g) and fractal (fa4g) growth models of which are shown in Figure \ref{fig:a4g} (the corresponding size distributions at 0.5 Myr can be found in Appendix \ref{sec:sizes}).  In this case, the bouncing and fragmentation barriers occur at much larger masses (sizes) which are mostly in the Stokes drag regime out to $\gtrsim 10-20$ AU. Inside the snowline, and as a result of the larger fragmentation mass (which consolidates more mass in fewer particles), we see considerably less growth beyond the fragmentation barrier in 
both models (rows 1 and 3) than for higher values of $\alpha_{\rm{t}}$. This is in spite of the larger bouncing barrier threshold which by itself would allow, at least early on, a wider range of feedstock to grow from. 
We believe the reason for this is that the rate of growth has slowed sufficiently 
(as feedstock at smaller sizes becomes more depleted) that a rough growth ``ceiling'' has been reached
where particles see little or no benefit from the velocity PDF or mass transfer, both of which are key for lucky particle growth. 

\begin{figure*}
\centering
\includegraphics[width=0.5\textwidth]{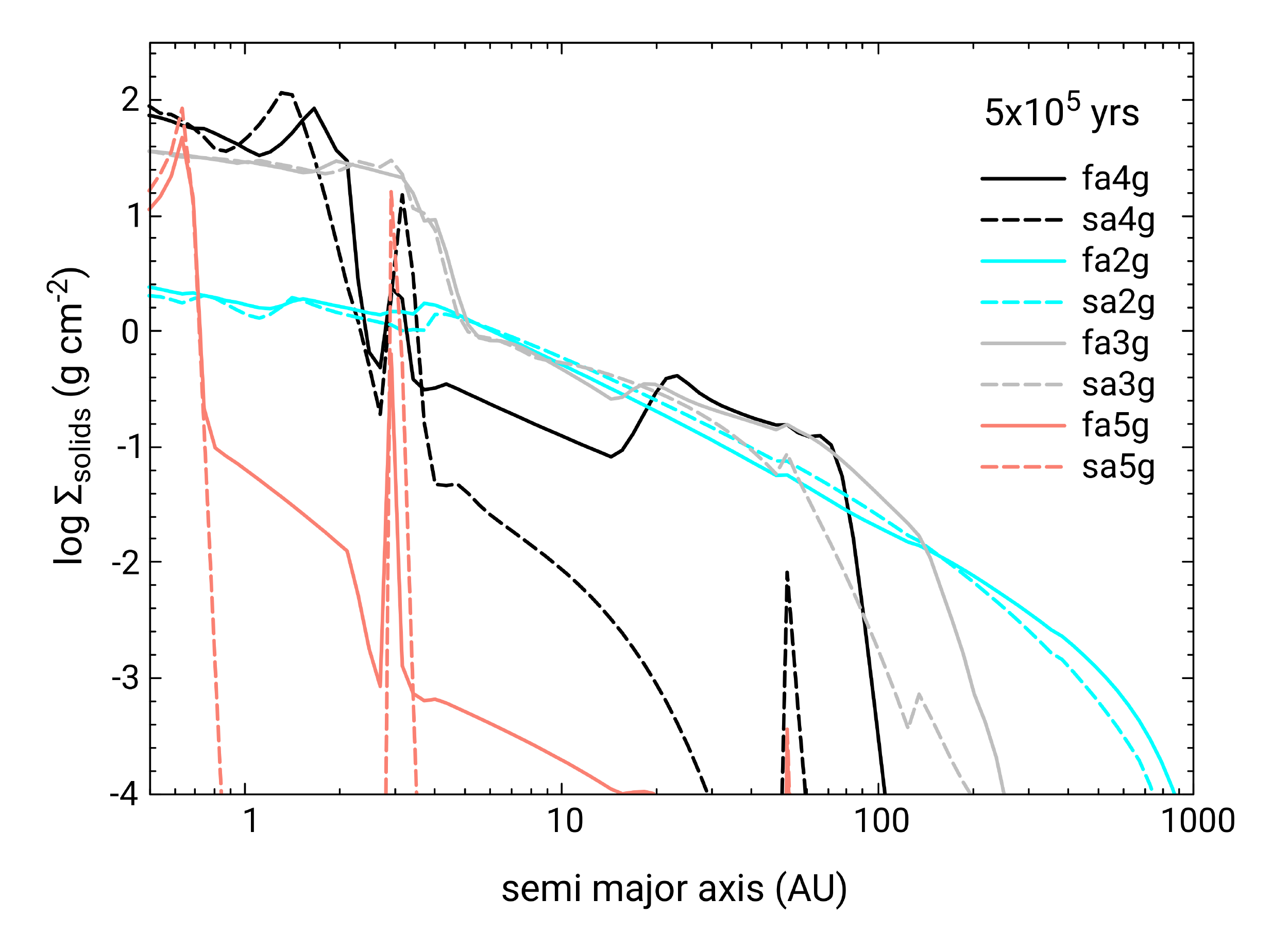}
\caption{Total solids surface density for the sa2g, sa4g and sa5g (dashed curves) and fa2g, fa4g and fa5g (solid curves) models after 0.5 Myr. Peaks in $\Sigma_{\rm{solids}}$ can be associated with solids enhancement outside EFs. Also plotted in grey are the fiducial models for direct comparison. 
\label{fig:agsurf}}
\end{figure*}

Unlike the model with $\alpha_{\rm{t}}=10^{-2}$, the outer disk beyond the water snowline in the compact growth case for $\alpha_{\rm{t}}=10^{-4}$ effectively becomes drift limited 
after 0.5 Myr. The radial drift curve for $m_{\rm{d}}$ has gradually evolved (over much of the region) well below the most massive particles in the distribution. 
For instance, at 0.1 Myr, the mass dominant compact particles are well below $m_{\rm{d}}$, but by 0.3 Myr (not shown) the $m_{\rm d}$ curve has dipped below the largest particles outside $\sim 40$ AU, and under these conditions we can say growth is ``drift limited''. Note that the St associated with $m_{\rm{d}}$ is now $\ll 1$. The reason of course is because the growth times have increased greatly due to the much lower $Z$ in the region. As mentioned in Section \ref{sec:fiducial}, particles can still continue to grow beyond $m_{\rm{d}}$, but 
material begins to drain (drift inwards) from the region more quickly than it grows. Particles Stokes numbers would need to exceed unity to reverse this trend, but they can never do so (Sec. \ref{sec:barriers}). Indeed, after 0.5 Myr
much of the disk material has migrated into the inner disk (see Fig. \ref{fig:agsurf}) 
and further growth is quashed well before reaching the fragmentation barrier.
Some enhancements in $\Sigma_{\rm{solids}}$  are seen outside the EFs for CO$_2$ ($\gtrsim 50$ AU) and CH$_4$ ($\gtrsim 125$ AU), which observationally may show up as supervolatile bands, particularly in young Class 0/I disks \citepalias[e.g., \citealt{SC20}; see][for more discussion]{Est21}. 
The systematic draining of the outer disk material into the inner disk is an expected result for low-$\alpha_{\rm{t}}$, compact  particle growth models \citepalias[see][and references therein]{Est16}. 

In the $\alpha_{\rm{t}}=10^{-4}$ fractal case, much like we saw for the fiducial model (fa3g), rapid growth just outside the H$_2$O EF leads to a sharp decrease in surface density in a band out to $\sim 10-20$ AU, much earlier on (third row, $10^5$ years), but much more material remains outside 20AU  compared to the compact growth model (see Fig. \ref{fig:agsurf}, solid and dashed black curves). Initially, in the outer disk most of the mass is not fragmentation limited, though by 0.5 Myr the most massive aggregates have approached $m_{\rm{f}}$ out to $\sim 15$ AU. At that same time a significant fraction of the particle mass distribution lies above the $m_{\rm d}$ curve (between $\sim 8-20$ AU), due to a dip in the solids surface density which significantly lengthens the growth times. 
Interestingly, the maximum particle masses achieved in the compact and fractal cases in the inner disk are roughly similar to each other (note that the fragmentation masses themselves only differ by $\sim 10$), 
but the maximum masses achieved outside the snowline differ by 4 orders of magnitude with the fractal case reaching particle masses as high as $\sim 10^9$ g. Although large, this is still not massive enough for compaction by self-gravity. Using our nominal particle density ($1.52$ g cm$^{-3}$, Sec. \ref{sec:collcomp}), this is a $\sim 75$ m size object with mass equivalent to roughly a 5 m radius 
 compact  particle, whereas in the  compact growth case, the largest radius particle is $\sim 25$ cm (see Appendix \ref{sec:sizes}, Fig. \ref{fig:wireframes}).

The similarity in the inner disk maximum particle masses can also be explained with the help of the distribution of Stokes numbers shown in Fig. \ref{fig:a4g} (2nd and 4th rows). Both the compact  and fractal case are in the Stokes regime, with $m_{\rm{f}} > m_\lambda$ and ${\rm{Re}}_{\rm{p}} < 1$, 
so neither the largest compact  nor fractal aggregate particles have any dependence on the local $T$ and $\Sigma$, and in fact the Stokes numbers are roughly the same for the two cases. 
Outside the snowline after 0.5 Myr, a typical mass-dominant particle Stokes number for the compact  case is ${\rm{St}} \sim 0.02-0.04$, but for the fractal case ${\rm{St}} \sim 0.1 - 0.15$. The disparity is because the maximum achieved mass in the compact case is severely drift limited, whereas the fractal case is mostly fragmentation limited, and drift only starts to become a limiting factor much later in the simulation. 

The fractal case still maintains a region outside of $\sim 20$ AU with a significant surface density of solids (see Fig. \ref{fig:agsurf}) that is increasingly below the fragmentation curve, but becomes drift limited again around $\sim 80$ AU. 
Material just outside the (steeply decreasing) radial drift curve in the rightmost panel of Figure \ref{fig:a4g} (3rd row) has significantly larger Stokes numbers than the material interior to it (see the inflection at $\sim 10^3$ g in the 4th row). Thus material is drifting in rapidly there, and corresponds nicely to a steep drop in $\Sigma_{\rm{solids}}$ (Fig. \ref{fig:agsurf}, black solid curve). This is not seen in the higher $\alpha_{\rm{t}}$ models where disk spreading (and smaller St) maintains all masses well below the radial drift curve (Fig. \ref{fig:a2g}, 3rd row). Even though aggregates have $m < m_{\rm{d}}$ in the region between $\sim 20-80$ AU, we expect that even this region 
will become depleted at later times. In both models, outside of $\sim 300$ AU, physically small particles remain trapped because of their large ${\rm{St}}$ (yellow colors), but these particles may eventually grow and drift as the gas disk slowly spreads outwards, decreasing their St as the local gas density increases. 

It should be noted that the high Stokes numbers outside the snowline combined with low $\alpha_{\rm{t}}=10^{-4}$ turbulence means that in this model eddy-crossing effects are just starting to become important (Sec. \ref{sec:barriers}). Indeed, the useful parameter ${\rm{St}}/\alpha_{\rm{t}}$ is $\sim 300$ (see Sec. \ref{sec:barriers}) for the compact  case, and $\sim 1200$ in the fractal case, both comparable or larger than the local headwind parameter $\beta^{-1}$, a necessary condition for $y^* > 1.6$ 
\citep[Fig. \ref{fig:ageta}, black curves; see Sec. \ref{sec:barriers} and also][for more discussion on the significance of ${\rm{St}}/\alpha_{\rm{t}}$]{OC07,Jac12,Jac14}. For the fractal case, we note that the fragmentation curve lies slightly above the particle mass distribution. This is because the eddy-crossing correction leads to a significantly different $y_*$ and collision speed such that 
the simple expression in Eq. \ref{equ:Stfrag} used in these fragmentation curves is an overestimate. 

\begin{figure}
\centering
\includegraphics[width=0.4\textwidth]{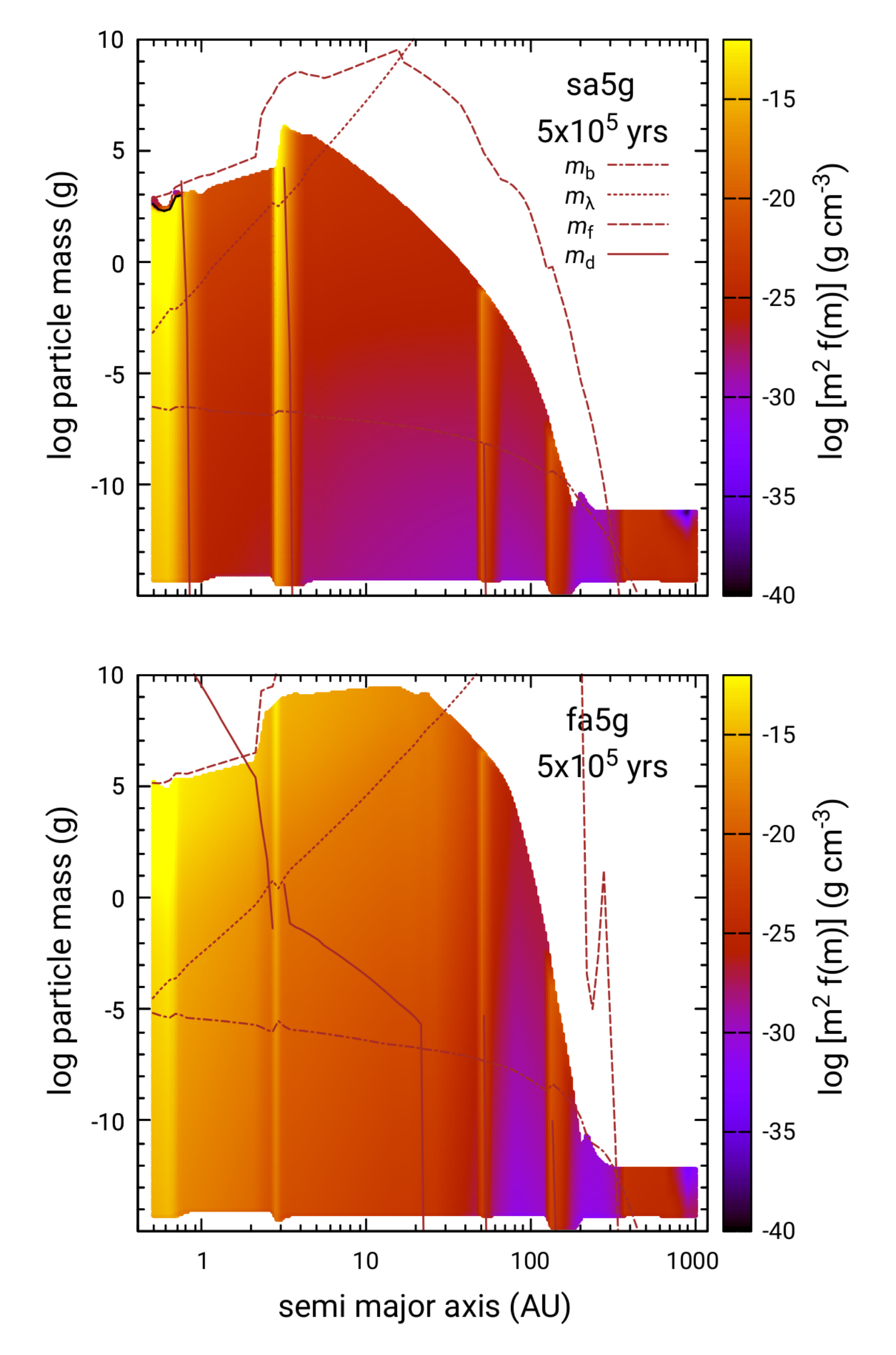}
\includegraphics[width=0.4\textwidth]{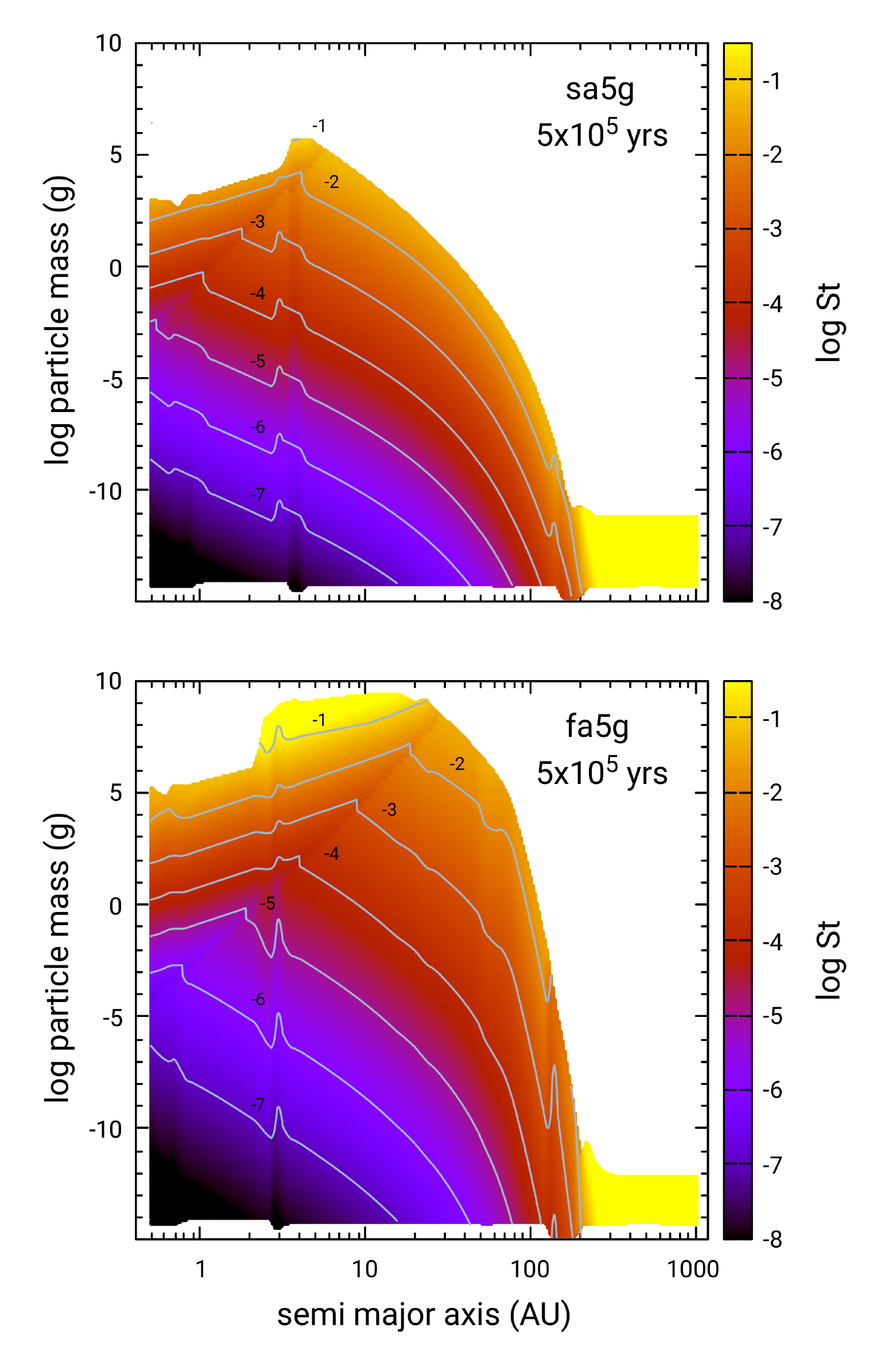}
\caption{Comparison of models with $\alpha_{\rm{t}}=10^{-5}$ for both compact  (top panel) and fractal (bottom panes) growth at $5\times 10^5$ years. Plotted in each panel is a snapshot of the particle mass distribution as a function of semi-major axis. Descriptions the same as in Fig. \ref{fig:a3g}.}
\label{fig:a5g}
\end{figure}

 \begin{figure*}
\centering
\includegraphics[width=0.5\textwidth]{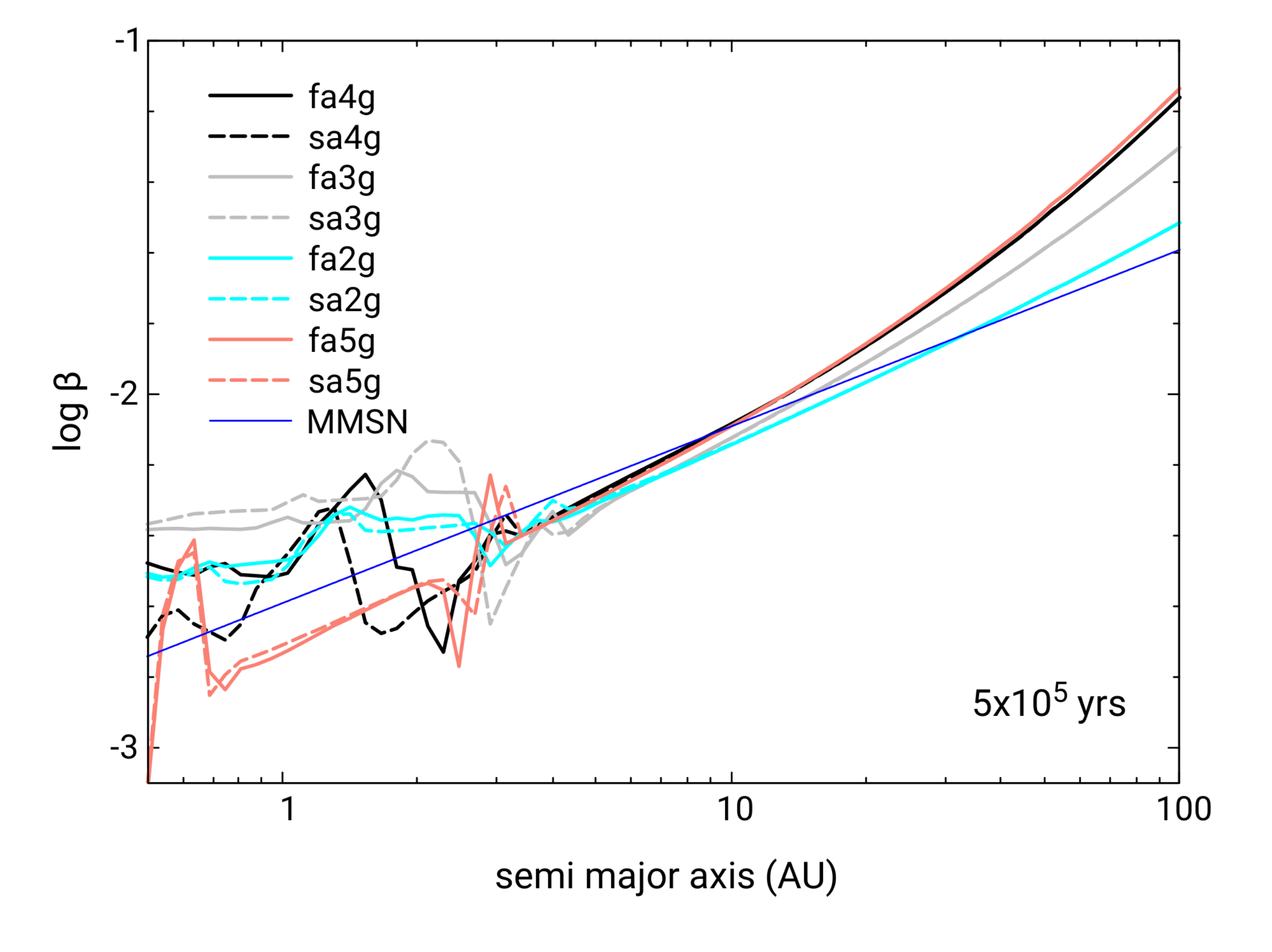}
\caption{Pressure gradient $\beta$ as a function of semi-major axis after 0.5 Myr for all models discussed thus far. Solid curves correspond to fractal models, and dashed to compact growth models. The blue curve corresponds to the MMSN nebula. Though the gradient shows considerable variation in the most inner disk regions, at no point is $\beta < 0$, so appreciable pressure bumps do not develop over the simulation time of these models. 
\label{fig:ageta}}
\end{figure*}

 \begin{figure*}
\centering
\includegraphics[width=0.5\textwidth]{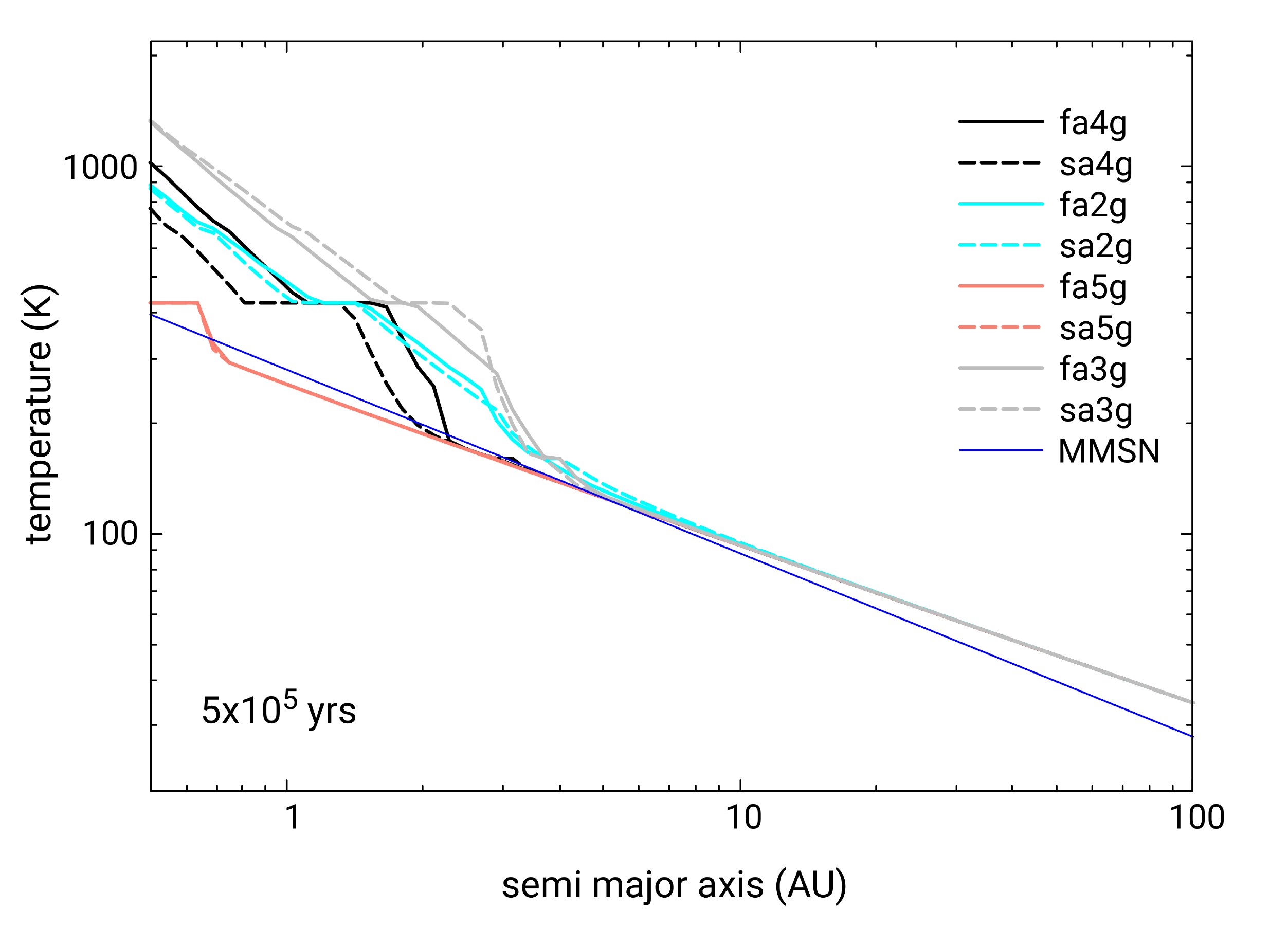}
\caption{Midplane temperature profiles after 0.5 Myr for all models discussed thus far.  Solid curves correspond to fractal models, and solid curves to compact models.  The blue curve corresponds to the MMSN nebula.
\label{fig:agtemp}}
\end{figure*}

We now turn to compact and fractal growth models for $\alpha_{\rm{t}}=10^{-5}$. The rapid loss of material via radial drift is even stronger as can be seen in Figure \ref{fig:a5g} where we show the solids mass volume density (left) and ${\rm{St}}$ (right) for the compact particle (top) and fractal growth (bottom) simulations after 0.5 Myr. In this case ${\rm{St}}/\alpha_{\rm{t}} \sim 1-3\times 10^4$ for the compact and fractal models, considerably larger than $\beta^{-1}$ (Fig. \ref{fig:ageta}, salmon curves) compared to the $\alpha_{\rm{t}}=10^{-4}$ case where ${\rm{St}}/\alpha_{\rm{t}}$ is smaller.  Here, by 0.5 Myr the compact particle growth simulation has lost almost all of the outer disk and much of the inner disk material, leaving distinct bands of trapped solids at the water and supervolatile EFs. The innermost region ($\lesssim 0.8$ AU) 
is a rich band of organics. This strong band arises, however, because we have treated organics evaporation and recondensation as a reversible process, which is likely not accurate \citepalias[see footnote 11 and][for more discussion of compositional evolution]{Est21}. In the fractal growth case, the solids surface density  profile changes considerably with time, looking similar to the $\alpha_{\rm{t}}=10^{-4}$ case at 0.2 Myr outside the snowline (0.2 Myr snapshots not shown for either $\alpha_{\rm t}$), but by 0.5 Myr most of this material has been lost inwards due to even faster drift than for $\alpha_{\rm t} = 10^{-4}$, and only leaving material remaining trapped in the bands outside their corresponding EFs (Fig. \ref{fig:agsurf}, salmon solid curve). 
In the region outside the snowline, fractal aggregates have achieved masses  $\gtrsim 10^9$ g, with ${\rm{St}} \sim 0.2-0.3$ (both about a factor of two larger than the $\alpha_{\rm{t}}=10^{-4}$ case), 
but the local $Z$ has decreased strongly to $Z\sim 2\times 10^{-4}$. 

In Figure \ref{fig:agtemp} 
we plot the midplane temperature profiles after 0.5 Myr for all the models discussed so far. 
The temperature profiles for the fiducial model (grey curves) are included for comparison. As one might expect with the stronger diffusion and advection associated with the higher $\alpha_{\rm{t}}=10^{-2}$ case (cyan curves), the temperature profile is less steep overall than in the  $\alpha_{\rm{t}}=10^{-4},10^{-3}$ models. In the latter cases, a steeper gradient in the temperature builds up inside the snowline (which lies further inward than in the  $\alpha_{\rm{t}}=10^{-2}$ case) due to the inward drift of material there that increases $\Sigma_{\rm{solids}}$ (see Fig. \ref{fig:agsurf}), and thus the local $Z$ and opacity. The pressure gradient is very much affected by the steep gradients in $T$, 
as can be seen by the variations in Fig. \ref{fig:ageta}. However, because aggregates are smaller and more coupled to the gas flow for $\alpha_{\rm{t}}=10^{-2}$, 
there is less variation in both $\Sigma_{{\rm{solids}}}$ and $\beta$ over the disk (cyan curves in Figs. \ref{fig:agsurf} and \ref{fig:ageta}). 
It should be noted that the variations in $\beta$ across the model set are not sufficient to lead to pressure bumps that could potentially trap particles. When comparing these temperatures across the range of $\alpha_{\rm t}$, the fiducial model (grey curves) is hotter inside the snowline than both the higher and lower turbulence cases after 0.5 Myr, which may seem counterintuitive. As it turns out, an optimal combination of solids enhancement across the snowline (which is highest in these models after 0.5 Myr), the mass/size of the dominant aggregate, and the ambient gas surface density  keeps the region inside the snowline hotter longer for the fiducial model. In the $\alpha_{\rm{t}}=10^{-2}$ case, the dominant aggregate masses are smaller than the fiducial model, increasing the opacity, but $Z$ is not much enhanced by drift, and $\Sigma$ is much lower because of fast viscous evolution. In the  $\alpha_{\rm{t}}=10^{-4}$ case as in the fiducial model, the enhancement in $Z$ in the inner disk is also high due to rapid radial drift, and  $\Sigma$ remains high due to slower viscous evolution, but the larger dominant aggregate masses allowed by $\alpha_{\rm t}=10^{-4}$ lead to lower opacities. Thus $\alpha_{\rm{t}}=10^{-3}$ seems to be a ``sweet spot'' from the standpoint of retaining a hot, massive inner nebula. In the $\alpha_{\rm{t}}=10^{-5}$ case (salmon curves), $Z$ is very low because of inward  drift out of the inner nebula such that the opacity is very small, other than in regions just outside an EF. In \citetalias{Est21} we discuss the EFs in more detail.

\begin{figure*}
\includegraphics[width=1.0\textwidth]{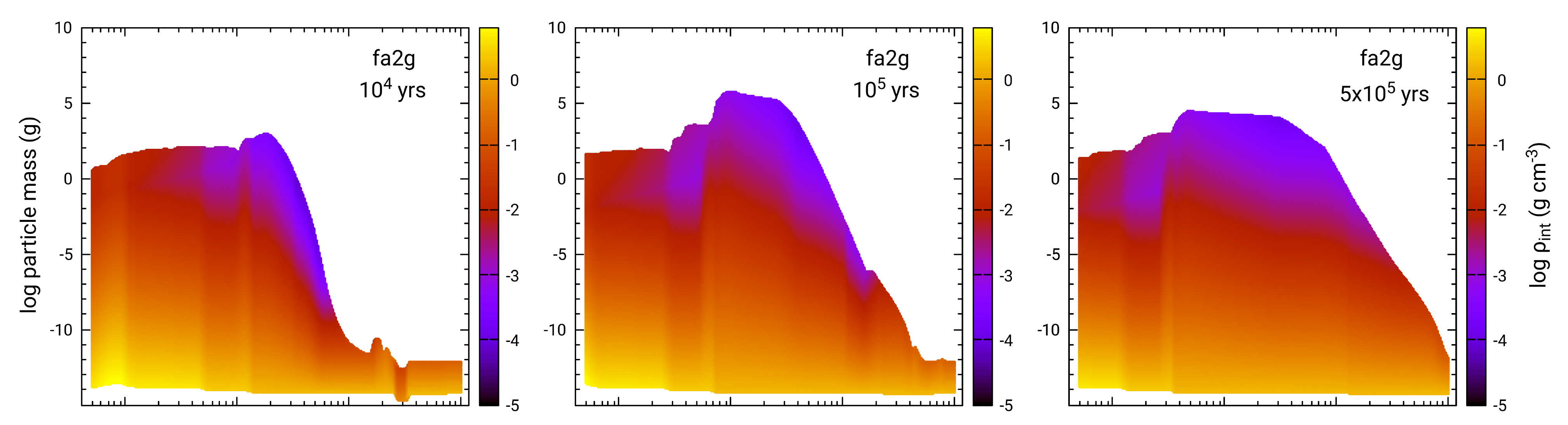}
\includegraphics[width=1.0\textwidth]{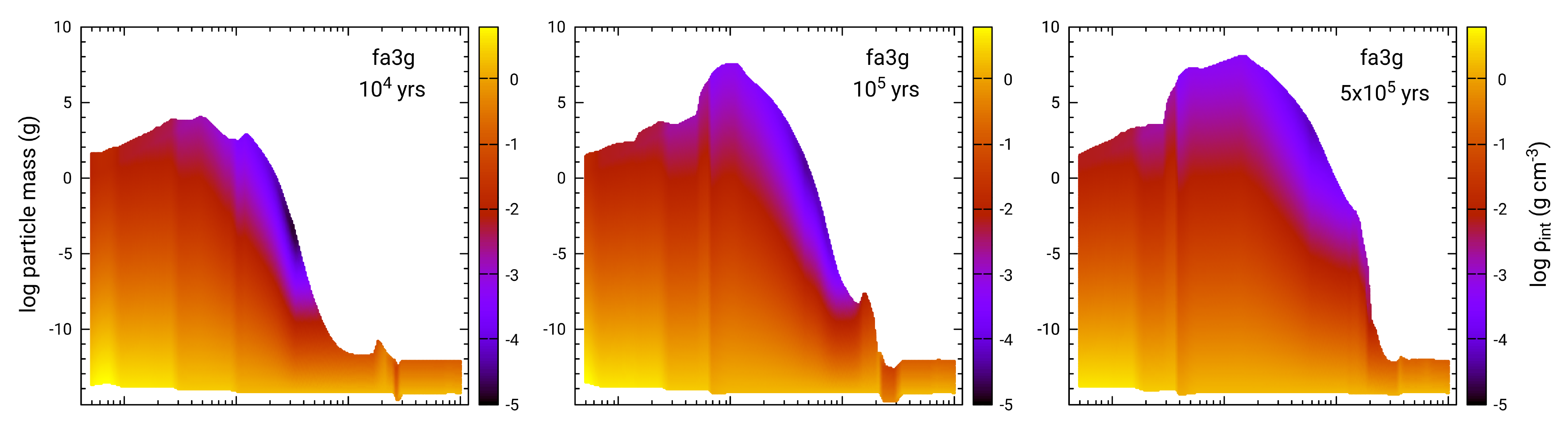}
\includegraphics[width=1.0\textwidth]{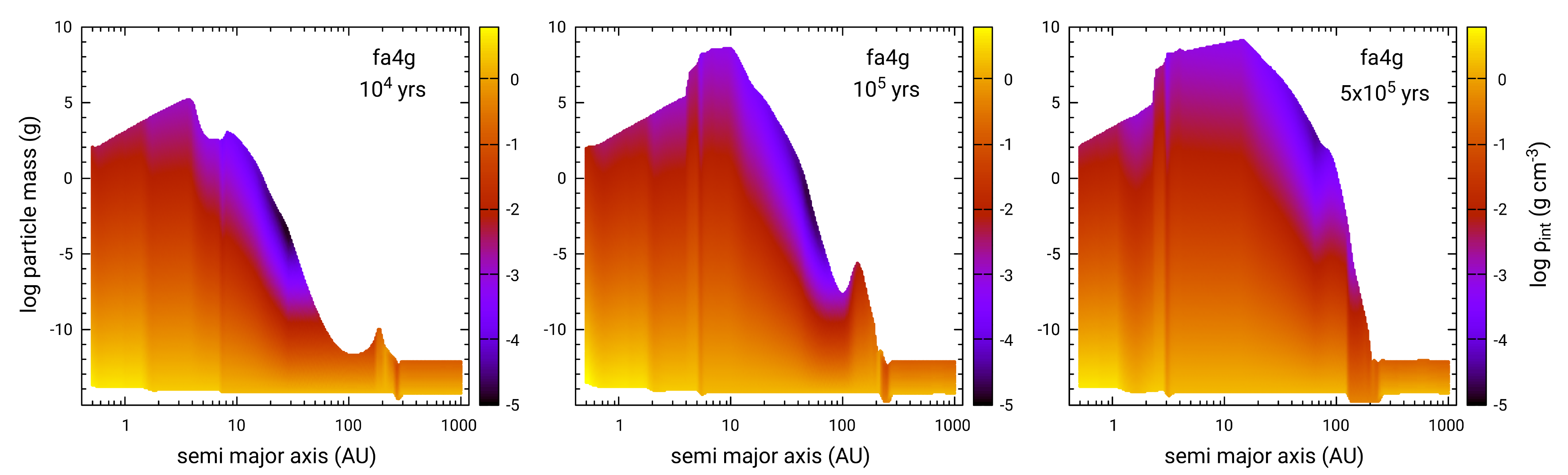}
\caption{Evolution of the {\it internal} density $\rho_{\rm{int}} = \rho_{\rm{p}}/\psi$ of aggregates for models fa2g, fa3g and fa4g shown in Figs. \ref{fig:a3g}, \ref{fig:a2g} and \ref{fig:a4g}. Despite being relatively massive ($\sim 10^5-10^9$ g), the large aggregates outside the snowline are quite underdense, though they continue to be compacted. In the inner disk, particles were initially underdense, but begin to be compacted earlier on and more vigorously. Transitions in density due to compositional changes at EFs can be seen as vertical  bands in the plots. 
Darker vertical bands (lower density particles) initially seen at $\sim 250$ AU are solid methane \citepalias[see][]{Est21}. This region of the disk also exhibits  rapid inward particle drift where ${\rm{St}}\sim 1$ (a location that moves outwards with time due to viscous spreading of the disk), which manifests as the peak in particle mass seen at 150-200 AU until $10^5$ years. The particle size distributions for these models after 0.5 Myr can be seen in Appendix \ref{sec:sizes}.
\label{fig:agdens}}
\end{figure*}

\subsection{Evolution of Internal Aggregate Densities}
\label{sec:dens}

In Figure \ref{fig:agdens} we plot the particle  internal density $\rho_{\rm{int}} = \rho_{\rm{p}}/\psi$ (recall $\psi$ is the enlargement factor, Sec. \ref{sec:fracgrow}) for our fiducial model (Sec. \ref{sec:fiducial}, middle row), model fa2g (top row) and model fa4g (bottom row) to better visualize the underdense nature of the fractal aggregates across these simulations. In all models, aggregates initially grow fractally with $D\simeq 2$ but eventually begin to compact as collisions become sufficiently energetic and $E_{\rm{roll}}$ is exceeded, and/or more often, gas ram pressure sets in (Sec. \ref{sec:collcomp}). In these plots only the monomer mass (the minimum mass in the distribution) has a compact  particle density $\rho_{\rm{p}}$, which is a function of semi-major axis. The locations of EFs where changes in aggregate density occur are in most cases discernible upon inspection (e.g, the water snowline is located between $\sim 3-4$ AU in these models after 0.5 Myr), even for CO (e.g., middle row at $\sim 350$ AU). \citetalias{Est21} discusses the radial distribution of aggregate bulk composition 
in more detail.

In the inner disk, compaction occurs much more quickly due to the faster dynamical times, higher gas surface density and a lower rolling energy threshold. Once the fragmentation barrier is reached in the inner disk, internal densities vary within a range of $\rho_{\rm{int}} \sim 10^{-3}-10^{-2}$ g cm$^{-3}$.  Outside the snowline where $E_{\rm{roll}}$ can be about an order of magnitude larger (Sec. \ref{sec:collcomp}, though see Sec. \ref{sec:varparm} regarding ``cold ice"), the most massive aggregates are more underdense with $\rho_{\rm{int}}\sim 10^{-4}-10^{-3}$ g cm$^{-3}$ (or even smaller). 
The maximum achieved aggregate masses inside the snowline differ by $\sim 1-2$ orders of magnitude between these fractal models, but the outer disk has maximum aggregate masses ranging from $\sim 10^5-10^9$ g. For the given densities, this means that the effective radius of these fractal aggregates ranges from $\sim 3-75$ m, with the upper value corresponding to model fa4g at $\sim 14$ AU. Even the high end of this range of aggregate mass is 
still too small for self-gravity to be important (note that for $\alpha_{\rm{t}}=10^{-5}$, the largest aggregate has a mass roughly twice this value). By equating $P_{\rm{c}} = P_{\rm{grav}}$ (see Sec. \ref{sec:collcomp}), we can estimate what the mass needs to be for self-gravity compaction if one knows the current fractal dimension $D$:

\begin{equation}
    \label{equ:mgrav}
    m_{\rm{grav}} \gtrsim \left[\left(\frac{3}{4\pi\rho_{\rm{p}}}\right)^{1/3}\frac{\pi E_{\rm{roll}}}{G}m_0^{5/D-8/3}\right]^{D/(5-D)}.
\end{equation}

\noindent
At 0.5 Myr and at 14 AU in the $\alpha_{\rm{t}}=10^{-4}$ model, $D \simeq 2.6$ and $E_{\rm{roll}} \sim 7.2\times 10^{-9}$ g cm$^2$ s$^{-2}$ which predicts $m_{\rm{grav}} \sim 7\times 10^{10}$ g. However, it is unlikely that icy aggregates will ever achieve such a mass, as they are not only fragmentation limited, but radial drift limited as well (Fig. \ref{fig:a4g}, 3rd row, 3rd panel).

While the aggregates outside the snowline appear quite large, and have very large cross sections, their rapid growth to such large sizes has come at the cost of rapidly decreasing $\Sigma_{\rm{solids}}$ of their ``feedstock", which slows down their growth rate. Despite fractal growth leading quickly to very underdense aggregates, which drift more slowly than a compact particle of equivalent mass, the local nebula conditions don't allow for continued incremental growth to larger sizes, and conditions are such that the local $Z$ is significantly reduced. 
In Section \ref{sec:obs}, we discuss how these highly porous aggregates relate to the observations.

\section{Discussion} 
\label{sec:discuss}

\subsection{Implications for Planetesimal Formation}\label{sec:pform}
\begin{figure*}
\includegraphics[width=1.0\textwidth]{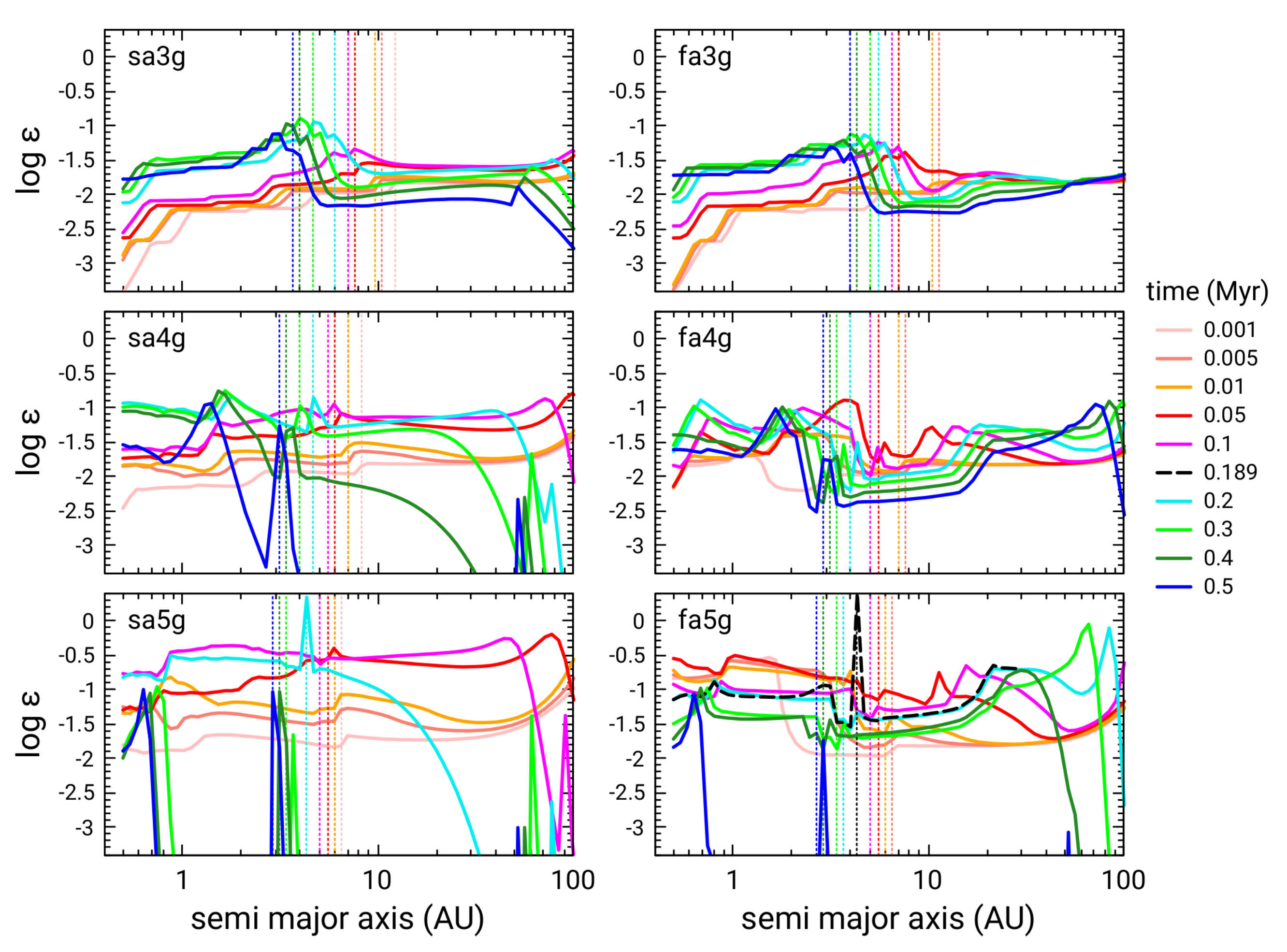}
\caption{Evolution of the midplane  solids-to-gas ratio $\epsilon =\rho_{\rm{solids}}/\rho$ over time, for compact particle (left panels) and fractal aggregate (right panels) growth models with $\alpha_{\rm{t}}=10^{-3}$, $10^{-4}$ and $10^{-5}$. The location of the water ice snowline at the corresponding simulation times is indicated by the vertical dotted lines. Only the lowest turbulence model sees a very short period where $\epsilon > 1$ at the snowline, with ${\rm{St}}\gtrsim 0.1$. Models with $\alpha_{\rm{t}}=10^{-2}$ are not shown because there is no significant enhancement in $\epsilon$ anywhere. Planetesimal formation by Streaming Instability requires midplane $\epsilon >1$ \citep[\citetalias{Est16};][]{Umu20}.
\label{fig:stog}}
\end{figure*}

Though we do not currently include an algorithm for planetesimal formation in our code, we can say with confidence that planetesimals do not likely form via incremental growth over the phase space we have explored. We can also explore whether conditions for leapfrog planetesimal formation by SI can be satisfied. These conditions are determined almost entirely by a combination of St and $\alpha_{\rm{t}}$ through the midplane solids-to-gas volume density ratio $\epsilon=\rho_{\rm{solids}}/\rho\approx Z\sqrt{(\alpha_{\rm{t}} + \rm{St})/\alpha_{\rm{t}}}$ \citep[and Paper I]{Umu20}. To that end, Figure \ref{fig:stog} shows a more detailed temporal evolution of $\epsilon(R)$ from 0.001 to 0.5 Myr, for a subset of compact particle (left panels) and fractal aggregate (right panels) growth models for turbulent intensities of $\alpha_{\rm{t}} = 10^{-3}$, $10^{-4}$ and $10^{-5}$. Models for $\alpha_{\rm{t}}=10^{-2}$ are not shown because for this large diffusivity, the well-coupled nature of the particles and aggregates to the nebula gas prevents any significant enhancements anywhere in the disk.  Indeed, the distinction in the evolution of nebular properties between fractal aggregate and compact  particle growth models (apart from their difference in particle mass) begins to disappear with increasing $\alpha_{\rm{t}}$  over the epoch simulated here.

The top panels show that, at first blush, the compact and fractal growth simulations for  $\alpha_{\rm{t}}=10^{-3}$ (our fiducial case)  initially look very similar. At a closer look, the more rapid growth of fractal aggregates in the fractal growth model fa3g just beyond (and to some degree, inside) the evolving snowline 
leads to small differences in both the inner and outer disks (Sec. \ref{sec:fiducial}). However, by $5\times 10^4$ years (the H$_2$O EF is located at $\sim 7$ AU) a deviation is seen where model fa3g has achieved a higher $\epsilon$ due to faster growth (and thus drift) of material from $\lesssim 20$ AU. By 0.1 Myr though, the compact growth model sa3g has largely caught up, and the variations of $\epsilon$ inside the snowline look relatively similar after that, except that the enhancements at the H$_2$O EF for model sa3g are somewhat larger. The reason for this has to do with the much lower ${\rm{St}}$ of fractal aggregates beyond $\gtrsim 20$ AU compared to compact particles, which allows mass to be retained in the outer disk for longer periods. Thus, in model fa3g, less material has been processed across the snowline because material beyond $\sim 20$ AU has not had time to drift in yet. 
On the other hand, model sa3g -  whose particle ${\rm{St}}$ are about an order of magnitude larger than in model fa3g outside of 20 AU has had a higher, more continuous inward mass flux of material, producing a larger enhancement both outside the snowline, but especially outside the organics EF (the higher of the two notable peaks, e.g., at $1.95$ AU after 0.5 Myr compared to $2.9$ AU for water, see also \citetalias{Est21}) which is further inward. 
Nevertheless, it is expected that the material in model fa3g will eventually drift in at later times. Overall the differences between the fractal and compact  growth cases are small and $\epsilon=1$ is never approached.

The simulations for $\alpha_{\rm{t}}=10^{-4}$ (middle panels) show a sharper deviation between compact and fractal growth. 
Lower turbulence values allow growth to larger St, and thus faster radial drift rates with more pronounced systematic loss of material from the outer disk to the inner disk, as is evident in model sa4g. Local peaks in $\epsilon$ near 100 AU and evolving inward caused by rapid inward particle drift are due to a combination of a strong pressure gradient and increasing St of material further outward.
By 0.5 Myr (blue curve), most of the material outside the snowline (located at $\sim 2.7$ AU) has been depleted in model sa4g, except for some distinct bands of material trapped at the H$_2$O EF and the supervolatile EFs at 50-80AU \citepalias[see][and Sec. \ref{sec:weak}]{Est21}. The more rapid drift and growth to larger sizes further out in the disk in the compact growth model explains the larger contrast in the enhancement in $\epsilon$ outside the snowline between these $\alpha_{\rm{t}}=10^{-4}$ models (and compared to sa3g and fa3g), which occurs simply as a result of more mass being processed across the water snowline in sa4g. In the fractal model, solid material is mostly being tapped from inside $\sim 20$ AU, so model fa4g still contains a significant amount of solid mass outside of $\sim 20$ AU (see Fig. \ref{fig:agsurf}) owing to smaller St numbers and slower drift. As a result, much larger enhancements are seen at the water snowline ($\epsilon \gtrsim 0.1$) in the compact  growth case with the largest peak around 0.2 Myr, and even after 0.5 Myr a strong peak in $\epsilon$ is maintained, but in the fractal model the magnitude of the peaks are notably less in comparison. 
Moreover, in the compact growth model there is a large reservoir of water vapor inside the snowline (see Fig. 6 of \citetalias{Est21}) that contributes to keeping the peak in solids strong (and becoming overwhelmingly water ice composition) via slow outward diffusion, but also as the disk cools and the H$_2$O EF evolves inwards. The reservoir of water vapor in fa4g inside the snowline is also substantial, but lower in magnitude than sa4g (and still of relatively diverse composition) again due to less mass being processed across the EF.
Finally, as in the fiducial  $\alpha_{\rm{t}}=10^{-3}$ model, the strongest enhancement occurs outside the organics EF (which is a questionable feature)\footnote{The enhancement of organic solids outside the organics EF is an artifact of our assumption that organics evaporate and recondense reversibly like the other volatiles. A more realistic model of irreversible decomposition requires a more subtle chemical model \citepalias[see Sec. \ref{sec:diskevol}, and][]{Est21}.}. 

Generally speaking, as $\alpha_{\rm{t}}$ decreases in our models, the resulting growth to larger aggregate or compact  particle masses with larger St, eventually leads to more rapid drift and depletion of much, if not all of the region outside the water snowline.
However, because particles (and their ${\rm{St}}$) are significantly smaller inside the snowline, drift is slower there, allowing for a much larger buildup at the (questionable) organics EF in the compact  case, or even the silicates EF as can be noted in the fractal model for simulation times $\lesssim 0.3$ Myr. 
Indeed, inside the water snowline, one finds sustained regions with $\epsilon \gtrsim 0.1$ over time, but St remain relatively small. Outside the water snowline, the $\epsilon$ can also increase to values exceeding 0.1 (in sa4g) early on as Stokes number growth outpaces any decrease in the local $Z$, but as the masses (and St) increase and with them the radial drift speeds, the local $Z$ {\it drops precipitously} (and eventually $\epsilon$ as well) to values considerably below 0.1 at the same time.
That is to say, there is never a situation where planetesimal formation by SI can occur at any of the EFs in the inner disk or at the water snowline in either model \citep[see][and below for more discussion]{Umu20}.

\begin{figure*}
\centering
\includegraphics[width=0.9\textwidth]{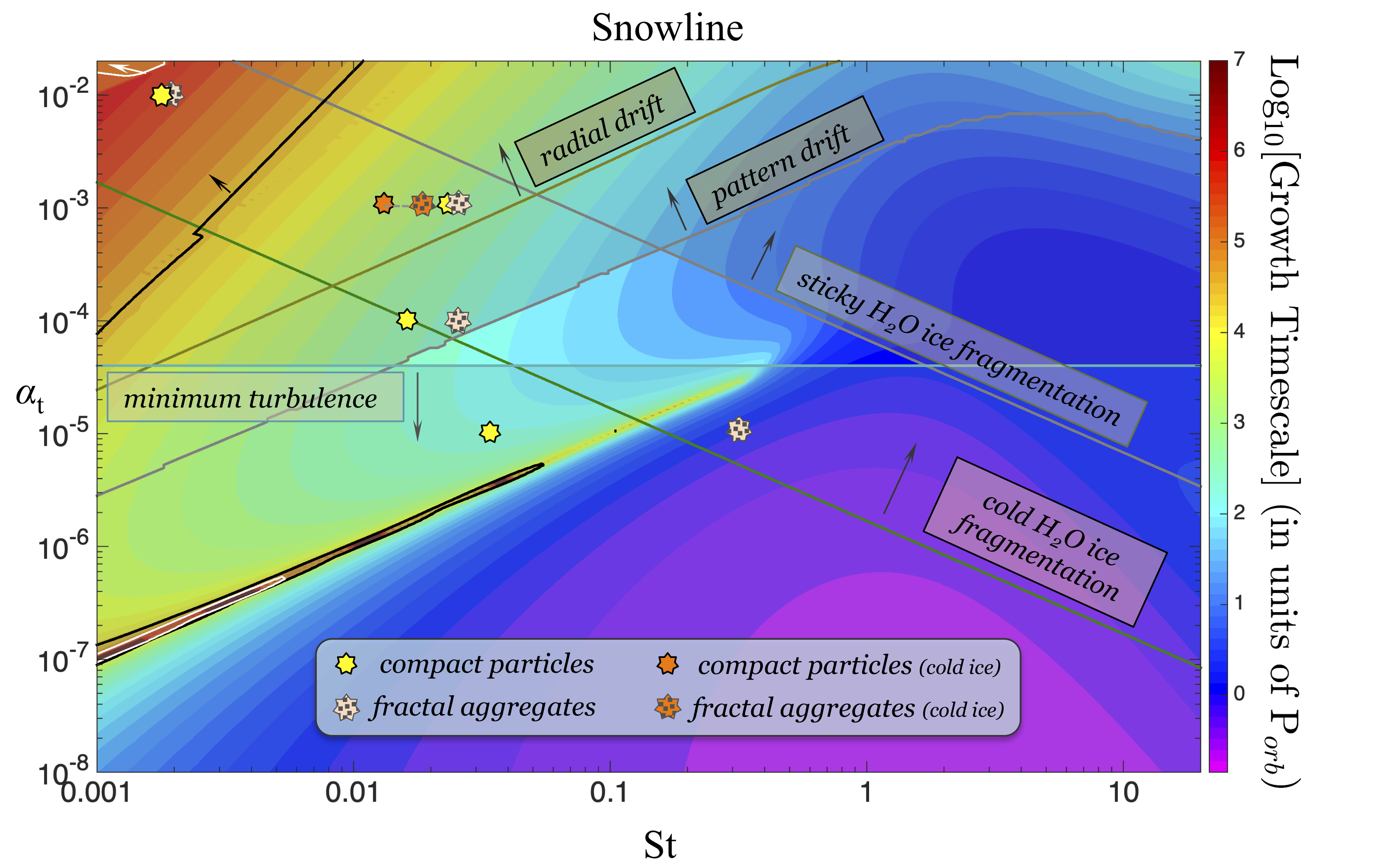}
\caption{Theoretically predicted SI growth timescales in a turbulent disk (colors) with superposed particle St from our models (symbols).  The growth timescales are for the {\it fastest} growing SI modes, and are based on the model of \citet{Umu20} and shown in shades of color in units of local orbit times. A sharp diagonal boundary in $(\alpha_t, {\rm St})$ phase space that cleanly separates the blue and green regions represents the  $\epsilon = 1$ line, with smaller values above the line. The values shown are specifically valid for a local $Z=0.01$ just outside the snowline ({\it i.e.}, $T \approx 160$ K). Various growth barriers
and turbulence constraints discussed in section \ref{sec:pform}  are shown as lines with arrows indicating putative forbidden regions in parameter space.  The plotted St symbols correspond to  particle radii dominating the particle mass, from our models for four different values of $\alpha_t$. Solid filled symbols represent non-porous particles while
stippled symbols represent porous fractal aggregates.  Two classes of particles are shown at $\alpha_{\rm t} = 10^{-3}$: the yellow symbols are for nominal ``sticky" water ice, and the orange colored
species denote particles whose fragmentation energies ($Q_{\rm{f}}$) are relatively low, representing cold water ices as discussed in section \ref{sec:varparm}  \citep[e.g.,][]{MW19}.  The particle St predicted in our growth results generally fall within the $\epsilon < 1$ region
of parameter space wherein SI growth times are predicted to be very long, in agreement with all published numerical simulations \citep{Umu20}. Only the fractal growth model (fa5g) for an extremely low  $\alpha_{\rm{t}}=10^{-5}$ falls below the $\epsilon = 1$ line. The compact  growth model (sa5g) can also satisfy this condition, but only at higher local $Z$ \citep[the $\epsilon = 1$ curve shifts upwards for increasing values of the metallicity, see][]{Umu20}.
\label{fig:orkfig}
}
\end{figure*} 

One interesting wrinkle appears when $\alpha_{\rm{t}}=10^{-5}$ (bottom panels). These extremely low global  turbulence models allow growth to much larger masses, which show an even faster radial drift. However, we find  for both the compact particle (sa5g) and fractal aggregate (fa5g) growth models that there is a brief period of time near $\sim 0.2$ Myr that lasts $\sim 2\times 10^3 - 2\times 10^4$ years during which a buildup in the solids surface density at the water snowline leads to $\epsilon > 1$. The corresponding Stokes numbers (particle radii) are $\sim 0.05$ (38 cm) and $\sim 0.35$ (150 m radius). These two conditions together would satisfy the condition for SI. Like the models for $\alpha_{\rm{t}}=10^{-4}$ though even more so, the spike in $\epsilon$ is due to the sustained high mass influx of material drifting in from the outer disk, and partially because of the inward migration of the water snowline as the disk cools. In sa5g and fa5g, the relatively faster cooling time allows for the even stronger enhancement of water vapor inside snowline to condense outside the snowline. For instance, in model fa5g, the local solids  mass density ratio $\epsilon$ increases by nearly two orders of magnitude (black dashed curve), to a composition of nearly pure water ice (the compacted densities are near $\sim 0.9$ g cm$^{-3}$). This pure ice composition is discussed in \citetalias{Est21}, where we explore the bulk composition of planetesimals if they were to form anywhere in the disk.
After this brief period where $\epsilon$ spikes, $\epsilon$ decreases falling below unity. 
By 0.3 Myr, most of the outer disk in model sa5g has collapsed into the inner region and is being lost to the central star, whereas significant solids mass remains in the disk even after 0.4 Myr in model fa5g. By 0.5 Myr though, most of the disk in the fractal aggregate model is depleted as well.

Some recent models of compact  particle growth have indeed assumed a turbulent intensity as low as $10^{-5}$, {\it but only for the particles}, while allowing for far larger turbulence driving disk evolution \citep{DD18}. However, these authors find planetesimal growth (based on satisfying the condition $\epsilon \ge 1$ and ${\rm{St}} \ge 0.01$) even for particle layer $\alpha_{\rm{t}}$ as high as $10^{-3}$, so while the behavior we see here could be connected to the fact that SI can form planetesimals at the snowline in those models, it is hard to judge the effects and viability of using different values for the turbulence for the particles and gas, and for such high values of $\alpha_{\rm{t}}$ where we do not see the conditions for SI satisfied\footnote{Note that \citet{DD18} use the symbol $\alpha_{\rm{t}}$ for the particles, and $\alpha_{\rm{v}}$ for the gas, whereas our single $\alpha_{\rm{t}}$ applies to both disk evolution and turbulent mixing of the particle layer.}.

Finally, we note that if conditions for SI {\it are} satisfied at the snowline,  the simulations may not properly reflect what happens afterwards since a period of planetesimal formation may produce bodies which are effectively immobile over long time scales. These planetesimals might sweep up inwardly drifting material, or form a planetary core, either of which could be a barrier to further inward mass flux. 
Thus the results for the extremely low value of $\alpha_{\rm{t}}=10^{-5}$ 
should only be cautiously extended beyond those times  
(see below).
These  results for different $\alpha_{\rm{t}}$ highlight the complexities to be found in these global evolutions, and the critical role played by $\alpha_{\rm t}$.


%
%

What do these predictions imply about SI activity?  In Figure \ref{fig:orkfig} we plot predicted SI growth rates under putative disk turbulent conditions, along with the largest value of ${\rm{St}}$ admitted by our modeling for several different values of $\alpha_{\rm t}$. The SI growth rates (colors) are based on the theory presented in \citet{Umu20} wherein turbulence is represented
by an enhanced viscosity and particle diffusivity quantified by $\alpha_{\rm t}$.  In this figure the growth rate of the {\it fastest} growing
SI mode is shown as a function of $\alpha_{\rm t}$ and ${\rm{St}}$ for a disk model with $Z=0.01$ and with a local opening angle $\delta \equiv H/R = 0.07$ 
representative of thermodynamic conditions appropriate to the snowline \citep[$T\lesssim 160K$;][also see their Table 6 and Fig. 12]{Umu20}\footnote{The St values for $\alpha_{\rm{t}}=10^{-2}$ were incorrectly input in Table 6 and Fig. 12 of \citet{Umu20}. The ${\rm{St}}$ for these particles are smaller, as shown in Fig. \ref{fig:orkfig}.}. All St points are from our models, and selected at times that closely satisfy these metallicity and temperature conditions.
As discussed at length in both \citet{Umu20} and \citet{CL20}, the diagonal $\epsilon = 1$ line 
(where $\epsilon = \rho_{{\rm solids}}/\rho_g \approx Z \sqrt{1+{\rm St}/\alpha_{\rm t}}$) 
divides the parameter space regarding SI activity. For local conditions
where $\epsilon$ falls below 1 (the region in Fig. \ref{fig:orkfig}  above the $\epsilon = 1$ line  and with  ${\rm{St}} \lesssim 0.3$) the SI is severely weakened, rendering it essentially inoperative. For $\epsilon > 1$ (below the $\epsilon = 1$ line  and/or ${\rm{St}} >0.3$), the SI  can grow relatively fast (1-10 orbit times), and planetesimals readily form in numerical simulations conducted assuming these conditions, as  discussed in \citet{Umu20}.
The graphic also shows other lines representing several familiar particle growth obstacles: cold and warm ice fragmentation, 
particle radial drift, and SI mode pattern drift \citep[the prescriptions for these constraints are detailed in Sections 7.2-7.4 of][]{Umu20}. The plot also shows a horizontal line representing the likely minimum level of 
globally driven HD disk turbulence with
$\alpha_{\rm t} > 4\times 10^{-5}$ \citep[e.g., the VSI, ][]{Flock_etal_2017,LU19}, which is likely to be widespread throughout the 
disk except possibly for singular narrow radial patches in which no linear instability is thought to operate \citep[e.g., see Figs. 1 and 10 of][]{Pfeil_Klahr_2019}.
 
As Fig. \ref{fig:orkfig} demonstrates, most of the self-consistent combinations of particle ${\rm{St}}$ and turbulent intensity $\alpha_{\rm{t}}$ that result from our growth modeling fall within the $\epsilon < 1$ zone of the $\alpha_{\rm{t}}$-St parameter space
in which the SI 
does not lead to planetesimal formation. Only particle growth models with $\alpha_{\rm{t}} = 10^{-5}$ show the possibility of $\epsilon > 1$ (and then for only a brief time, as discussed earlier in this section)\footnote{Because of the restriction to plot values for $Z=0.01$, the compact  growth model point for $\alpha_{\rm{t}}=10^{-5}$ appear above the $\epsilon = 1$ line even though conditions were satisfied for SI in a brief window. However, unlike the fractal growth model as discussed earlier in this section, there is no time where the metallicity and temperature conditions would place the point below the $\epsilon = 1$ line. Rather, this unstable point would fall on a similar plot with higher (instantaneous) $Z$ \citep[e.g. see Fig. 4, ][]{Umu20}.}, 
where SI growth can 
lead to planetesimal formation.  But, such low levels of global turbulence are currently thought
to be 
unexpected because of currently known routes to linear instability in protoplanetary disks (section \ref{sec:intro}). 

\subsection{Observations of Particle Porosity}\label{sec:obs}



There is some tension between the low internal densities of the fractal aggregates in our simulations \citep[Sec. \ref{sec:dens}, and as found by others,][]{Oku12,Kri15} and the $\sim 0.1-0.5$ g cm$^{-3}$ internal densities found in comets and cometary particles  \citep[e.g.,][]{Sie15,Ful16,Pat16}. Another source of tension is a possible discrepancy between the high porosities of our particles and ALMA mm-cm wavelength observations of protoplanetary disks, which most analyses treat as compact. No systematic attempt has yet been made to actually constrain particle porosity from ALMA  observations, although it would seem  desirable to try based on the maturity of the field  \citep{CarrascoGonzalezetal2019,Maciasetal2021, Sierraetal2021}. This is partly because the radiative transfer problem involved has recently become more challenging.  First, there is a growing recognition of the importance of scattering by wavelength-sized particles, in addition to pure emission \citep{Kataokaetal2014,Kataokaetal2015, Zhuetal2019}, especially for large disk optical depths. Second, treating the scattering properties of highly porous particles with the current approximate techniques may be problematic in some ways \citep{TazakiTanaka2018}.  \citet{Bir18} published  calculations of extinction efficiencies for porous particles using a traditional set of approximations, but did not systematically compare them with observations along with results for compact particles, while allowing for possibly degenerate parameters (specifically, the also-unknown actual particle size), so the case remains open. A proper analysis of mm-cm observations for improved constraints on particle porosity would be quite valuable. 

\begin{figure*}
\centering
\includegraphics[width=0.5\textwidth]{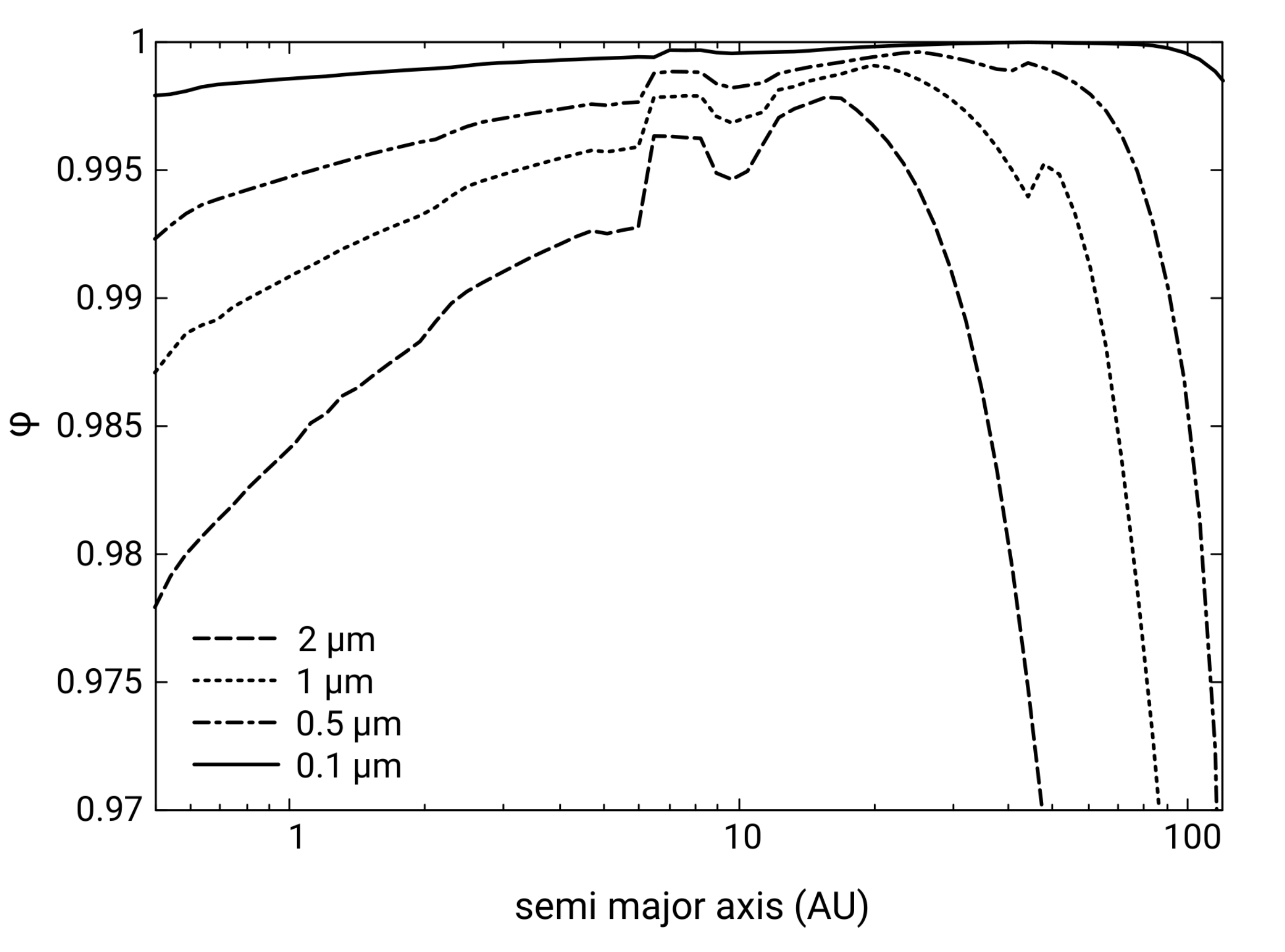}
\caption{Variation of the monomer size for model fa3gQ (cold H$_2$O ice model) after 0.1 Myr. In the context of our adopted fractal growth model \citep{Oku12}, larger monomer sizes do lead to slightly more compact particles compared to the nominal model (solid curve), but porosities are still much larger than typical cometary IDPs. 
\label{fig:por_comp}}
\end{figure*}

In addition to improving the radiative transfer modeling, there are several other ways one might reconcile the porous particle models and the observations. The particle growth theory used here uses the most advanced current treatment of the porosity of growing, colliding aggregates, but even this treatment is based on certain assumptions and some parameters are uncertain. One of these is the assumed monomer size which we have taken to be $0.1$ $\mu$m in this work. In Figure \ref{fig:por_comp}, we show that increasing the monomer size to values as large as 2 $\mu$m \citep[e.g.,][]{Oku19} in our adopted fractal growth model \citep{Suy12,Oku12,Kat13a} makes some difference, but not nearly enough to bring porosities to values consistent with most cometary IDPs. 

Perhaps a more straightforward way to reconcile the  observations of actual disk particles with the models of truly primary growth presented here is to consider compaction processes that might occur after sizeable planetesimals form. Aggregates of sufficient mass can compact due to self-gravity, for example. Using Eqns. (\ref{equ:Pc}) and (\ref{equ:Pgasgrav}) one can estimate that for an icy aggregate with such low internal density to compact sufficiently to an internal density of 0.1 g cm$^{-3}$ would require a mass of $\sim 5\times 10^{18}$ g, equivalent to a $\sim 20-25$ km radius body. Thus formation of planetesimals via whatever means to at least this mass could be a pathway to resolve this conundrum. The debris from subsequent breakup of such planetesimals (which may make up the dust content at later times) should then retain lower porosities. So, if the particles in disks studied to date are really low-porosity, it might imply they are secondary, rather than primary, products. Observations of earlier disks (Types 0/I) would also be revealing.

\section{Summary} 
\label{sec:sum}

In this work, we have conducted simulations that compare compact and fractal aggregate particle growth models in globally evolving early stage protoplanetary disks (ages up to 0.5 Myr, covering the formation age of the first planetesimals). We have studied most closely the effects of varying the turbulent intensity $\alpha_{\rm{t}}$, but have also explored a subset of the potential parameter space for other nebula or particle growth properties. We can draw some general observations from our results.

\begin{enumerate}
    \item Fractal aggregates grow to much larger masses than their compact  particle counterparts, and reach those  masses in a much shorter time. This is mostly because of the larger cross sections that characterize fluffy, underdense aggregates. Inside the snowline, growth is generally fragmentation limited across all models, with fragmentation masses being larger by up to several orders of magnitude for fractal particles than compact ones of the same mass, because of the smaller St (and lower collision speed) of the porous particles. Growth beyond the fragmentation barrier, due to mass transfer  and a velocity PDF \citep{Win12a,Gar13}, diminishes with decreasing turbulent intensity for both growth models, with no ``lucky particles" seen for $\alpha_{\rm{t}}=10^{-5}$ in either case. This comes about because of a reduced 
    production rate (by fragmentation) of  particle ``feedstock"  at the smaller sizes ($m < m_{\rm{b}}$) suitable for sweepup by larger particles, so that beyond the fragmentation limit the mutual collisional destruction rate dominates over growth by sweepup. Especially beyond the snowline, where the fragmentation mass limit is much higher (with the exception of ``cold ice" model fa3Qg), growth rarely  reaches the fragmentation mass $m_{\rm{f}}$, except at the highest $\alpha_{\rm{t}}$ compact growth model (sa2g), where fragmentation can once again become an effective production source of small particles to feed ``lucky" large ones. Overall, the turbulent intensity $\alpha_{\rm t}$ is one of the most important factors determining particle growth outcomes and thus global nebula radial structure.  
    
    \item Our simulations confirm that radial drift is a less influential factor for fractal aggregates than for compact particles, for a given $\alpha_{\rm{t}}$, during much of their growth phase. 
    The Stokes numbers for small compact  particles initially increase more quickly compared to fractal aggregates (despite the faster growth rate of fractal particles overall) so porous particles have slower radial drift at early times. By $0.1$ Myr or less though, in almost all models, fractal aggregates just outside the snowline where growth is most rapid achieve larger Stokes numbers than the corresponding compact particles  and begin to suffer strong radial drift. Meanwhile, beyond $\sim 20$ AU the Stokes numbers of porous aggregates typically remain smaller by up to an order of magnitude relative to compact particles even after 0.5 Myr. Thus, a characteristic of fractal growth models is that they retain significant amounts of mass in the outer disk for longer periods, while carving out broad, deep gaps in particle surface mass density from the snowline to tens of AU. Interestingly though, as the turbulent intensity increases to higher values ($\alpha_{\rm{t}} \gtrsim 10^{-2}$), this distinction between fractal and compact  growth models begins to lessen considerably.
    
    \item Despite their larger masses (and sizes) and longer retention in the outer disk, we conclude that for our given compaction model and its assumptions, even porous fractal aggregates will also eventually be lost {\it en masse} to the inner disk and the central star due to radial drift, without the intervention of a process or processes of planetesimal formation that would allow  solids to be retained for much longer. However, the longer retention time for the fractal aggregates would allow more time for planetesimals to potentially form, mitigating material loss  compared to compact particle growth models. 
 
    \item We find that even after collisional compaction sets in, for either Gaussian or Maxwellian (Appendix \ref{sec:vpdf}) collision velocity PDFs, it generally does not lead to increasing internal density as aggregates grow in mass, in agreement with previous work \citep[e.g.,][]{Oku12,Kri15}. Compaction is instead dominated by gas ram pressure, with internal particle density increasing along with increasing relative particle-to-gas velocity $\rho_{\rm{int}} \propto (\Delta v_{\rm{pg}})^{1/3}$ (Sec. \ref{sec:collcomp}). Self-gravitational compaction is not achieved in our models because the largest aggregates fall just short of the required critical mass.

    \item We tested models in which the fragmentation energy threshold of ice-rich particles $Q_{\rm{f}}$ is significantly larger than silicates (as is usually assumed) {\it only over a limited range of temperature about the snowline} \citep{MW19}, which we refer to as cold H$_2$O ice models \citep[e.g., see][]{Umu20}. These models (which also apply to supervolatile ices) retain material outside the snowline for much longer periods, because the icy  aggregates and their  Stokes numbers are then an order of magnitude smaller than under the usual ``strong water ice" assumptions throughout the outer nebula, except for in a narrow band just outside the snowline. As a result, inward drift  transport of condensibles is diminished, and both  the depletion of solids outside the snowline, and the enhancement of solids $Z$ inside the snowline, are muted (see Appendix \ref{sec:varparm}, and \citetalias{Est21}).
    
    \item Simulations in which we vary the initial disk mass are qualitatively similar in how they evolve compared to the fiducial model, with predictably lower disk temperatures for decreasing disk mass. The more extended disk model $R_0=60$ AU (Appendix \ref{sec:varparm})  maintains a higher metallicity $Z$ in the planet forming region (and out to $\sim 200$ AU) because the model starts with more mass distributed in the outer disk. Differences in ${\rm{St}}$ are most apparent outside of $\sim 20$ AU, but inside $20$ AU, Stokes numbers are remarkably similar.
 
    \item Evaporation fronts serve as locations where solids can become trapped over long periods, even after the bulk of solid material has largely been lost; this retention effect appears especially dramatic for $\alpha_{\rm{t}} \leq 10^{-4}$. The effect is characterized by narrow bands of  volatiles, each lying stably just outside its EF. In the case of supervolatiles, the bands could continue to be fed by outwardly advecting and/or diffusing gas, that is partly composed of the supervolatile's vapor phase, recondensing outside the respective EF \citepalias[see][for more discussion]{Est21}. It is intriguing to compare these structures with narrow bands seen in multiple ALMA images of disks \citep[e.g.,][]{And18}, especially Class 0/I \citep[e.g.,][]{SC20}. This effect may be amplified by decomposition of refractory organics at their ``EF", and conversion into simpler supervolatiles. 
 
    \item We find that particles in our disks beyond $R \gtrsim 200-300$ AU (where $\Sigma$ drops off sharply, and the gas radial velocity $v_{\rm{g}}$ 
    increases rapidly) have increasingly large ${\rm{St}}$ so that the combination of radial drift and small $\rho_{\rm{solids}}$ means growth times are exceedingly long, at least until the nebula gas expands outwards. However, this may be an artifact of our initial conditions, and the effect is lessened with larger $R_0$ or $\alpha_t$. 
    
    \item Quite generally, our simulations do not achieve conditions necessary for planetesimal formation by streaming instability (SI),  with almost all models resulting in combinations of ${\rm{St}}$, $\alpha_{\rm{t}}$ (and $Z$) that place them firmly in the ``turbulent regime" where the local particle-to-gas mass volume  density ratio $\epsilon = \rho_{\rm{solids}}/\rho <1$ \citep[referred to as Zone II,][]{Umu20}; indeed this basic criterion has been a known requirement for SI since the earliest studies \citep{YG05}. We find that only for the lowest turbulent intensity ($\alpha_{\rm{t}}=10^{-5}$) models is there a relatively short period of time lasting $\sim 2\times 10^3-2\times 10^4$ years, and only at the snow line, where $\epsilon > 1$ \citep[laminar regime, Zone I,][]{Umu20}. This $\alpha_{\rm{t}}$ and corresponding Stokes numbers ($\sim 0.05$ and $\sim 0.4$ for sa5g and fa5g, respectively) are consistent with critical values necessary for SI to lead directly to planetesimal formation as is usually assumed \citep{Umu20}. However, for the higher turbulent intensities of $\alpha_{\rm t} \sim 10^{-4}-10^{-3}$  commonly found in simulations of critically triggered hydrodynamic turbulence \citep[for a review, see][]{LU19,Les22}, 
    this  moment of instability is never achieved, because particle growth to large enough sizes (masses) is not fast enough to keep up with the radial drift of material across the snowline. The local $Z$ drops precipitously to such low values that the mechanism never becomes viable and/or the particles are too large and rare to drive the gas velocities as required for SI. 

\end{enumerate}

%

\acknowledgments

We thank the NASA Ames Research Center's Origins Group for numerous conversations, and specifically useful discussions with Denbanjan Sengupta, Uma Gorti and Maxime Ruaud as well as Carlos Carrasco-Gonzalez, Enrique Macias, and Anibal Sierra. We thank Thomas Hartlep and Diane Wooden for thorough internal reviews, and we thank an anonymous reviewer for very helpful comments to improve the paper's exposition. P.R.E. and J.N.C. acknowledge support from the NASA Emerging Worlds Program grant NNX17AL60A. JNC acknowledges support from the NASA Astrobiology Institute (WBS 132379.04.05.09.03).  J.N.C and O.M.U acknowledge the NASA PSD ISFM for Planet Formation and Exoplanets Theory at NASA/Ames for financial and computational support. O.M.U. acknowledges support from NASA Astrophysics Theory Program grant NNX17AK59G.

\appendix
\renewcommand{\thefigure}{A\arabic{figure}}
\setcounter{figure}{0}

\section{Particle size distributions}
\label{sec:sizes}

\begin{figure}
\gridline{\fig{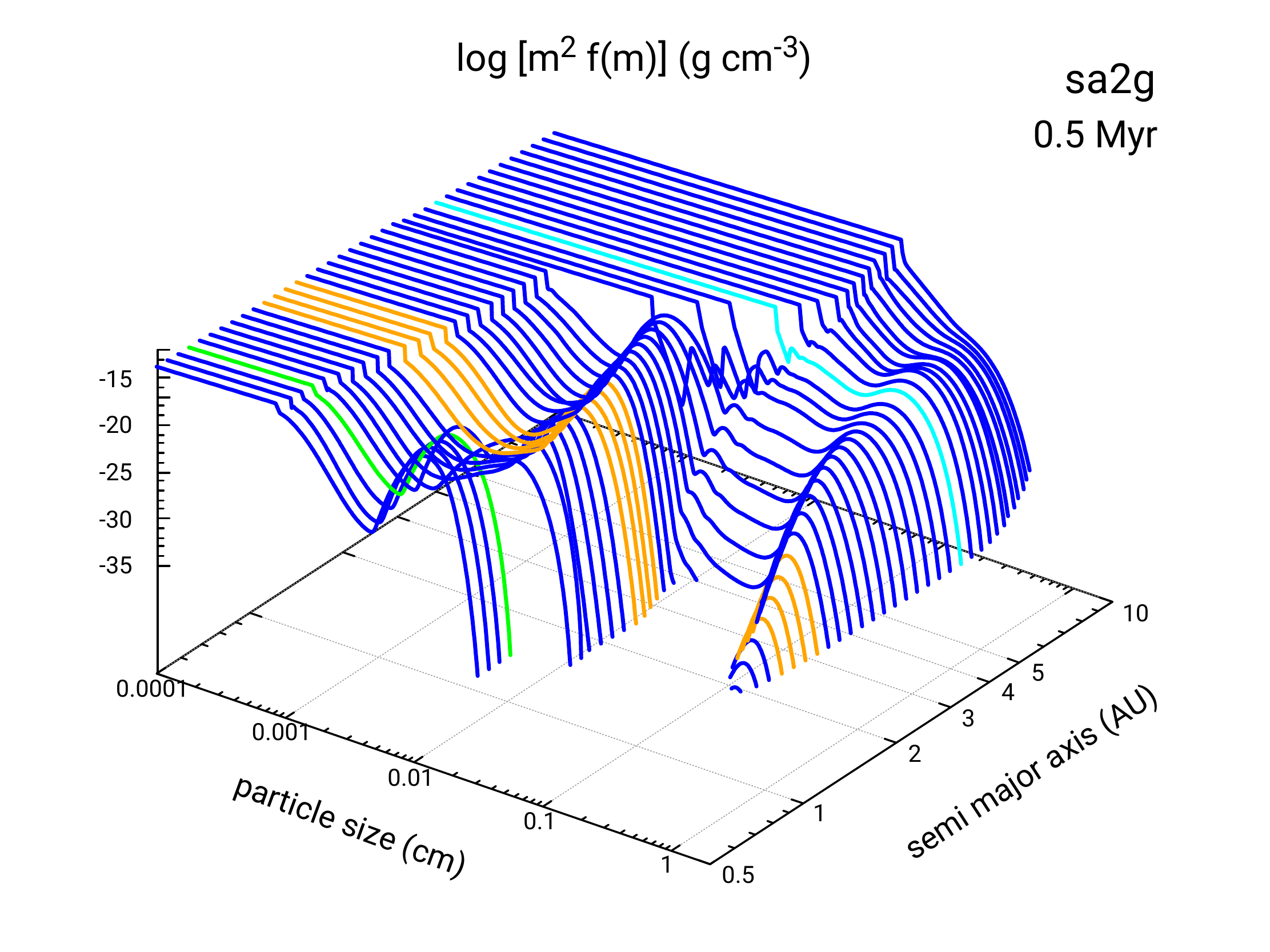}{0.33\textwidth}{}
          \fig{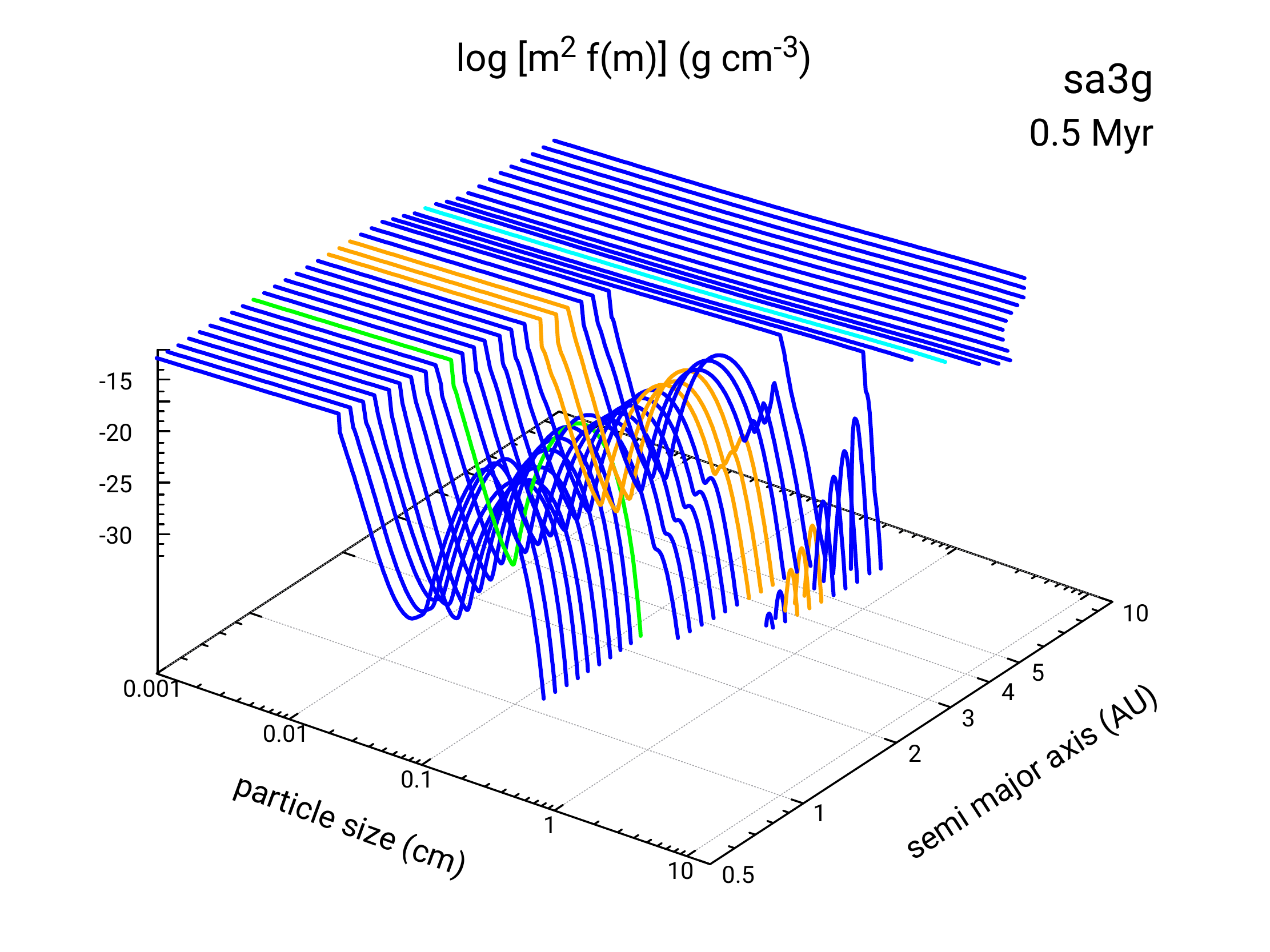}{0.33\textwidth}{}
          \fig{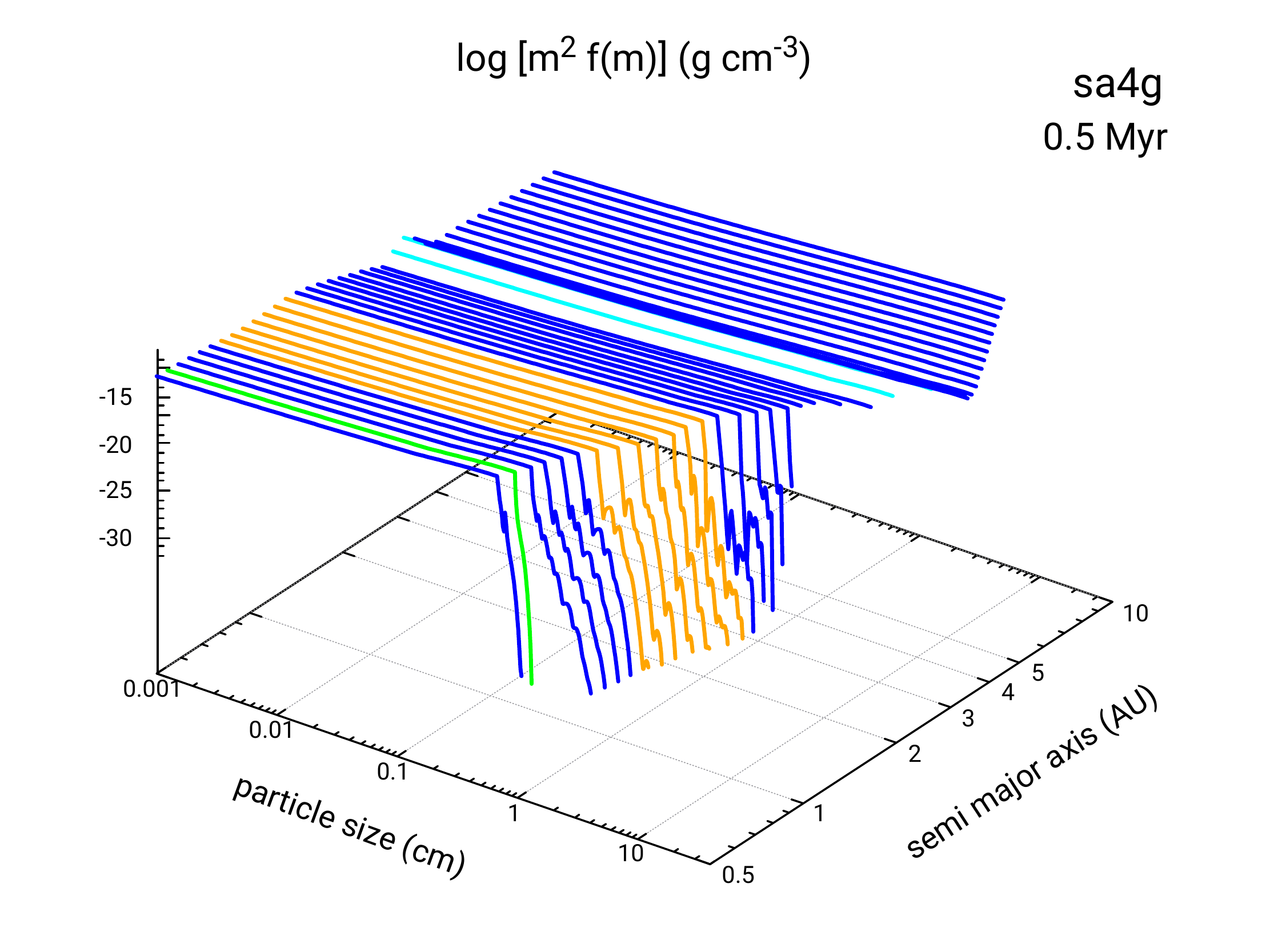}{0.33\textwidth}{}
          }
\vspace{-0.4in}
\gridline{\fig{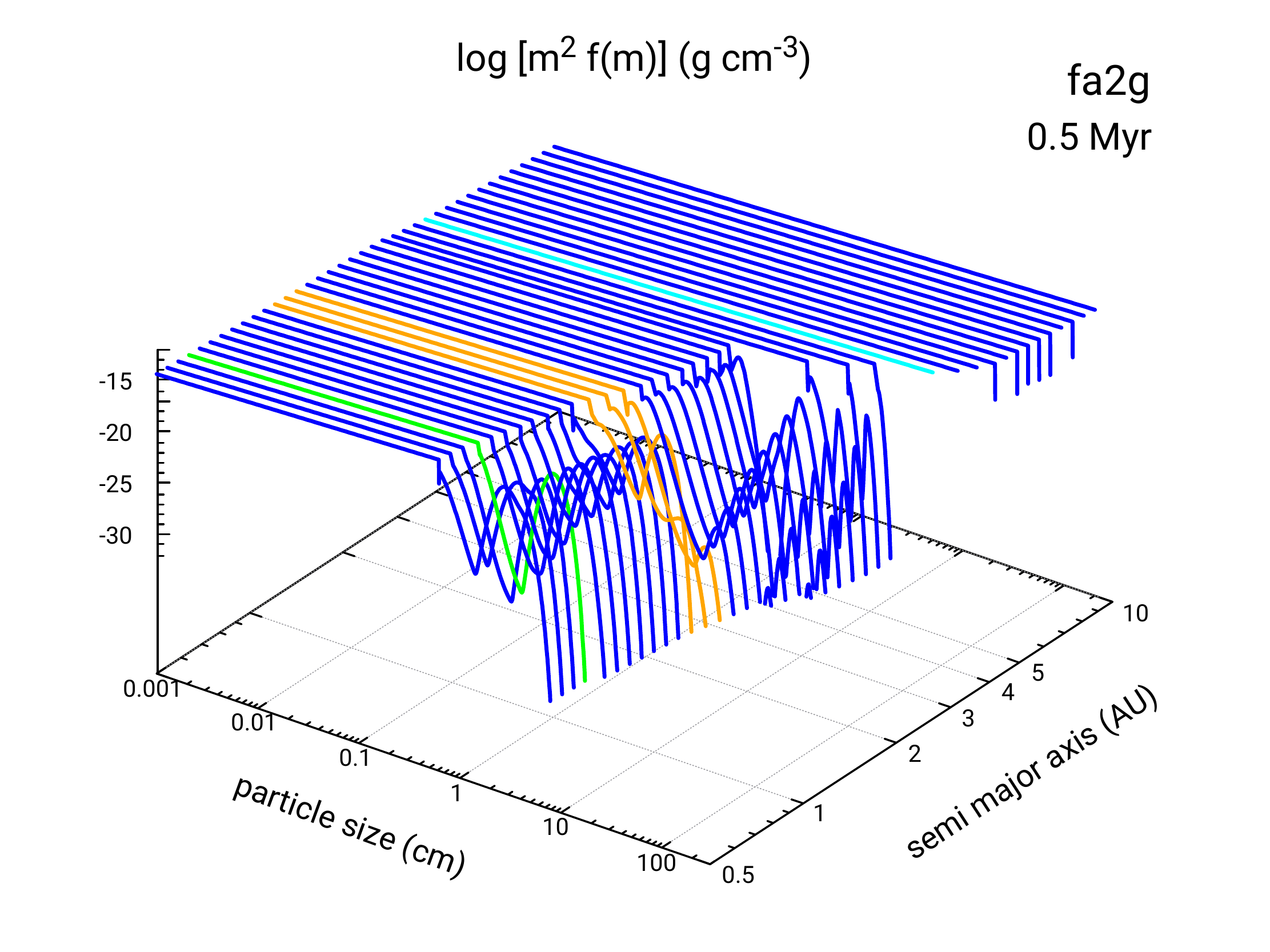}{0.33\textwidth}{}
          \fig{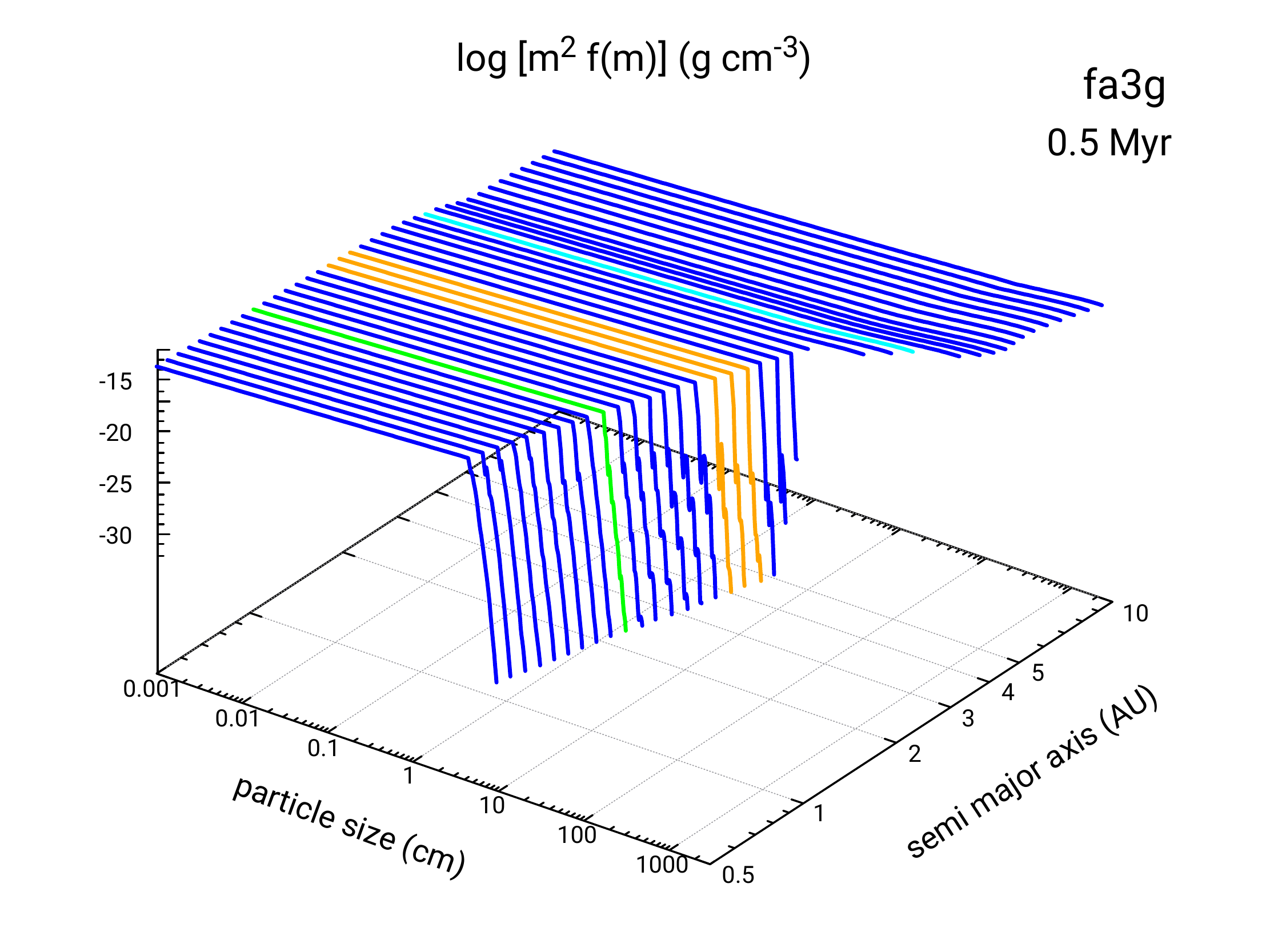}{0.33\textwidth}{}
          \fig{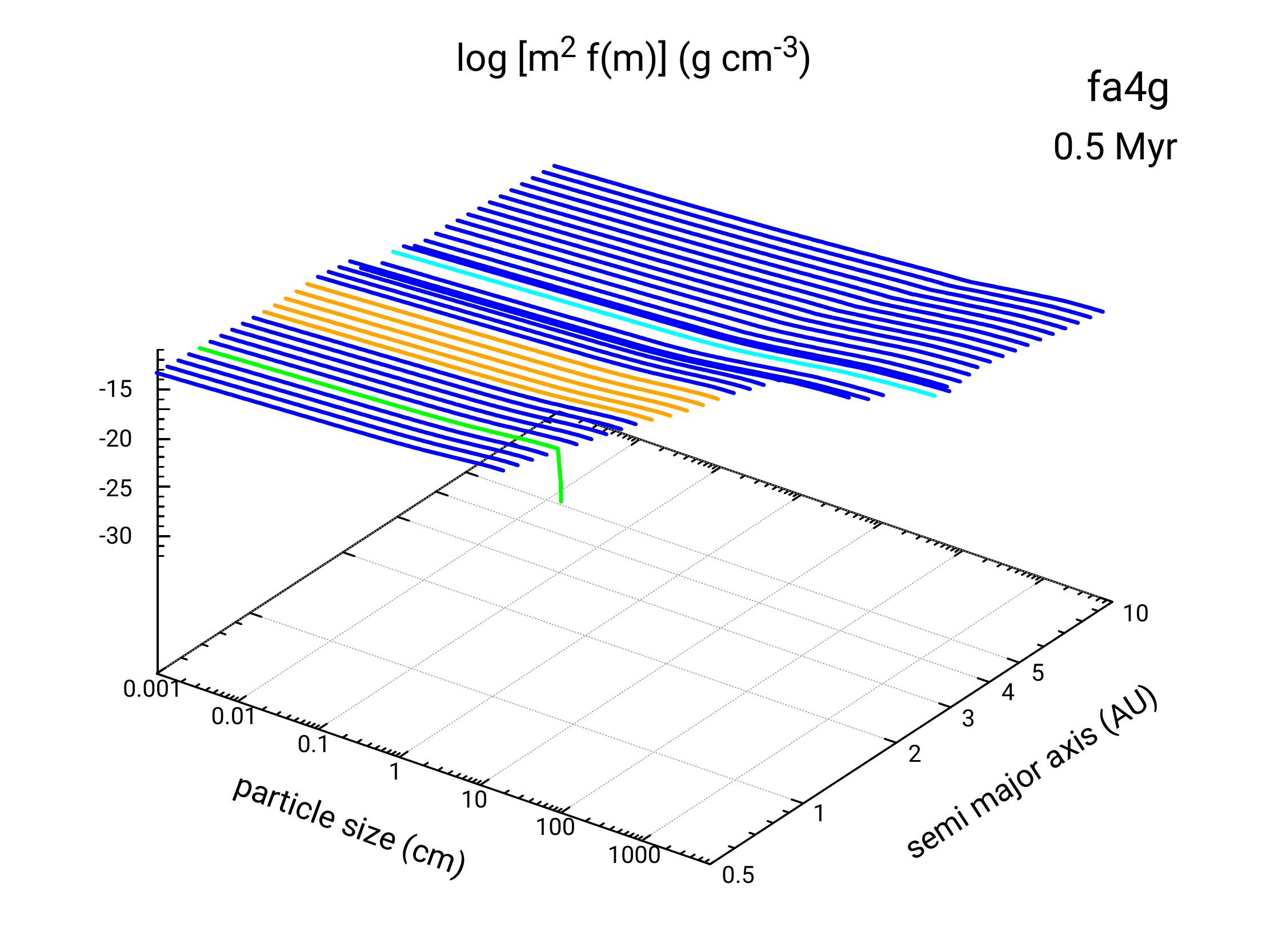}{0.33\textwidth}{}
          }
\caption{Particle size distributions for compact particle (top panels) and fractal aggregate growth models (bottom panels) plotted on roughly the same scale in solids mass volume density $m^2 f(m)$ after 0.5 Myr for the indicated simulations. These wireframes (plotted out to 10 AU) can be compared with the rightmost panels of the particle mass distributions in the main text. Several EFs are indicated by the cyan (water), orange (organics) and green (troilite) curves. Regions that appear flat at this scale correspond to particle sizes that have not reached the fragmentation barrier. Recall that in the moments method, the ``dust" population is modeled as a power law up to the fragmentation barrier $r_{\rm{f}}$.}
\label{fig:wireframes}
\end{figure}

Figure \ref{fig:wireframes} shows as a reference the particle size distributions for compact  particle and fractal aggregate growth models with $\alpha_{\rm{t}}= 10^{-4}$, $10^{-3}$ and $10^{-2}$ after 0.5 Myr as a function of semi-major axis and solids mass volume density $m^2 f(m)$. In these panels are also highlighted the locations of various EFs in the inner 10 AU of the disk after this simulation time. These are the water snowline (cyan), the organics EF (orange) and the troilite EF (green). As noted in the text, other EFs were present at earlier times but have since migrated beyond the computational grid inside 0.5 AU as the disk cools. ``Flat'' regions seen in these distributions correspond to particle sizes that are below the fragmentation size $r_{\rm{f}}$. Recall that in our model the ``dust'' distribution (defined by $r < r_{\rm{f}}$) is modeled using the moments method \citepalias[see Sec. \ref{sec:partgrow}, and][]{Est16}, while particle growth is followed explicitly for $r \ge r_{\rm{f}}$. The non-flat regions can thus be associated with ``lucky'' particle growth beyond the fragmentation barrier seen in the related mass distribution plots for the given $\alpha_{\rm{t}}$ (section \ref{sec:results}).

The leftmost panels show snapshots for the sa2g (top) and fa2g (bottom) models presented in Sec. \ref{sec:strong}. For $\alpha_{\rm{t}}=10^{-2}$, particle sizes are smallest compared to lower turbulence models because the fragmentation and bouncing barriers occur at much smaller sizes. For instance, $r_{\rm{f}} \sim 0.001$ cm at 1 AU for sa2g, whereas it is $\sim 1$ cm for fa2g. As a result, there is substantial growth beyond the fragmentation barrier in the inner disk for both models, and for the compact  case,  also outside the water snowline. This figure also shows clearly the secondary and even tertiary growth peaks  that were seen in Fig. \ref{fig:a2g} (rightmost panel of top row), which are due to a combination of mass transfer and our velocity PDF (see Sec. \ref{sec:fiducial}). Model fa2g also has additional peaks beyond the secondary, more easily seen here than in the corresponding particle mass distribution plot. The tertiary peaks decrease in magnitude, however, as particles are ground down  - likely due to increasing destruction probabilities further inward in the disk. The largest particle sizes achieved are roughly 1 cm for the compact  particle growth model, but as much as $2-3$ m outside the water snowline in the fractal aggregate model. Note that even the secondary mass peaks contain far less mass than at $r_{\rm f}$ (see also Paper I).

The middle panels show snapshots of our fiducial models sa3g (top) and fa3g (bottom), with  $\alpha_{\rm{t}}=10^{-3}$. These cases display a more stark contrast between fractal aggregate and compact growth models regarding lucky particles. The compact particle case still shows significant growth beyond the fragmentation barrier out to the water snowline with a secondary and tertiary growth peak (both clearly visible in Fig. \ref{fig:a3g}, right upper panel), while in the fractal aggregate case no secondary peak develops and the mass and size of the mass-dominant aggregate ($m_{\rm{M}}, r_{\rm{M}}$) remain near, but are slightly larger than, the values for fragmentation (black curve, lower right panel of Fig. \ref{fig:a3g}). Due to the lower $\alpha_{\rm{t}}$, the fragmentation sizes are considerably larger than the $\alpha_{\rm{t}} = 10^{-2}$ model. At 1 AU, the fragmentation size for model sa3g is $\sim 0.03$ cm, whereas it is $\sim 13$ cm for model fa3g. The largest particle sizes achieved outside the water snowline are $\sim 15$ cm for the compact particle case, but $\sim 27$ m for the fractal aggregate model. Though more difficult to see in this orientation, a systematic decrease in the solid mass volume densities can be seen outside the water ice EF.

Finally, the rightmost panels show snapshots for the sa4g (top) and fa4g (bottom) models shown in Fig. \ref{fig:a4g} of Sec. \ref{sec:weak}. In this case there are no secondary growth peaks. In the compact particle model, growth beyond the fragmentation size is very limited and the mass dominant particle is still near the fragmentation barrier (upper right panel, Fig. \ref{fig:a4g}), while lucky particle growth is completely stymied in the fractal aggregate model (see discussion, Sec. \ref{sec:weak}). Aggregates that grow beyond $r_{\rm{f}}$ find only relatively brief existence before being destroyed. As expected with such a low $\alpha_{\rm{t}}$, the fragmentation sizes and largest particle sizes achieved are larger than in the models discussed above. At 1 AU, the fragmentation size in model sa4g is $\sim 2.3$ cm, and is $\sim 54$ cm in model fa4g. Outside the water snowline, the largest sizes achieved are $\sim 38$ cm and $\sim 64$ m, respectively. As in the fiducial case, a systematic drop in the solid mass volume densities can be seen outside the H$_2$O EF that can be associated with the steep drop in solids surface density (Fig. \ref{fig:agsurf}).

\renewcommand{\thefigure}{B\arabic{figure}}
\setcounter{figure}{0}

\section{Variation with Some Initial Disk Parameters}
\label{sec:varparm}

\begin{figure}
\gridline{\fig{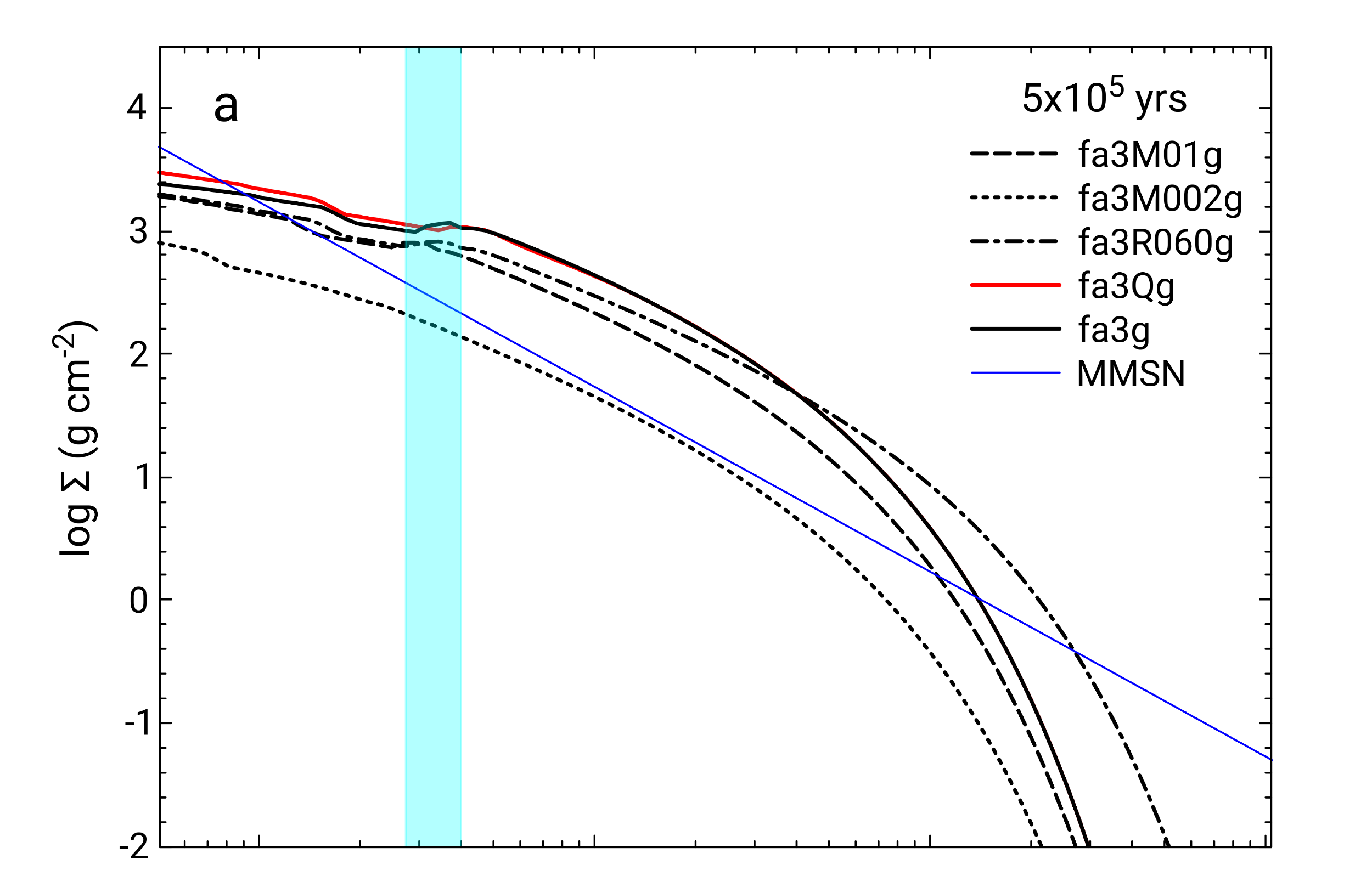}{0.45\textwidth}{}
          \fig{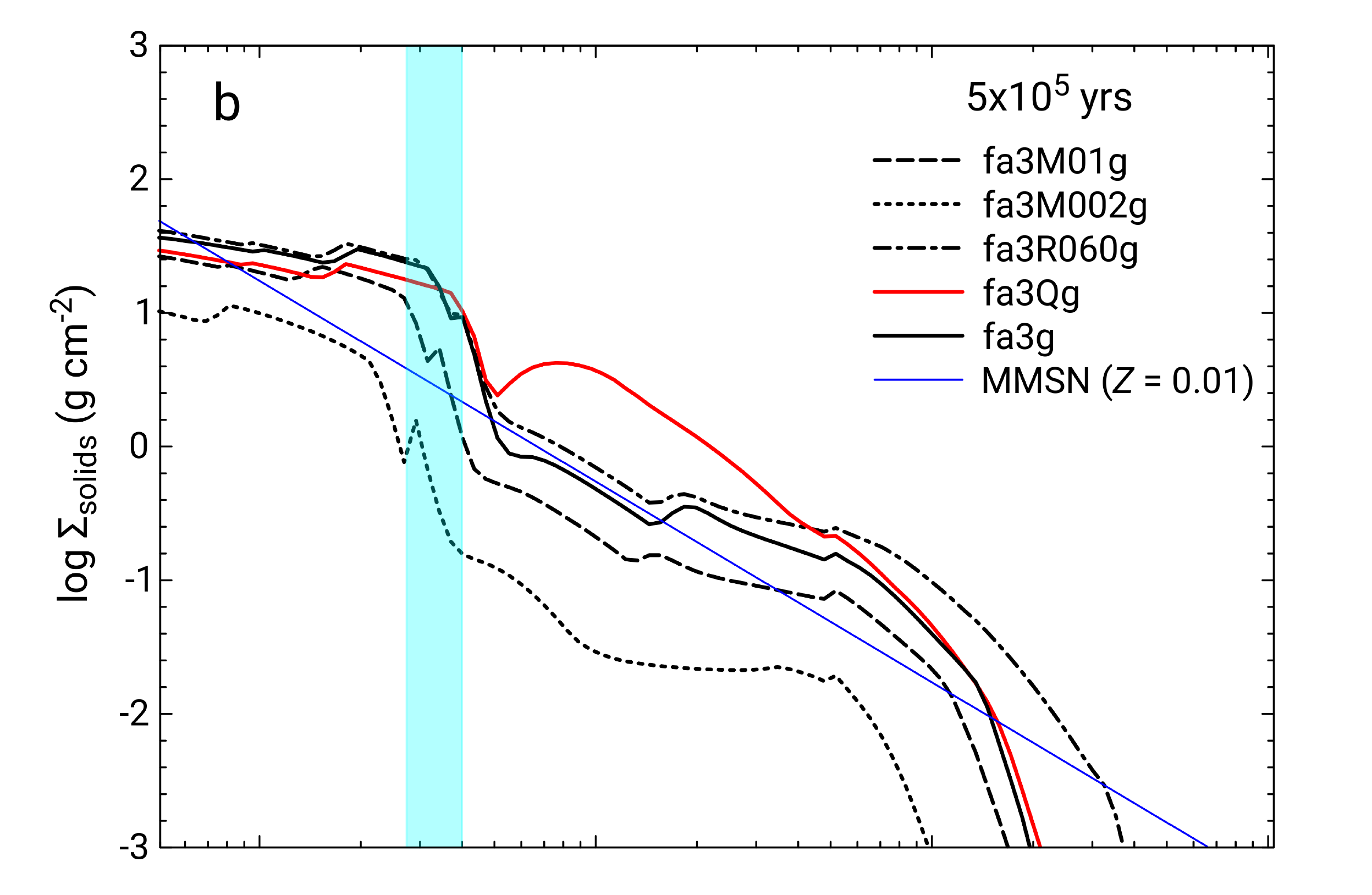}{0.45\textwidth}{}
          }
\vspace{-0.4in}
\gridline{\fig{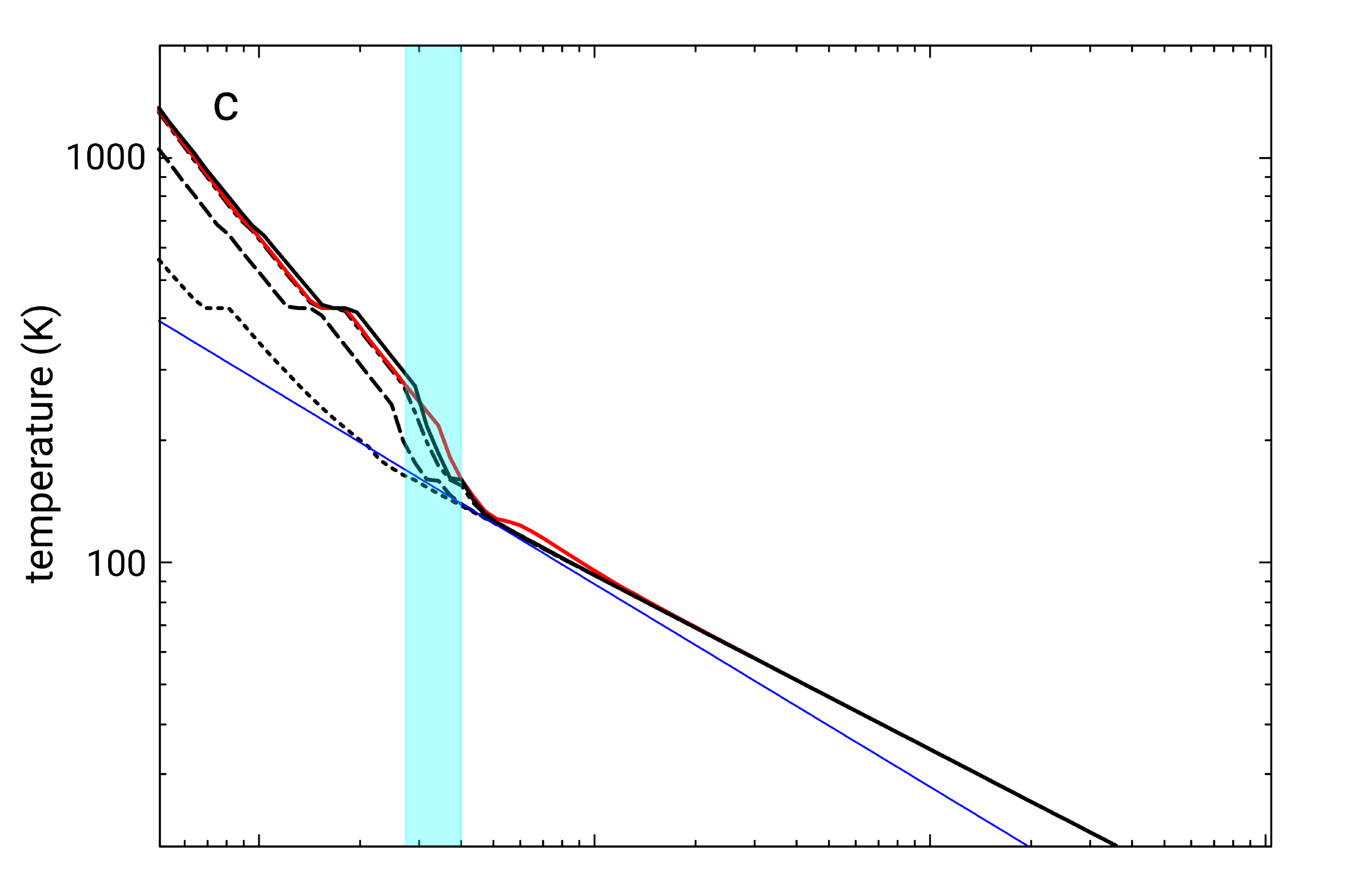}{0.45\textwidth}{}
          \fig{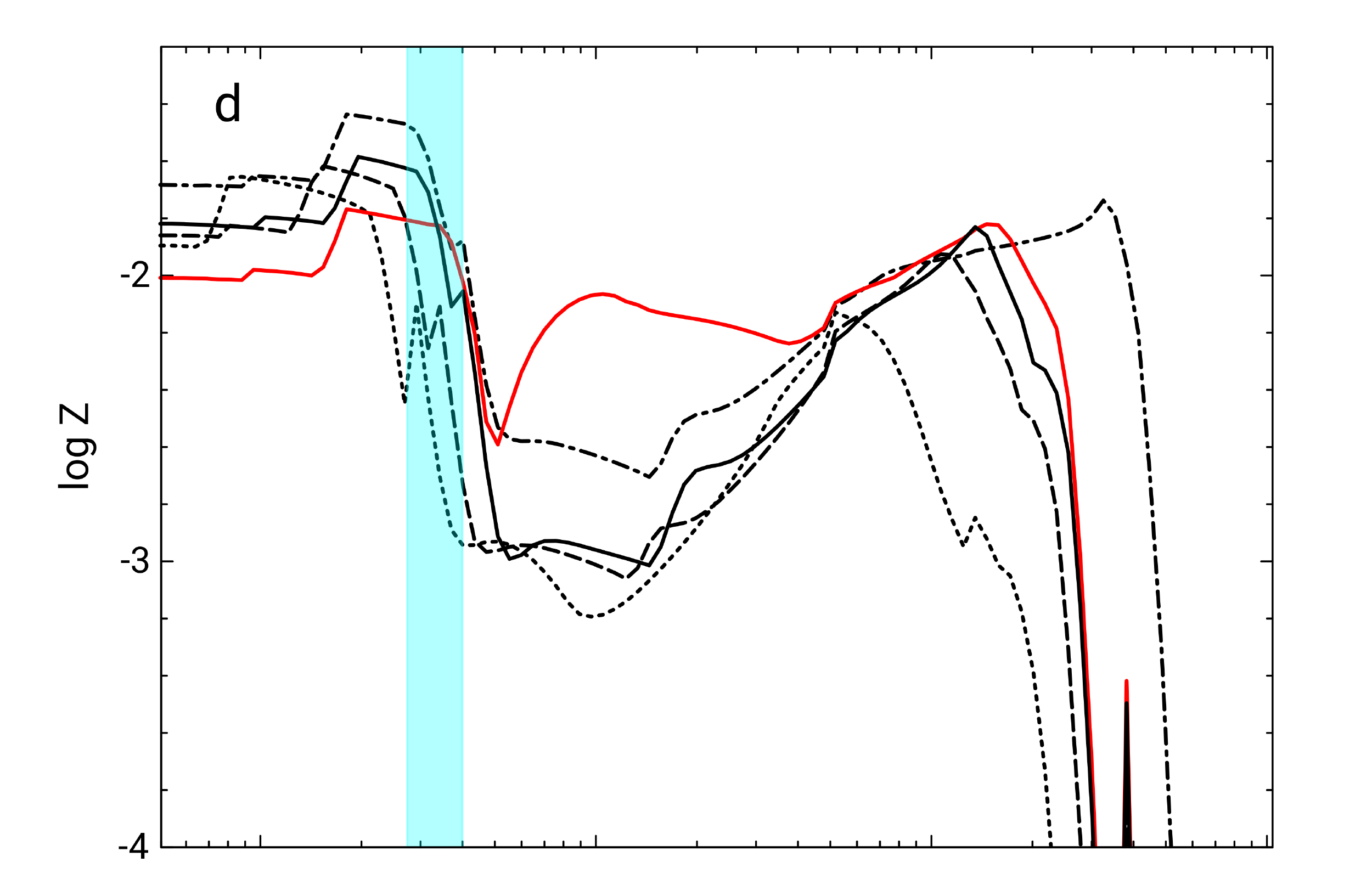}{0.45\textwidth}{}
          }
\vspace{-0.4in}
\gridline{\fig{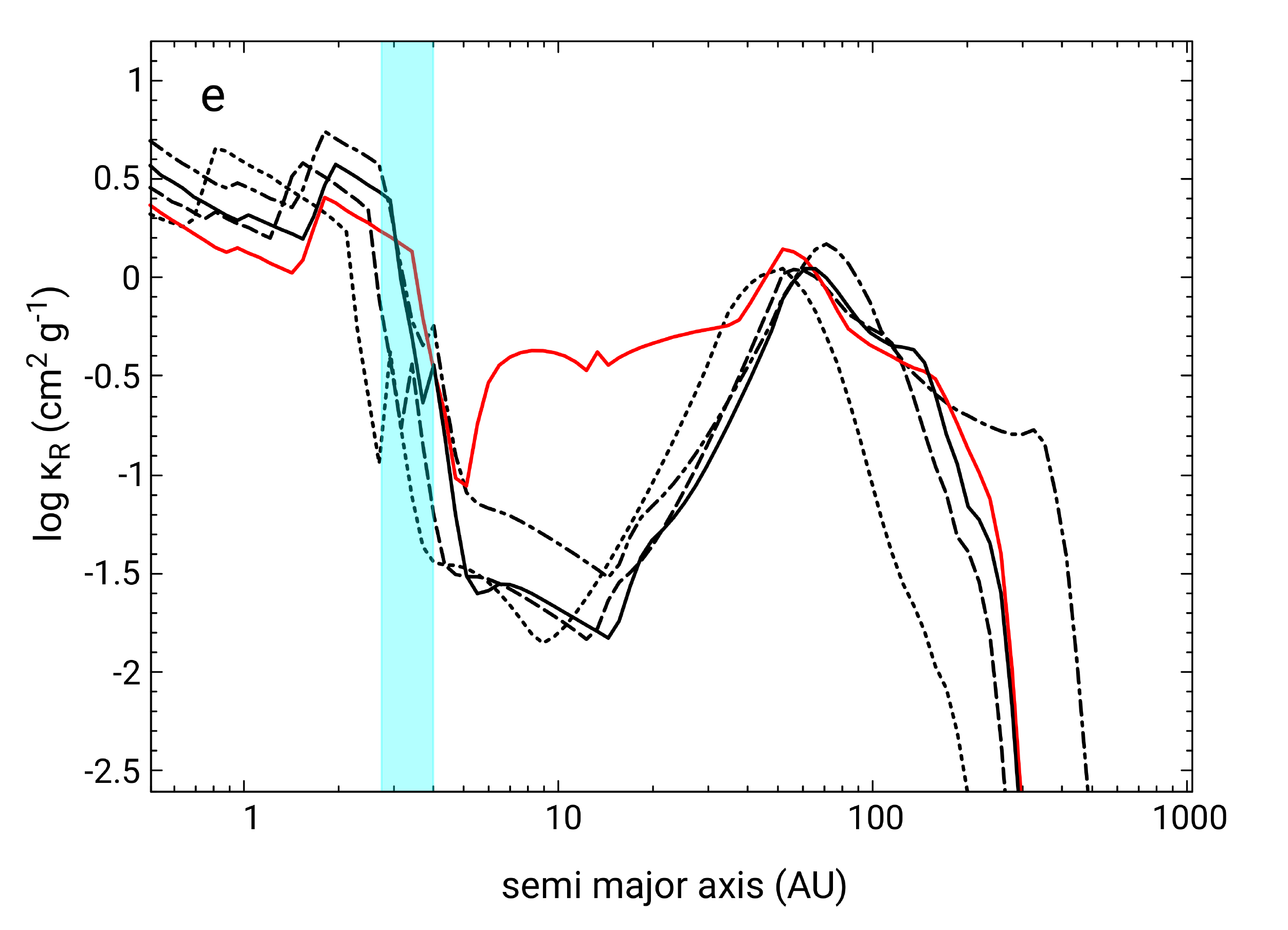}{0.45\textwidth}{}
          \fig{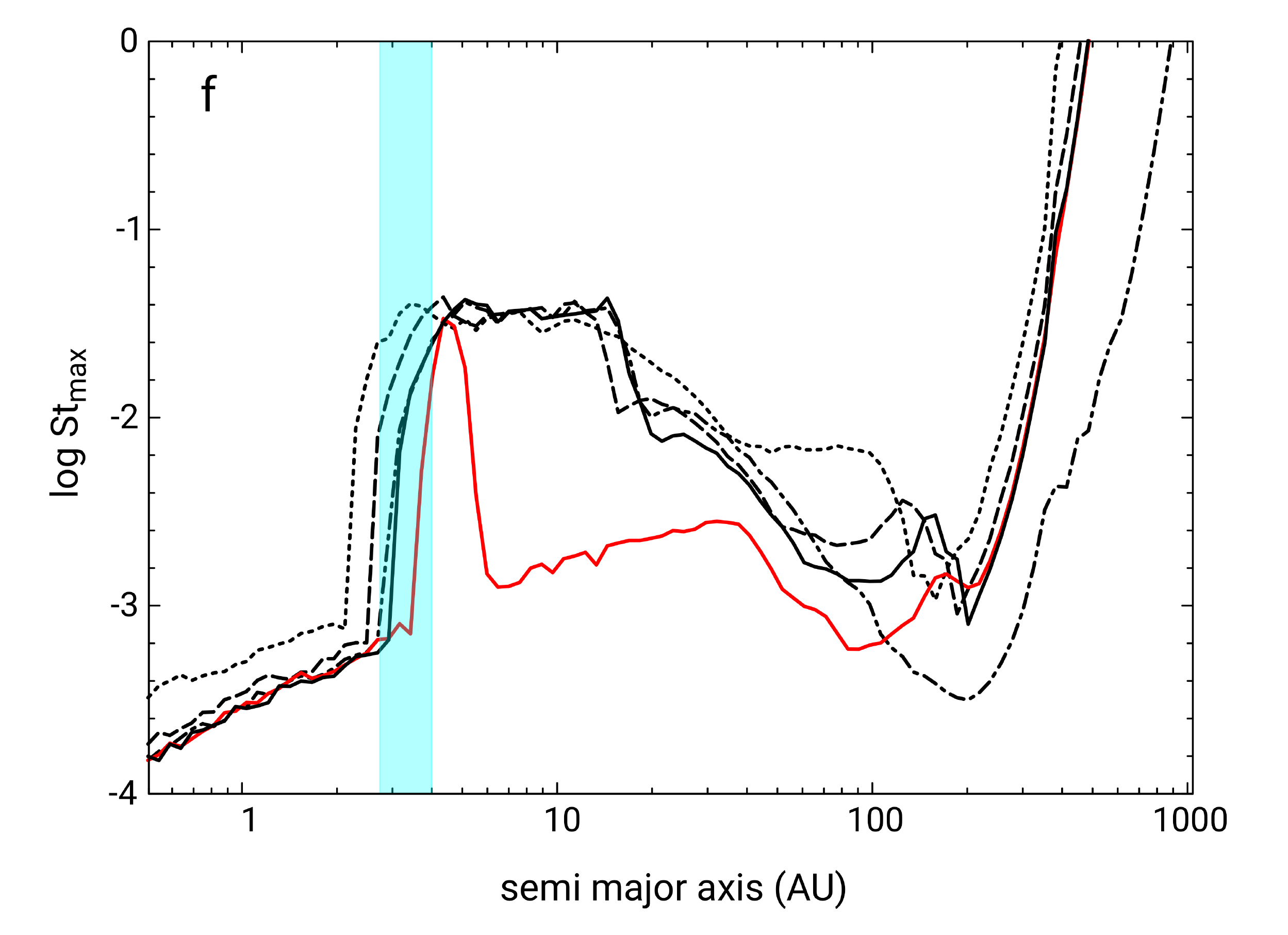}{0.45\textwidth}{}
          }
\caption{Fiducial model after 0.5 Myr (black solid curves) compared to a suite of models where we vary the disk mass $M_{\rm{D}}$, disk scaling parameter $R_0$, and fragmentation threshold $Q_{\rm{f}}$ as listed in Table \ref{tab:models}. (a) Gas surface densities; (b) total solids surface densities; (c) midplane temperatures; (d) disk metallicities; (e) Rosseland mean opacities; and (f) Stokes numbers of the mass dominant aggregates, all as a function of semi-major axis. The solid blue curves that appear in panels (a-c) are for a standard MMSN. In each panel, the cyan region demarcates the range in location of the water snowline for these models (between $\sim 2.7-4$ AU).
\label{fig:a3models}}
\end{figure}


In this appendix, we compare our fiducial model for fractal aggregate growth with $\alpha_{\rm{t}}=10^{-3}$ (fa3g, Sec. \ref{sec:fiducial}) to simulations with different  disk mass $M_{\rm{D}}$, disk size scaling parameter $R_0$, and fragmentation threshold $Q_{\rm{f}}$ to assess the effects on the particle mass distribution. These simulations vary only a subset  of parameters, and of the parameters highlighted only a 
fraction of their parameter space is explored here. 
As summarized in Table \ref{tab:models}, specifically we compare models with initial disk masses of 0.1M$_\odot$  (fa3M01g) and 0.02M$_\odot$ (fa3M002g), a disk scaling parameter $R_0 = 60$ AU (fa3R060g), and perhaps of greatest interest a model in which water ice is only ``sticky" in a narrow range about the snowline (fa3Qg) as suggested from the recent work of \citet{MW19}. This model is referred to as the ``cold ice" model in \citet{Umu20}. The lowest mass model fa3M002g corresponds most closely with the standard MMSN\footnote{Strictly speaking, the \citet{Hay81} MMSN model is constructed to account for the radial distribution of planetary masses, and thus cuts off at $\sim 36$ AU.} which has surface density $\propto R^{-3/2}$, and temperature $\propto R^{-1/2}$, while the gas surface densities in the higher mass models (including our fiducial model) though more massive, are generally in agreement with the models of \citet{Des07} in radial extent hypothesized for the same timeframe. 
In Figure \ref{fig:a3models} (panels a-f) we plot a subset of modeled quantities at 0.5 Myr as a function of semi-major axis for the suite of fractal aggregate growth simulations described above. In all panels our fiducial model (Sec. \ref{sec:fiducial}) is given by the solid black curve. The solid blue curves that appear in panels a-c refer to a standard MMSN disk. 

Focusing first on models in which the disk mass is varied (fa3M01g, fa3M002g), the differences between these and the fiducial model are intuitive - $\Sigma$ is lower overall and evolves similarly to the fiducial case, although the lower mass model is much cooler in the inner disk (dotted curve in panel c, closer to MMSN). 
The variations in $\Sigma$ as a function of semi-major axis seen between $1-4$ AU 
 in the higher mass models are due to strong gradients in $T$, associated with the water and organics snowlines, that lead to fluctuations in the local viscosity $\nu$ 
 and thus relative velocity. This contributes (along with the smaller fragmentation threshold inside the snowline) to the sharp decrease in particle ${\rm{St}}$ (panel f) near $\sim 3$ AU, which then coincides with a pileup in $\Sigma_{\rm{solids}}$ (panel b) and thus an enhanced $Z$ (panel d). These effects are muted in the lowest mass model (fa3M002g; dotted curve, panel c) at this point in the simulation because, owing to its initially low mass, this model cools quickly. The lower $T$ in fa3M002g means that the fragmentation limit ${\rm{St}}_{\rm f}$, 
 which is in the Stokes regime (${\rm{Re}}_{\rm{p}} < 1$), is higher than for the fiducial case by a factor of several in the inner disk (recall from Eq. \ref{equ:Stfrag} that ${\rm{St}}_{\rm{f}} \propto T^{-1}$), and leads to generally higher Stokes numbers (panel f) so that solid material drains away at a faster rate.  Beyond the snowline on the other hand, particle masses in this low mass disk are also smaller than in the fiducial model fa3g, leaving ${\rm{St}}$ comparable (or slightly smaller), so particles experience similar radial drift speeds. Although the evolution of $\Sigma_{\rm{solids}}$ appears to follow the evolutionary trend of higher mass models, as can be seen in panel e, $Z$ is generally smaller everywhere compared to the fiducial model. In the inner nebula, model fa3M01g is also cooler than the fiducial case, but less so than fa3M002g so that differences with the fiducial model are less distinctive.

Model fa3R060g (in which the disk scaling parameter is set to 60 AU rather than 20 AU) initially distributes the mass such that  the surface density is smaller  in the inner disk ($\sim 2-2.5$ times less at 0.5 AU), and larger in the outer disk ($\sim 10$ times higher at 100 AU). 
This leads to $Z$ (panel d) having its steep cutoff much further out in the disk $\sim 300$ AU. In other models, this region is where the Stokes numbers begin to increase dramatically - causing rapid depletion of material within the region, and isolation of material outside of it where ${\rm{St}} \gg 1$ (e.g., right panel of Fig. \ref{fig:a3gtemp}, and Fig. \ref{fig:a3gst}, bottom panel). This behavior still occurs in this model at early times, but after 0.5 Myr, Stokes numbers in the outermost portion of the disk are mostly ${\rm{St}} \lesssim 1$, leading to relatively rapid radial drift with very little material left isolated \citepalias[see][]{Est21}. Generally the temperature and Stokes numbers are comparable to the fiducial model, but one notable difference is that in fa3R060g  there is more material outside the snowline (which is reflected in the plot of $Z$ in Fig. \ref{fig:a3models}), so the opacity there is higher as well (panel e). This is  simply because the disk starts out with much more mass at larger radial distance. The more extended disk retains more mass for the same evolutionary time, and in fact there is about $\sim 2.5$ times more total mass remaining in this model compared to the fiducial case \citepalias{Est21}.

Finally, we look at a cold H$_2$O ice model in which $Q_{\rm{f}}$ is much larger than for silicates only in a region about the snowline (fa3Qg, red solid curves), compared to all other models in which water ice is sticky everywhere it is solid - that is, everywhere outside the water ice EF.  The surface density of solids outside the snowline (located at $4$ AU in this and the fiducial model) does not decrease nearly as much as the fiducial model (which is factor $\sim 2.5$ times lower at $\sim 5$ AU)  
because aggregates are considerably smaller for this $Q_{\rm f}$ (they are fragmentation limited), so their radial drift times are longer. Interior to the snowline, the surface density becomes lower than in the fiducial case because less material has been able to drift inwards across the snowline, 
so aggregates (and their ${\rm{St}}$) are smaller there as well. The decrease in $\Sigma_{\rm solids}$ outside the water snowline still correlates with the substantial decrease (minimum at $\sim 5$ AU) in the opacity, but outside and just inside the opacity is higher which leads to the modest increase in temperatures seen about the snowline in panel (c). 
The clearest indication of the effects of the lower $Q_{\rm{f}}$ across most of the outer nebula, however, can be seen in the distribution of Stokes numbers with semi-major axis (panel f) where only a very narrow region near the snowline displays the peak values of other models  (${\rm{St}} \sim 0.03-0.04$), while outside $5-6$ AU the Stokes numbers drop to $\sim 10^{-3}$ and then increase slightly going outwards. The behavior is similar to a constant $Q_{\rm{f}}=10^3$ model in \citetalias{Est16}. The spike in ${\rm{St}}$ just outside the snowline in panel (f) still coincides with a dip in $Z$ (panel d), though the drop is less dramatic than in the fiducial model. Like every other fractal model, this model also retains more mass in the outer disk, but because of the low $Q_{\rm{f}}$, the surface density of solids (and $Z$) remains considerably higher from $\sim 6-70$ AU (panels b and d). 


\renewcommand{\thefigure}{C\arabic{figure}}
\setcounter{figure}{0}

\section{Effects of the choice of velocity pdf}
\label{sec:vpdf}

Here, we examine the effect of the velocity PDF on growth beyond the fragmentation barrier. In Section \ref{sec:destprob} (also, see Sec. \ref{sec:relvel}) we discussed the destruction probabilities we utilize to model growth beyond the fragmentation barrier which, combined with mass transfer, determine the largest mass a compact or aggregate particle can grow to beyond $m_{\rm{f}}$. We found in our simulations that the dominant limitation in growing beyond the fragmentation barrier is the mass $m_{\rm f}$ at which the barrier is reached, which determines the amount of ``dust" remaining for lucky particles with $m>m_{\rm f}$ to grow from. The value of $m_{\rm f}$ is largest for the smallest $\alpha_{\rm{t}}$, but also is larger for fractal aggregates than for compact particles at the same $\alpha_{\rm{t}}$ (e.g., see Sec. \ref{sec:fiducial}). Thus, models with $\alpha_{\rm{t}}\lesssim 10^{-4}$ saw little if any growth beyond $m_{\rm{f}}$, while $\alpha_{\rm{t}}=10^{-2}$ saw the most because there is still ample feedstock (especially $m < m_{\rm{b}}$) to grow from (see also Appendix \ref{sec:sizes}). Despite the difference in fragmentation mass, the largest particle mass $m_{\rm{L}}$ achieved inside the snowline, and its variation with semi-major axis, are similar for most all of our simulations, 
roughly ranging from $\sim 10-1000$ g. 
An obvious exception is the lowest $\alpha_{\rm{t}}=10^{-5}$ fractal aggregate model  (Fig. \ref{fig:a5g}, lower left panel). Note that when $m_{\rm{L}} > m_{\rm{f}}$ ({\it i.e.}, lucky particles), they generally carry a negligible amount of the mass, whereas outside the snowline, aggregates and particles generally do not reach or exceed $m_{\rm{f}}$ and so $m_{\rm{L}}$ is the mass dominant particle. 

\begin{figure}
\gridline{\fig{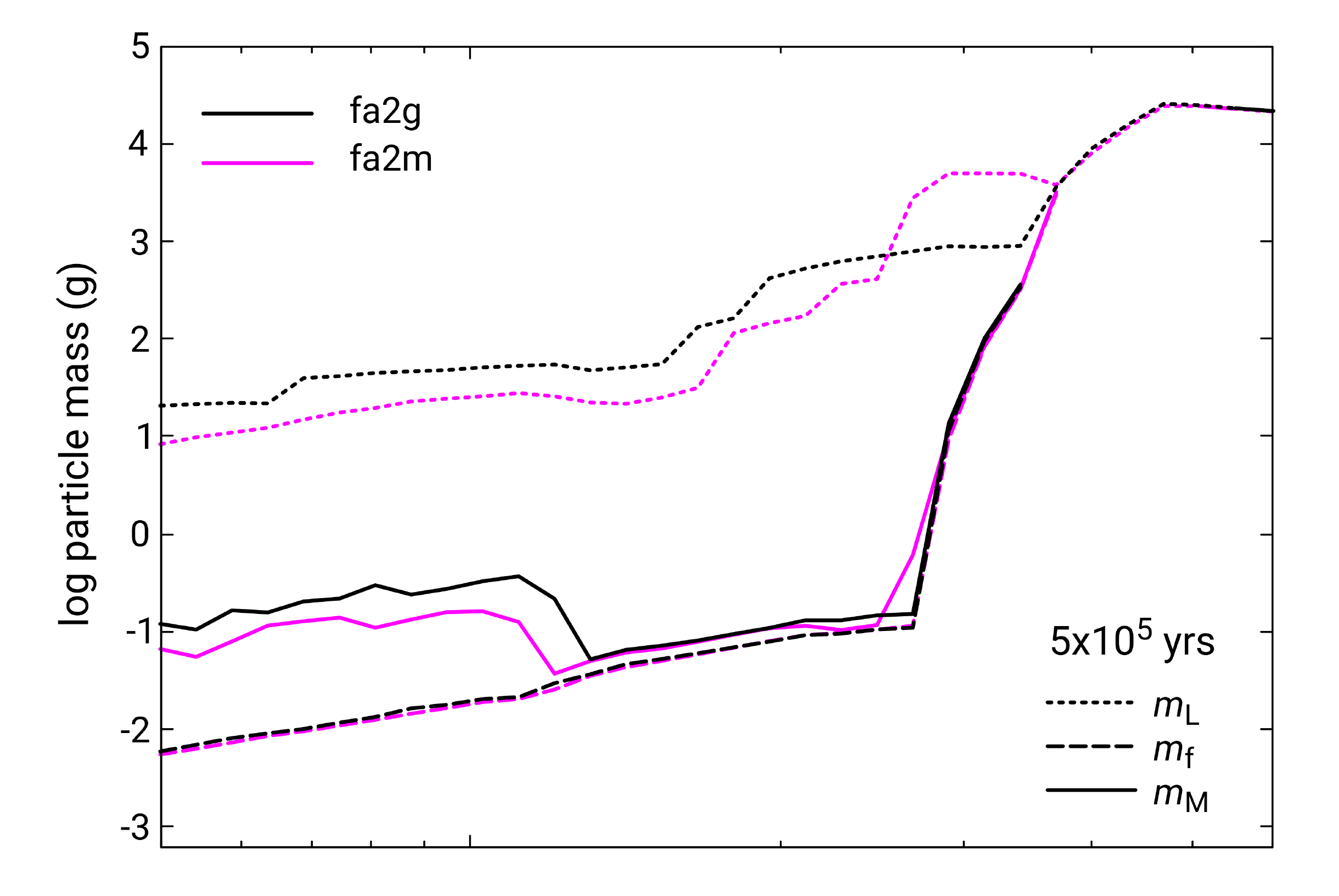}{0.45\textwidth}{}
          }
\vspace{-0.4in}
\gridline{\fig{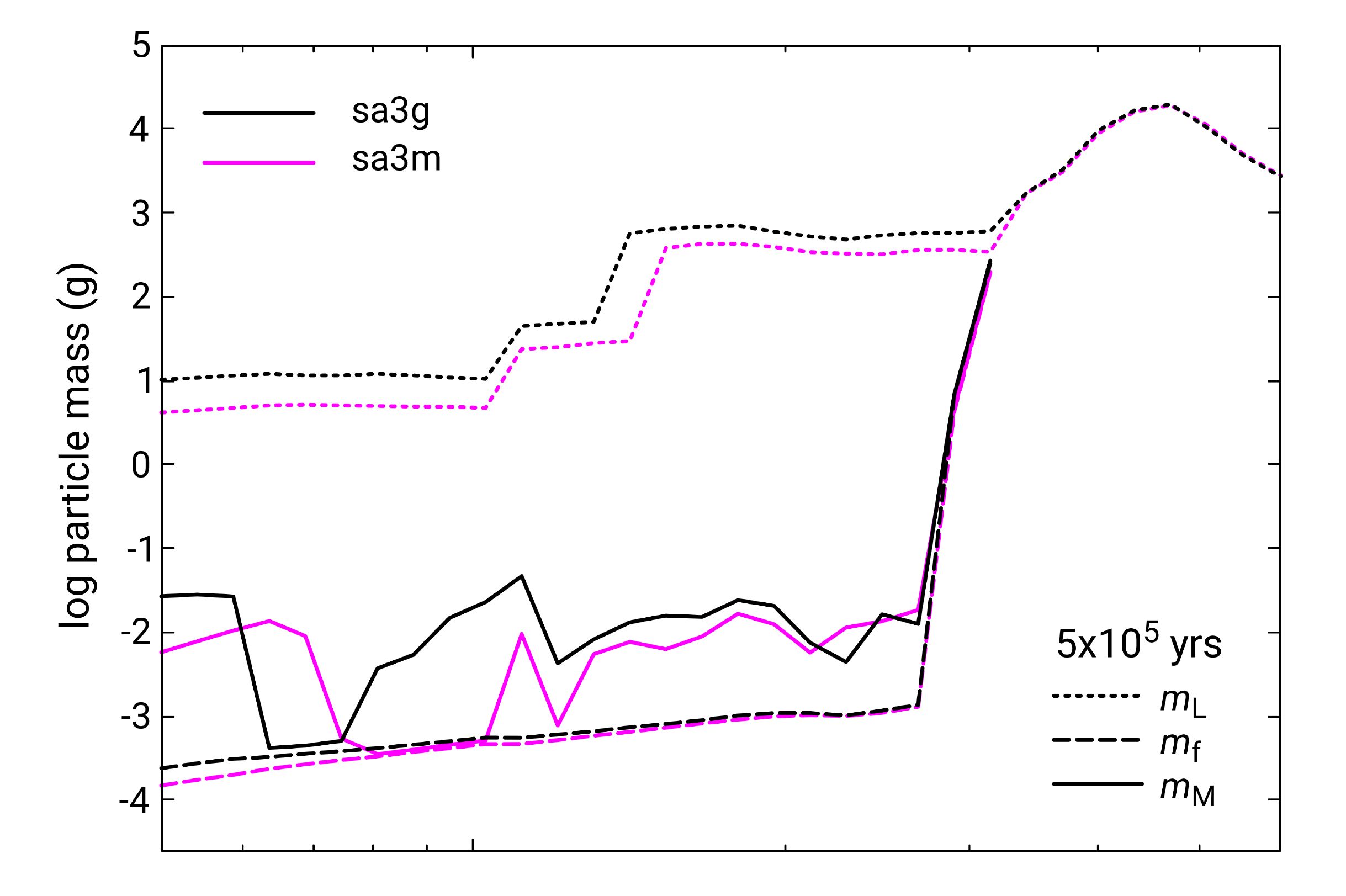}{0.45\textwidth}{}
          }
\vspace{-0.4in}
\gridline{\fig{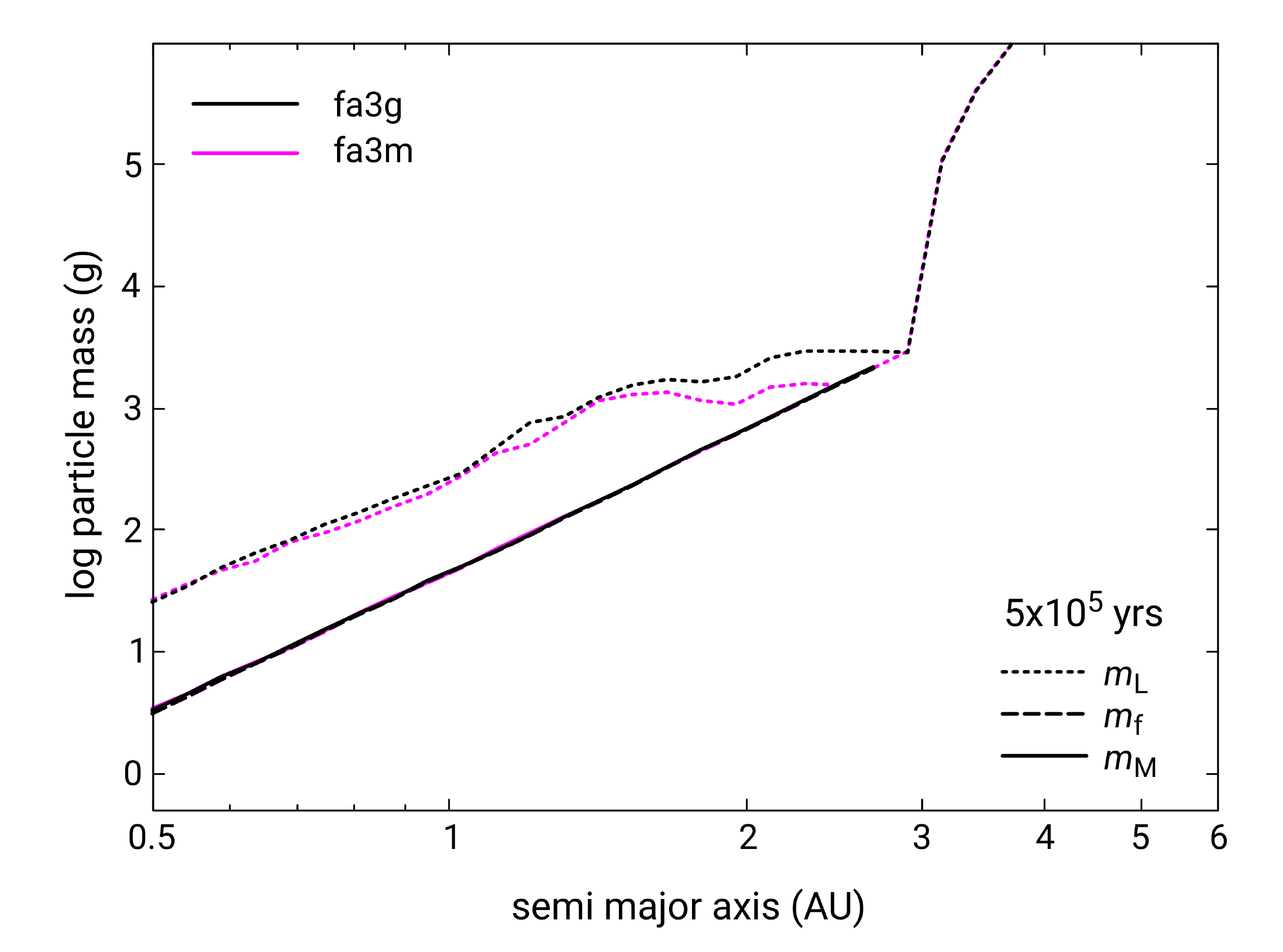}{0.45\textwidth}{}
          }
\caption{Comparison of simulations that use Gaussian (g; Sec. \ref{sec:destprob}) and Maxwellian (m) velocity PDFs . These highlight subtle differences in the particle mass distributions, and in particular the difference in the mass of the ``lucky particles'' $m_{\rm{L}}$, which are the largest mass to which particles or aggregates can grow beyond the fragmentation barrier $m_{\rm{f}}$ (dashed curves). The solid curves are the dominant mass particle $m_{\rm{M}}$ of the distribution. Sharp variations in $m_{\rm{M}}$ and $m_{\rm{L}}$ especially seen in the top two panels are associated with crossing EFs. For example, the organics EF is at $\sim 1.5$ AU, and the water-ice EF is located at $\sim 4$ AU, outside of which $Q_{\rm{f}}$ increases so aggregates can begin to grow larger. 
\label{fig:mgcomp}}
\end{figure}  

To compare directly with our Gaussian PDF approach, we also employ the model of \citet{Gar13} which more rigorously takes into account that particles both have deterministic and stochastic velocities. In our notation, 

\begin{equation}
    \label{equ:garaudpdf}
    \zeta_{\rm{G}}(m^\prime,m) = \frac{1}{\sqrt{2\pi}\sigma\Delta v_{\rm{D}}}
    \int_{v_c(m)}^\infty v^\prime \left[e^{-\frac{(v^\prime-\Delta v_{\rm{D}})^2}
    {2\sigma^2}} - e^{-\frac{(v^\prime+\Delta v_{\rm{D}})^2}{2\sigma^2}}\right]\,dv^\prime,
\end{equation}

\noindent
where $\Delta v_{\rm{D}}(m^\prime,m)$ is the deterministic relative velocity as defined in Sec. \ref{sec:relvel} between $m$ and $m^\prime$, and 

\begin{equation}
    \label{equ:sigij}
    \sigma^2 = \frac{\pi}{8}\left[(\Delta v_{\rm{B}})^2+(\Delta v_{\rm{t}})^2\right],
\end{equation}
 
\noindent
which ensures that the mean velocity of the distribution is what would be obtained when the stochastic velocities dominate the motions. The integrand in Eq. (\ref{equ:garaudpdf}) reduces to a Maxwellian distribution in the limit that $\Delta v_{\rm{D}} \ll \sigma$, whereas in the opposite limit when $\Delta v_{\rm{D}} \gg \sigma$, it reduces to a Gaussian with a significant tail for large values of the velocity. In the latter limit, the deterministic velocities are much larger than their rms velocities.
Integration of Eq. (\ref{equ:garaudpdf}) gives

\begin{equation}
    \label{equ:garaudint}
    \zeta_{\rm{G}}(m^\prime,m) = 1 - \frac{1}{2}\left[{\rm{erf}}\left(\frac{v_c-\Delta v_{\rm{D}}}{\sqrt{2}\sigma}\right) + {\rm{erf}}\left(\frac{v_c+\Delta v_{\rm{D}}}{\sqrt{2}\sigma}\right)\right] + \frac{1}{\sqrt{2\pi}}\frac{\sigma}{\Delta v_{\rm{D}}}\left[e^{-\frac{(v_c-\Delta v_{\rm{D}})^2}{2\sigma^2}} - e^{-\frac{(v_c+\Delta v_{\rm{D}})^2}{2\sigma^2}}\right].
\end{equation}

\noindent
In the limit where $\Delta v_{\rm{D}}$ is very small compared to $\sigma$ (e.g., for same-sized particles), the 3rd term on the RHS of Eq. (\ref{equ:garaudint}) reduces to $\sqrt{2/\pi}(v_c/\sigma){\rm{exp}}[-v_c^2/2\sigma^2]$. As noted by \citet{Gar13}, the mean collisional velocity is not the same as the traditional construct (Eq. [\ref{equ:dVpp}]), but rather

\begin{equation}
    \label{equ:dVppG}
    \Delta v_{\rm{pp}}(m^\prime,m) = \frac{\left(\Delta v_{\rm{D}}\right)^2 +\sigma^2}
    {\Delta v_{\rm{D}}}{\rm{erf}}\left(\frac{\Delta v_{\rm{D}}}{\sqrt{2}\sigma}\right) + 
    \sqrt{\frac{2}{\pi}}\sigma e^{-\frac{(\Delta v_{\rm{D}})^2}{2\sigma^2}},
\end{equation}

\noindent
which we adopt for 
simulations that utilize the Garaud et al. scheme.

Figure \ref{fig:mgcomp} shows a comparison between our Gaussian PDF for the relative velocities and destruction probabilities between particles (Eqns. \ref{equ:dVpp} and \ref{equ:gausspdf}, and designated by ``g''), or a Maxwellian PDF (Eqns. \ref{equ:dVppG} and  \ref{equ:garaudint}, labeled as ``m'') for several simulations as indicated in Table \ref{tab:models}. The Maxwellian approach is more rigorous in the limit that systematic velocities $\Delta v_{\rm{D}}\ll \Delta v_{\rm{t}}$  whereas in the opposite limit, the PDF is Gaussian-like but with a long tail. Our standard approach used here \citepalias[and also in][]{Est16} is Gaussian in both limits, thus the possibility of significant differences arises at least in the limit where (stochastic) turbulence-induced velocities dominate. The top panel of Fig. \ref{fig:mgcomp} shows that the difference between the two fractal aggregate growth models with our highest $\alpha_{\rm{t}}$ (=$10^{-2}$) simulations after 0.5 Myr is not considerable. Most notable is  that the mass $m_{\rm{L}}$ of the largest  or luckiest particle tends to be smaller when the PDF is Maxwellian. The corresponding  mass of the mass-dominant particle $m_{\rm{M}}$  appears smaller for the Maxwellian PDF in some cases, or comparable in others.

A more direct PDF comparison between compact and fractal growth models, for the fiducial case  $\alpha_{\rm{t}}=10^{-3}$, can be seen in the lower two panels of Fig. \ref{fig:mgcomp}. The compact growth models show the most notable differences in $m_{\rm{L}}$ and $m_{\rm{M}}$ (the Maxwellian values again tending to be smaller) and  the fractal growth simulations show smaller differences (bottom panel). We note that the difference in the fragmentation mass between models sa3g and sa3m inside of 1 AU is due to modest differences in the temperature profiles, while somewhat sharp variations in $m_{\rm{M}}$ and $m_{\rm{L}}$ are due to crossing EFs, notably the organics EF at $\sim 1.5$ AU in the top two panels. These PDF differences effectively vanish for smaller $\alpha_{\rm{t}}$ as the turbulence-induced velocities become less effective on the more massive particles or aggregates at the fragmentation limit,  and growth beyond the barrier mostly stalls. Thus we find, at least for these simulation times and disk parameters, that between the two PDFs, the differences are not significant enough to affect the qualitative nature of the results. Execution of the \citeauthor{Gar13} scheme in our code is slightly more involved, however, which led to roughly 20\% longer simulation times for the same parameter set.

\bibliography{my.bib}

\begin{thebibliography}{}
\expandafter\ifx\csname natexlab\endcsname\relax\def\natexlab#1{#1}\fi
\providecommand{\url}[1]{\href{#1}{#1}}

\bibitem[{{Andrews} {et~al.}(2018){Andrews}, {Huang}, {P{\'e}rez}, {Isella},
  {Dullemond}, {Kurtovic}, {Guzm{\'a}n}, {Carpenter}, {Wilner}, {Zhang}, {Zhu},
  {Birnstiel}, {Bai}, {Benisty}, {Hughes}, {{\"O}berg}, \& {Ricci}}]{And18}
{Andrews}, S.~M., {Huang}, J., {P{\'e}rez}, L.~M., {et~al.} 2018, \apjl, 869,
  L41

\bibitem[{{Bai} \& {Stone}(2013)}]{BS13}
{Bai}, X.-N., \& {Stone}, J.~M. 2013, \apj, 769, 76

\bibitem[{{Baraffe} {et~al.}(1998){Baraffe}, {Chabrier}, {Allard}, \&
  {Hauschildt}}]{Bar98}
{Baraffe}, I., {Chabrier}, G., {Allard}, F., \& {Hauschildt}, P.~H. 1998, \aap,
  337, 403

\bibitem[{{Baraffe} {et~al.}(2002){Baraffe}, {Chabrier}, {Allard}, \&
  {Hauschildt}}]{Bar02}
---. 2002, \aap, 382, 563

\bibitem[{{Barranco} {et~al.}(2018){Barranco}, {Pei}, \& {Marcus}}]{Bar18}
{Barranco}, J.~A., {Pei}, S., \& {Marcus}, P.~S. 2018, \apj, 869, 127

\bibitem[{{Beitz} {et~al.}(2012){Beitz}, {G{\"u}ttler}, {Weidling}, \&
  {Blum}}]{Bei12}
{Beitz}, E., {G{\"u}ttler}, C., {Weidling}, R., \& {Blum}, J. 2012, \icarus,
  218, 701

\bibitem[{{Birnstiel} {et~al.}(2010){Birnstiel}, {Dullemond}, \&
  {Brauer}}]{Bir10}
{Birnstiel}, T., {Dullemond}, C.~P., \& {Brauer}, F. 2010, \aap, 513, A79

\bibitem[{{Birnstiel} {et~al.}(2012){Birnstiel}, {Klahr}, \&
  {Ercolano}}]{Bir12}
{Birnstiel}, T., {Klahr}, H., \& {Ercolano}, B. 2012, \aap, 539, A148

\bibitem[{{Birnstiel} {et~al.}(2018){Birnstiel}, {Dullemond}, {Zhu}, {Andrews},
  {Bai}, {Wilner}, {Carpenter}, {Huang}, {Isella}, {Benisty}, {P{\'e}rez}, \&
  {Zhang}}]{Bir18}
{Birnstiel}, T., {Dullemond}, C.~P., {Zhu}, Z., {et~al.} 2018, \apjl, 869, L45

\bibitem[{{Bischoff} {et~al.}(2020){Bischoff}, {Kreuzig}, {Haack}, {Gundlach},
  \& {Blum}}]{Bis20}
{Bischoff}, D., {Kreuzig}, C., {Haack}, D., {Gundlach}, B., \& {Blum}, J. 2020,
  \mnras, 497, 2517

\bibitem[{{Bitsch} {et~al.}(2019){Bitsch}, {Izidoro}, {Johansen}, {Raymond},
  {Morbidelli}, {Lambrechts}, \& {Jacobson}}]{Bit19}
{Bitsch}, B., {Izidoro}, A., {Johansen}, A., {et~al.} 2019, \aap, 623, A88

\bibitem[{{Blum} \& {M{\"u}nch}(1993)}]{BM93}
{Blum}, J., \& {M{\"u}nch}, M. 1993, \icarus, 106, 151

\bibitem[{{Blum} \& {Wurm}(2000)}]{BW00}
{Blum}, J., \& {Wurm}, G. 2000, \icarus, 143, 138

\bibitem[{{Bodenheimer} {et~al.}(2018){Bodenheimer}, {Stevenson}, {Lissauer},
  \& {D'Angelo}}]{Bod18}
{Bodenheimer}, P., {Stevenson}, D.~J., {Lissauer}, J.~J., \& {D'Angelo}, G.
  2018, \apj, 868, 138

\bibitem[{{Brauer} {et~al.}(2008){Brauer}, {Dullemond}, \& {Henning}}]{Bra08}
{Brauer}, F., {Dullemond}, C.~P., \& {Henning}, T. 2008, \aap, 480, 859

\bibitem[{{Carballido} {et~al.}(2010){Carballido}, {Cuzzi}, \& {Hogan}}]{Car10}
{Carballido}, A., {Cuzzi}, J.~N., \& {Hogan}, R.~C. 2010, \mnras, 405, 2339

\bibitem[{{Carrasco-Gonz{\'a}lez} {et~al.}(2019){Carrasco-Gonz{\'a}lez},
  {Sierra}, {Flock}, {Zhu}, {Henning}, {Chandler}, {Galv{\'a}n-Madrid},
  {Mac{\'\i}as}, {Anglada}, {Linz}, {Osorio}, {Rodr{\'\i}guez}, {Testi},
  {Torrelles}, {P{\'e}rez}, \& {Liu}}]{CarrascoGonzalezetal2019}
{Carrasco-Gonz{\'a}lez}, C., {Sierra}, A., {Flock}, M., {et~al.} 2019, \apj,
  883, 71

\bibitem[{{Chambers}(2014)}]{Cha14}
{Chambers}, J.~E. 2014, {Planet Formation}, ed. A.~M. {Davis}, Vol.~2, 55--72

\bibitem[{{Chambers} {et~al.}(2010){Chambers}, {O'Brien}, \& {Davis}}]{Cha10}
{Chambers}, J.~E., {O'Brien}, D.~P., \& {Davis}, A.~M. 2010, {Accretion of
  Planetesimals and the Formation of Rocky Planets}, ed. D.~A. {Apai} \& D.~S.
  {Lauretta}, 299--335

\bibitem[{{Charnoz} {et~al.}(2021){Charnoz}, {Avice}, {Hyodo}, {Pignatale}, \&
  {Chaussidon}}]{Cha21}
{Charnoz}, S., {Avice}, G., {Hyodo}, R., {Pignatale}, F.~C., \& {Chaussidon},
  M. 2021, \aap, 652, A35

\bibitem[{{Chen} \& {Lin}(2020)}]{CL20}
{Chen}, K., \& {Lin}, M.-K. 2020, \apj, 891, 132

\bibitem[{{Chiang} \& {Goldreich}(1997)}]{CG97}
{Chiang}, E.~I., \& {Goldreich}, P. 1997, \apj, 490, 368

\bibitem[{{Ciesla} \& {Cuzzi}(2006)}]{CC06}
{Ciesla}, F.~J., \& {Cuzzi}, J.~N. 2006, \icarus, 181, 178

\bibitem[{{Cuzzi} {et~al.}(2003){Cuzzi}, {Davis}, \& {Dobrovolskis}}]{Cuz03}
{Cuzzi}, J.~N., {Davis}, S.~S., \& {Dobrovolskis}, A.~R. 2003, \icarus, 166,
  385

\bibitem[{{Cuzzi} {et~al.}(2014){Cuzzi}, {Estrada}, \& {Davis}}]{Cuz14}
{Cuzzi}, J.~N., {Estrada}, P.~R., \& {Davis}, S.~S. 2014, \apjs, 210, 21

\bibitem[{{Cuzzi} {et~al.}(2010){Cuzzi}, {Hogan}, \& {Bottke}}]{Cuz10}
{Cuzzi}, J.~N., {Hogan}, R.~C., \& {Bottke}, W.~F. 2010, \icarus, 208, 518

\bibitem[{{Cuzzi} {et~al.}(2001){Cuzzi}, {Hogan}, {Paque}, \&
  {Dobrovolskis}}]{Cuz01}
{Cuzzi}, J.~N., {Hogan}, R.~C., {Paque}, J.~M., \& {Dobrovolskis}, A.~R. 2001,
  \apj, 546, 496

\bibitem[{{Cuzzi} \& {Weidenschilling}(2006)}]{CW06}
{Cuzzi}, J.~N., \& {Weidenschilling}, S.~J. 2006, {Particle-Gas Dynamics and
  Primary Accretion}, ed. D.~S. {Lauretta} \& H.~Y. {McSween}, 353

\bibitem[{{D'Antona} \& {Mazzitelli}(1994)}]{DM94}
{D'Antona}, F., \& {Mazzitelli}, I. 1994, \apjs, 90, 467

\bibitem[{{Desch}(2007)}]{Des07}
{Desch}, S.~J. 2007, \apj, 671, 878

\bibitem[{{Desch} {et~al.}(2017){Desch}, {Estrada}, {Kalyaan}, \&
  {Cuzzi}}]{Des17}
{Desch}, S.~J., {Estrada}, P.~R., {Kalyaan}, A., \& {Cuzzi}, J.~N. 2017, \apj,
  840, 86

\bibitem[{{di Criscienzo} {et~al.}(2009){di Criscienzo}, {Ventura}, \&
  {D'Antona}}]{DiC09}
{di Criscienzo}, M., {Ventura}, P., \& {D'Antona}, F. 2009, \aap, 496, 223

\bibitem[{{Dominik} \& {Tielens}(1997)}]{DT97}
{Dominik}, C., \& {Tielens}, A.~G.~G.~M. 1997, \apj, 480, 647

\bibitem[{{Draine}(2003)}]{Dra03}
{Draine}, B.~T. 2003, \araa, 41, 241

\bibitem[{{Dra{\.z}kowska} {et~al.}(2016){Dra{\.z}kowska}, {Alibert}, \&
  {Moore}}]{Dra16}
{Dra{\.z}kowska}, J., {Alibert}, Y., \& {Moore}, B. 2016, \aap, 594, A105

\bibitem[{{Dra{\.z}kowska} \& {Dullemond}(2018)}]{DD18}
{Dra{\.z}kowska}, J., \& {Dullemond}, C.~P. 2018, \aap, 614, A62

\bibitem[{{Dubrulle} {et~al.}(1995){Dubrulle}, {Morfill}, \& {Sterzik}}]{Dub95}
{Dubrulle}, B., {Morfill}, G., \& {Sterzik}, M. 1995, \icarus, 114, 237

\bibitem[{{Estrada} \& {Cuzzi}(2008)}]{EC08}
{Estrada}, P.~R., \& {Cuzzi}, J.~N. 2008, \apj, 682, 515

\bibitem[{{Estrada} \& {Cuzzi}(2009)}]{EC09}
{Estrada}, P.~R., \& {Cuzzi}, J.~N. 2009, in Lunar and Planetary Science
  Conference, Lunar and Planetary Science Conference, 1241

\bibitem[{{Estrada} \& {Cuzzi}(2022)}]{Est21}
---. 2022, \apj

\bibitem[{{Estrada} {et~al.}(2016){Estrada}, {Cuzzi}, \& {Morgan}}]{Est16}
{Estrada}, P.~R., {Cuzzi}, J.~N., \& {Morgan}, D.~A. 2016, \apj, 818, 200

\bibitem[{{Flock} {et~al.}(2017){Flock}, {Nelson}, {Turner}, {Bertrang},
  {Carrasco-Gonz{\'a}lez}, {Henning}, {Lyra}, \& {Teague}}]{Flock_etal_2017}
{Flock}, M., {Nelson}, R.~P., {Turner}, N.~J., {et~al.} 2017, \apj, 850, 131

\bibitem[{{Freedman} {et~al.}(2014){Freedman}, {Lustig-Yaeger}, {Fortney},
  {Lupu}, {Marley}, \& {Lodders}}]{Fre14}
{Freedman}, R.~S., {Lustig-Yaeger}, J., {Fortney}, J.~J., {et~al.} 2014, \apjs,
  214, 25

\bibitem[{{Fulle} {et~al.}(2016){Fulle}, {Della Corte}, {Rotundi},
  {Rietmeijer}, {Green}, {Weissman}, {Accolla}, {Colangeli}, {Ferrari},
  {Ivanovski}, {Lopez-Moreno}, {Epifani}, {Morales}, {Ortiz}, {Palomba},
  {Palumbo}, {Rodriguez}, {Sordini}, \& {Zakharov}}]{Ful16}
{Fulle}, M., {Della Corte}, V., {Rotundi}, A., {et~al.} 2016, \mnras, 462, S132

\bibitem[{{Galvagni} {et~al.}(2011){Galvagni}, {Mayer}, \& {Saha}}]{Gal11}
{Galvagni}, M., {Mayer}, L., \& {Saha}, P. 2011, in AAS/Division for Extreme
  Solar Systems Abstracts, AAS/Division for Extreme Solar Systems Abstracts,
  33.06

\bibitem[{{Garaud}(2007)}]{Gar07}
{Garaud}, P. 2007, \apj, 671, 2091

\bibitem[{{Garaud} {et~al.}(2013){Garaud}, {Meru}, {Galvagni}, \&
  {Olczak}}]{Gar13}
{Garaud}, P., {Meru}, F., {Galvagni}, M., \& {Olczak}, C. 2013, \apj, 764, 146

\bibitem[{{Gressel} {et~al.}(2011){Gressel}, {Nelson}, \&
  {Turner}}]{Gresseletal2011}
{Gressel}, O., {Nelson}, R.~P., \& {Turner}, N.~J. 2011, \mnras, 415, 3291

\bibitem[{{Gressel} {et~al.}(2015){Gressel}, {Turner}, {Nelson}, \&
  {McNally}}]{Gre15}
{Gressel}, O., {Turner}, N.~J., {Nelson}, R.~P., \& {McNally}, C.~P. 2015,
  \apj, 801, 84

\bibitem[{{Gundlach} {et~al.}(2011){Gundlach}, {Kilias}, {Beitz}, \&
  {Blum}}]{Gun11}
{Gundlach}, B., {Kilias}, S., {Beitz}, E., \& {Blum}, J. 2011, \icarus, 214,
  717

\bibitem[{{G{\"u}ttler} {et~al.}(2010){G{\"u}ttler}, {Blum}, {Zsom}, {Ormel},
  \& {Dullemond}}]{Gut10}
{G{\"u}ttler}, C., {Blum}, J., {Zsom}, A., {Ormel}, C.~W., \& {Dullemond},
  C.~P. 2010, \aap, 513, A56

\bibitem[{{Hartlep} \& {Cuzzi}(2020)}]{HC20}
{Hartlep}, T., \& {Cuzzi}, J.~N. 2020, \apj, 892, 120

\bibitem[{{Hartmann} {et~al.}(1998){Hartmann}, {Calvet}, {Gullbring}, \&
  {D'Alessio}}]{Har98}
{Hartmann}, L., {Calvet}, N., {Gullbring}, E., \& {D'Alessio}, P. 1998, \apj,
  495, 385

\bibitem[{{Hayashi}(1981)}]{Hay81}
{Hayashi}, C. 1981, Progress of Theoretical Physics Supplement, 70, 35

\bibitem[{{Heim} {et~al.}(1999){Heim}, {Blum}, {Preuss}, \& {Butt}}]{Hei99}
{Heim}, L.-O., {Blum}, J., {Preuss}, M., \& {Butt}, H.-J. 1999, \prl, 83, 3328

\bibitem[{{Homma} \& {Nakamoto}(2018)}]{HN18}
{Homma}, K., \& {Nakamoto}, T. 2018, \apj, 868, 118

\bibitem[{{Homma} {et~al.}(2019){Homma}, {Okuzumi}, {Nakamoto}, \&
  {Ueda}}]{Hom19}
{Homma}, K.~A., {Okuzumi}, S., {Nakamoto}, T., \& {Ueda}, Y. 2019, in
  AAS/Division for Extreme Solar Systems Abstracts, Vol.~51, AAS/Division for
  Extreme Solar Systems Abstracts, 317.03

\bibitem[{{Hubbard}(2012)}]{Hub12}
{Hubbard}, A. 2012, \mnras, 426, 784

\bibitem[{{Hughes} \& {Armitage}(2012)}]{HA12}
{Hughes}, A. L.~H., \& {Armitage}, P.~J. 2012, \mnras, 423, 389

\bibitem[{{Ida} {et~al.}(2008){Ida}, {Guillot}, \& {Morbidelli}}]{Idaetal2008}
{Ida}, S., {Guillot}, T., \& {Morbidelli}, A. 2008, \apj, 686, 1292

\bibitem[{{Ishihara} {et~al.}(2018){Ishihara}, {Kobayashi}, {Enohata},
  {Umemura}, \& {Shiraishi}}]{Ish18}
{Ishihara}, T., {Kobayashi}, N., {Enohata}, K., {Umemura}, M., \& {Shiraishi},
  K. 2018, \apj, 854, 81

\bibitem[{{Jacquet}(2014)}]{Jac14}
{Jacquet}, E. 2014, Comptes Rendus Geoscience, 346, 3

\bibitem[{{Jacquet} {et~al.}(2012){Jacquet}, {Gounelle}, \& {Fromang}}]{Jac12}
{Jacquet}, E., {Gounelle}, M., \& {Fromang}, S. 2012, \icarus, 220, 162

\bibitem[{{Jessberger} {et~al.}(1988){Jessberger}, {Christoforidis}, \&
  {Kissel}}]{Jessbergeretal1988}
{Jessberger}, E.~K., {Christoforidis}, A., \& {Kissel}, J. 1988, \nat, 332, 691

\bibitem[{{Jessberger} \& {Kissel}(1991)}]{JessbergerKissel1991}
{Jessberger}, E.~K., \& {Kissel}, J. 1991, in Astrophysics and Space Science
  Library, Vol. 167, IAU Colloq. 116: Comets in the post-Halley era, ed.
  J.~{Newburn}, R.~L., M.~{Neugebauer}, \& J.~{Rahe}, 1075

\bibitem[{{Johansen} {et~al.}(2014){Johansen}, {Blum}, {Tanaka}, {Ormel},
  {Bizzarro}, \& {Rickman}}]{Joh14}
{Johansen}, A., {Blum}, J., {Tanaka}, H., {et~al.} 2014, in Protostars and
  Planets VI, ed. H.~{Beuther}, R.~S. {Klessen}, C.~P. {Dullemond}, \&
  T.~{Henning}, 547

\bibitem[{{Johansen} {et~al.}(2007){Johansen}, {Oishi}, {Mac Low}, {Klahr},
  {Henning}, \& {Youdin}}]{Joh07}
{Johansen}, A., {Oishi}, J.~S., {Mac Low}, M.-M., {et~al.} 2007, \nat, 448,
  1022

\bibitem[{{Kary} \& {Lissauer}(1994)}]{KL94}
{Kary}, D.~M., \& {Lissauer}, J.~J. 1994, {Numerical simulations of planetary
  growth.}, 364

\bibitem[{{Kataoka} {et~al.}(2014){Kataoka}, {Okuzumi}, {Tanaka}, \&
  {Nomura}}]{Kataokaetal2014}
{Kataoka}, A., {Okuzumi}, S., {Tanaka}, H., \& {Nomura}, H. 2014, \aap, 568,
  A42

\bibitem[{{Kataoka} {et~al.}(2013{\natexlab{a}}){Kataoka}, {Tanaka}, {Okuzumi},
  \& {Wada}}]{Kat13a}
{Kataoka}, A., {Tanaka}, H., {Okuzumi}, S., \& {Wada}, K. 2013{\natexlab{a}},
  \aap, 554, A4

\bibitem[{{Kataoka} {et~al.}(2013{\natexlab{b}}){Kataoka}, {Tanaka}, {Okuzumi},
  \& {Wada}}]{Kat13b}
---. 2013{\natexlab{b}}, \aap, 557, L4

\bibitem[{{Kataoka} {et~al.}(2015){Kataoka}, {Muto}, {Momose}, {Tsukagoshi},
  {Fukagawa}, {Shibai}, {Hanawa}, {Murakawa}, \& {Dullemond}}]{Kataokaetal2015}
{Kataoka}, A., {Muto}, T., {Momose}, M., {et~al.} 2015, \apj, 809, 78

\bibitem[{{Kenyon} \& {Bromley}(2010)}]{KB10}
{Kenyon}, S.~J., \& {Bromley}, B.~C. 2010, \apjs, 188, 242

\bibitem[{{Kenyon} \& {Hartmann}(1987)}]{KH87}
{Kenyon}, S.~J., \& {Hartmann}, L. 1987, \apj, 323, 714

\bibitem[{{Kokubo} \& {Ida}(2002)}]{KI02}
{Kokubo}, E., \& {Ida}, S. 2002, \apj, 581, 666

\bibitem[{{Kothe} {et~al.}(2010){Kothe}, {G{\"u}ttler}, \& {Blum}}]{Kot10}
{Kothe}, S., {G{\"u}ttler}, C., \& {Blum}, J. 2010, \apj, 725, 1242

\bibitem[{{Kouchi} {et~al.}(2002){Kouchi}, {Kudo}, {Nakano}, {Arakawa},
  {Watanabe}, {Sirono}, {Higa}, \& {Maeno}}]{Kou02}
{Kouchi}, A., {Kudo}, T., {Nakano}, H., {et~al.} 2002, \apjl, 566, L121

\bibitem[{{Krijt} {et~al.}(2014){Krijt}, {Dominik}, \& {Tielens}}]{Kri14}
{Krijt}, S., {Dominik}, C., \& {Tielens}, A.~G.~G.~M. 2014, Journal of Physics
  D Applied Physics, 47, 175302

\bibitem[{{Krijt} {et~al.}(2015){Krijt}, {Ormel}, {Dominik}, \&
  {Tielens}}]{Kri15}
{Krijt}, S., {Ormel}, C.~W., {Dominik}, C., \& {Tielens}, A.~G.~G.~M. 2015,
  \aap, 574, A83

\bibitem[{{Krijt} {et~al.}(2018){Krijt}, {Schwarz}, {Bergin}, \&
  {Ciesla}}]{Kri18}
{Krijt}, S., {Schwarz}, K.~R., {Bergin}, E.~A., \& {Ciesla}, F.~J. 2018, \apj,
  864, 78

\bibitem[{{Kruijer} {et~al.}(2017){Kruijer}, {Burkhardt}, {Budde}, \&
  {Kleine}}]{Kruijeretal2017}
{Kruijer}, T.~S., {Burkhardt}, C., {Budde}, G., \& {Kleine}, T. 2017,
  Proceedings of the National Academy of Science, 114, 6712

\bibitem[{{Lambrechts} \& {Johansen}(2012)}]{LJ12}
{Lambrechts}, M., \& {Johansen}, A. 2012, \aap, 544, A32

\bibitem[{{Lambrechts} \& {Johansen}(2014)}]{LJ14}
---. 2014, \aap, 572, A107

\bibitem[{{Lawler} \& {Brownlee}(1992)}]{LawlerBrownlee1992}
{Lawler}, M.~E., \& {Brownlee}, D.~E. 1992, \nat, 359, 810

\bibitem[{{Lesur} {et~al.}(2022){Lesur}, {Ercolano}, {Flock}, {Lin}, {Yang},
  {Barranco}, {Benitez-Llambay}, {Goodman}, {Johansen}, {Klahr}, {Laibe},
  {Lyra}, {Marcus}, {Nelson}, {Squire}, {Simon}, {Turner}, {Umurhan}, \&
  {Youdin}}]{Les22}
{Lesur}, G., {Ercolano}, B., {Flock}, M., {et~al.} 2022, arXiv e-prints,
  arXiv:2203.09821

\bibitem[{{Lissauer} {et~al.}(2011){Lissauer}, {Ragozzine}, {Fabrycky},
  {Steffen}, {Ford}, {Jenkins}, {Shporer}, {Holman}, {Rowe}, {Quintana},
  {Batalha}, {Borucki}, {Bryson}, {Caldwell}, {Carter}, {Ciardi}, {Dunham},
  {Fortney}, {Gautier}, {Howell}, {Koch}, {Latham}, {Marcy}, {Morehead}, \&
  {Sasselov}}]{Lis11}
{Lissauer}, J.~J., {Ragozzine}, D., {Fabrycky}, D.~C., {et~al.} 2011, \apjs,
  197, 8

\bibitem[{{Lodders}(2003)}]{Lod03}
{Lodders}, K. 2003, \apj, 591, 1220

\bibitem[{{Lorek} {et~al.}(2018){Lorek}, {Lacerda}, \& {Blum}}]{Lor18}
{Lorek}, S., {Lacerda}, P., \& {Blum}, J. 2018, \aap, 611, A18

\bibitem[{{Lynden-Bell} \& {Pringle}(1974)}]{LP74}
{Lynden-Bell}, D., \& {Pringle}, J.~E. 1974, \mnras, 168, 603

\bibitem[{{Lyra}(2014)}]{Lyr14}
{Lyra}, W. 2014, \apj, 789, 77

\bibitem[{{Lyra} \& {Umurhan}(2019)}]{LU19}
{Lyra}, W., \& {Umurhan}, O.~M. 2019, \pasp, 131, 072001

\bibitem[{{Mac{\'\i}as} {et~al.}(2021){Mac{\'\i}as}, {Guerra-Alvarado},
  {Carrasco-Gonz{\'a}lez}, {Ribas}, {Espaillat}, {Huang}, \&
  {Andrews}}]{Maciasetal2021}
{Mac{\'\i}as}, E., {Guerra-Alvarado}, O., {Carrasco-Gonz{\'a}lez}, C., {et~al.}
  2021, \aap, 648, A33

\bibitem[{{Marcus} {et~al.}(2015){Marcus}, {Pei}, {Jiang}, {Barranco},
  {Hassanzadeh}, \& {Lecoanet}}]{Mar15}
{Marcus}, P.~S., {Pei}, S., {Jiang}, C.-H., {et~al.} 2015, \apj, 808, 87

\bibitem[{{Marcus} {et~al.}(2013){Marcus}, {Pei}, {Jiang}, \&
  {Hassanzadeh}}]{Mar13}
{Marcus}, P.~S., {Pei}, S., {Jiang}, C.-H., \& {Hassanzadeh}, P. 2013, \prl,
  111, 084501

\bibitem[{{Morbidelli} {et~al.}(2009){Morbidelli}, {Bottke}, {Nesvorn{\'y}}, \&
  {Levison}}]{Mor09}
{Morbidelli}, A., {Bottke}, W.~F., {Nesvorn{\'y}}, D., \& {Levison}, H.~F.
  2009, \icarus, 204, 558

\bibitem[{{Mumma} \& {Charnley}(2011)}]{MummaCharnley2011}
{Mumma}, M.~J., \& {Charnley}, S.~B. 2011, \araa, 49, 471

\bibitem[{{Musiolik} {et~al.}(2016){Musiolik}, {Teiser}, {Jankowski}, \&
  {Wurm}}]{Mus16}
{Musiolik}, G., {Teiser}, J., {Jankowski}, T., \& {Wurm}, G. 2016, \apj, 827,
  63

\bibitem[{{Musiolik} \& {Wurm}(2019)}]{MW19}
{Musiolik}, G., \& {Wurm}, G. 2019, \apj, 873, 58

\bibitem[{{Nakagawa} {et~al.}(1986){Nakagawa}, {Sekiya}, \& {Hayashi}}]{Nak86}
{Nakagawa}, Y., {Sekiya}, M., \& {Hayashi}, C. 1986, \icarus, 67, 375

\bibitem[{{Nakamoto} \& {Nakagawa}(1994)}]{NN94}
{Nakamoto}, T., \& {Nakagawa}, Y. 1994, \apj, 421, 640

\bibitem[{{Nelson} \& {Gressel}(2010)}]{NelsonGressel2010}
{Nelson}, R.~P., \& {Gressel}, O. 2010, \mnras, 409, 639

\bibitem[{{Nelson} {et~al.}(2013){Nelson}, {Gressel}, \& {Umurhan}}]{Nel13}
{Nelson}, R.~P., {Gressel}, O., \& {Umurhan}, O.~M. 2013, \mnras, 435, 2610

\bibitem[{{Okuzumi} {et~al.}(2012){Okuzumi}, {Tanaka}, {Kobayashi}, \&
  {Wada}}]{Oku12}
{Okuzumi}, S., {Tanaka}, H., {Kobayashi}, H., \& {Wada}, K. 2012, \apj, 752,
  106

\bibitem[{{Okuzumi} {et~al.}(2009){Okuzumi}, {Tanaka}, \& {Sakagami}}]{Oku09}
{Okuzumi}, S., {Tanaka}, H., \& {Sakagami}, M.-a. 2009, \apj, 707, 1247

\bibitem[{{Okuzumi} {et~al.}(2011){Okuzumi}, {Tanaka}, {Takeuchi}, \&
  {Sakagami}}]{Oku11}
{Okuzumi}, S., {Tanaka}, H., {Takeuchi}, T., \& {Sakagami}, M.-a. 2011, \apj,
  731, 95

\bibitem[{{Okuzumi} \& {Tazaki}(2019)}]{Oku19}
{Okuzumi}, S., \& {Tazaki}, R. 2019, \apj, 878, 132

\bibitem[{{Ormel} \& {Cuzzi}(2007)}]{OC07}
{Ormel}, C.~W., \& {Cuzzi}, J.~N. 2007, \aap, 466, 413

\bibitem[{{Ormel} \& {Klahr}(2010)}]{OK10}
{Ormel}, C.~W., \& {Klahr}, H.~H. 2010, \aap, 520, A43

\bibitem[{{Ormel} \& {Okuzumi}(2013)}]{OrmelOkuzumi2013}
{Ormel}, C.~W., \& {Okuzumi}, S. 2013, \apj, 771, 44

\bibitem[{{Ormel} {et~al.}(2007){Ormel}, {Spaans}, \& {Tielens}}]{Orm07}
{Ormel}, C.~W., {Spaans}, M., \& {Tielens}, A.~G.~G.~M. 2007, \aap, 461, 215

\bibitem[{{Palla} \& {Stahler}(1999)}]{PS99}
{Palla}, F., \& {Stahler}, S.~W. 1999, \apj, 525, 772

\bibitem[{{Pan} \& {Padoan}(2013)}]{PP13}
{Pan}, L., \& {Padoan}, P. 2013, \apj, 776, 12

\bibitem[{{Pan} \& {Padoan}(2015)}]{PP15}
---. 2015, \apj, 812, 10

\bibitem[{{P{\"a}tzold} {et~al.}(2016){P{\"a}tzold}, {Andert}, {Hahn}, {Asmar},
  {Barriot}, {Bird}, {H{\"a}usler}, {Peter}, {Tellmann}, {Gr{\"u}n},
  {Weissman}, {Sierks}, {Jorda}, {Gaskell}, {Preusker}, \& {Scholten}}]{Pat16}
{P{\"a}tzold}, M., {Andert}, T., {Hahn}, M., {et~al.} 2016, \nat, 530, 63

\bibitem[{{Pfeil} \& {Klahr}(2019)}]{Pfeil_Klahr_2019}
{Pfeil}, T., \& {Klahr}, H. 2019, \apj, 871, 150

\bibitem[{{Podolak} {et~al.}(1988){Podolak}, {Pollack}, \& {Reynolds}}]{Pod88}
{Podolak}, M., {Pollack}, J.~B., \& {Reynolds}, R.~T. 1988, \icarus, 73, 163

\bibitem[{{Pollack} {et~al.}(1994){Pollack}, {Hollenbach}, {Beckwith},
  {Simonelli}, {Roush}, \& {Fong}}]{Pol94}
{Pollack}, J.~B., {Hollenbach}, D., {Beckwith}, S., {et~al.} 1994, \apj, 421,
  615

\bibitem[{{Pringle}(1981)}]{Pri81}
{Pringle}, J.~E. 1981, \araa, 19, 137

\bibitem[{{Ros} \& {Johansen}(2013)}]{RJ13}
{Ros}, K., \& {Johansen}, A. 2013, \aap, 552, A137

\bibitem[{{Ruden} \& {Pollack}(1991)}]{RP91}
{Ruden}, S.~P., \& {Pollack}, J.~B. 1991, \apj, 375, 740

\bibitem[{{Schlichting} {et~al.}(2013){Schlichting}, {Fuentes}, \&
  {Trilling}}]{Sch13}
{Schlichting}, H.~E., {Fuentes}, C.~I., \& {Trilling}, D.~E. 2013, \aj, 146, 36

\bibitem[{{Schoonenberg} \& {Ormel}(2017)}]{SO17}
{Schoonenberg}, D., \& {Ormel}, C.~W. 2017, \aap, 602, A21

\bibitem[{{Segura-Cox} {et~al.}(2020){Segura-Cox}, {Schmiedeke}, {Pineda},
  {Stephens}, {Fern{\'a}ndez-L{\'o}pez}, {Looney}, {Caselli}, {Li}, {Mundy},
  {Kwon}, \& {Harris}}]{SC20}
{Segura-Cox}, D.~M., {Schmiedeke}, A., {Pineda}, J.~E., {et~al.} 2020, \nat,
  586, 228

\bibitem[{{Sengupta} {et~al.}(2019){Sengupta}, {Dodson-Robinson}, {Hasegawa},
  \& {Turner}}]{Sen19}
{Sengupta}, D., {Dodson-Robinson}, S.~E., {Hasegawa}, Y., \& {Turner}, N.~J.
  2019, \apj, 874, 26

\bibitem[{{Sengupta} {et~al.}(2022){Sengupta}, {Estrada}, {Cuzzi}, \&
  {Humayun}}]{Sen21}
{Sengupta}, D., {Estrada}, P.~R., {Cuzzi}, J.~N., \& {Humayun}, M. 2022, \apj

\bibitem[{{Sierks} {et~al.}(2015){Sierks}, {Barbieri}, {Lamy}, {Rodrigo},
  {Koschny}, {Rickman}, {Keller}, {Agarwal}, {A'Hearn}, {Angrilli}, {Auger},
  {Barucci}, {Bertaux}, {Bertini}, {Besse}, {Bodewits}, {Capanna}, {Cremonese},
  {Da Deppo}, {Davidsson}, {Debei}, {De Cecco}, {Ferri}, {Fornasier}, {Fulle},
  {Gaskell}, {Giacomini}, {Groussin}, {Gutierrez-Marques}, {Guti{\'e}rrez},
  {G{\"u}ttler}, {Hoekzema}, {Hviid}, {Ip}, {Jorda}, {Knollenberg}, {Kovacs},
  {Kramm}, {K{\"u}hrt}, {K{\"u}ppers}, {La Forgia}, {Lara}, {Lazzarin},
  {Leyrat}, {Lopez Moreno}, {Magrin}, {Marchi}, {Marzari}, {Massironi},
  {Michalik}, {Moissl}, {Mottola}, {Naletto}, {Oklay}, {Pajola}, {Pertile},
  {Preusker}, {Sabau}, {Scholten}, {Snodgrass}, {Thomas}, {Tubiana}, {Vincent},
  {Wenzel}, {Zaccariotto}, \& {P{\"a}tzold}}]{Sie15}
{Sierks}, H., {Barbieri}, C., {Lamy}, P.~L., {et~al.} 2015, Science, 347,
  aaa1044

\bibitem[{{Sierra} {et~al.}(2021){Sierra}, {P{\'e}rez}, {Zhang}, {Law},
  {Guzm{\'a}n}, {Qi}, {Bosman}, {{\"O}berg}, {Andrews}, {Long}, {Teague},
  {Booth}, {Walsh}, {Wilner}, {M{\'e}nard}, {Cataldi}, {Czekala}, {Bae},
  {Huang}, {Bergner}, {Ilee}, {Benisty}, {Le Gal}, {Loomis}, {Tsukagoshi},
  {Liu}, {Yamato}, \& {Aikawa}}]{Sierraetal2021}
{Sierra}, A., {P{\'e}rez}, L.~M., {Zhang}, K., {et~al.} 2021, \apjs, 257, 14

\bibitem[{{Siess} {et~al.}(2000){Siess}, {Dufour}, \& {Forestini}}]{Sie00}
{Siess}, L., {Dufour}, E., \& {Forestini}, M. 2000, \aap, 358, 593

\bibitem[{{Stewart} \& {Leinhardt}(2009)}]{SL09}
{Stewart}, S.~T., \& {Leinhardt}, Z.~M. 2009, \apjl, 691, L133

\bibitem[{{Stoll} {et~al.}(2017){Stoll}, {Kley}, \& {Picogna}}]{Sto17}
{Stoll}, M. H.~R., {Kley}, W., \& {Picogna}, G. 2017, \aap, 599, L6

\bibitem[{{Suyama} {et~al.}(2008){Suyama}, {Wada}, \& {Tanaka}}]{Suy08}
{Suyama}, T., {Wada}, K., \& {Tanaka}, H. 2008, \apj, 684, 1310

\bibitem[{{Suyama} {et~al.}(2012){Suyama}, {Wada}, {Tanaka}, \&
  {Okuzumi}}]{Suy12}
{Suyama}, T., {Wada}, K., {Tanaka}, H., \& {Okuzumi}, S. 2012, \apj, 753, 115

\bibitem[{{Takeuchi} \& {Lin}(2002)}]{TL02}
{Takeuchi}, T., \& {Lin}, D.~N.~C. 2002, \apj, 581, 1344

\bibitem[{{Tanaka} {et~al.}(2005){Tanaka}, {Himeno}, \& {Ida}}]{Tan05}
{Tanaka}, H., {Himeno}, Y., \& {Ida}, S. 2005, \apj, 625, 414

\bibitem[{{Tazaki} \& {Tanaka}(2018)}]{TazakiTanaka2018}
{Tazaki}, R., \& {Tanaka}, H. 2018, \apj, 860, 79

\bibitem[{{Tognelli} {et~al.}(2011){Tognelli}, {Prada Moroni}, \&
  {Degl'Innocenti}}]{Tog11}
{Tognelli}, E., {Prada Moroni}, P.~G., \& {Degl'Innocenti}, S. 2011, \aap, 533,
  A109

\bibitem[{{Triaud} {et~al.}(2017){Triaud}, {Neveu-VanMalle}, {Lendl},
  {Anderson}, {Collier Cameron}, {Delrez}, {Doyle}, {Gillon}, {Hellier},
  {Jehin}, {Maxted}, {S{\'e}gransan}, {Smalley}, {Queloz}, {Pollacco},
  {Southworth}, {Tregloan-Reed}, {Udry}, \& {West}}]{Tri17}
{Triaud}, A. H.~M.~J., {Neveu-VanMalle}, M., {Lendl}, M., {et~al.} 2017,
  \mnras, 467, 1714

\bibitem[{{Turner} {et~al.}(2014){Turner}, {Fromang}, {Gammie}, {Klahr},
  {Lesur}, {Wardle}, \& {Bai}}]{Tur14}
{Turner}, N.~J., {Fromang}, S., {Gammie}, C., {et~al.} 2014, in Protostars and
  Planets VI, ed. H.~{Beuther}, R.~S. {Klessen}, C.~P. {Dullemond}, \&
  T.~{Henning}, 411

\bibitem[{{Umurhan} {et~al.}(2020){Umurhan}, {Estrada}, \& {Cuzzi}}]{Umu20}
{Umurhan}, O.~M., {Estrada}, P.~R., \& {Cuzzi}, J.~N. 2020, \apj, 895, 4

\bibitem[{{V\"olk} {et~al.}(1980){V\"olk}, {Jones}, {Morfill}, \&
  {Roeser}}]{Vol80}
{V\"olk}, H.~J., {Jones}, F.~C., {Morfill}, G.~E., \& {Roeser}, S. 1980, \aap,
  85, 316

\bibitem[{{Wada} {et~al.}(2013){Wada}, {Tanaka}, {Okuzumi}, {Kobayashi},
  {Suyama}, {Kimura}, \& {Yamamoto}}]{Wad13}
{Wada}, K., {Tanaka}, H., {Okuzumi}, S., {et~al.} 2013, \aap, 559, A62

\bibitem[{{Wada} {et~al.}(2008){Wada}, {Tanaka}, {Suyama}, {Kimura}, \&
  {Yamamoto}}]{Wad08}
{Wada}, K., {Tanaka}, H., {Suyama}, T., {Kimura}, H., \& {Yamamoto}, T. 2008,
  \apj, 677, 1296

\bibitem[{{Wada} {et~al.}(2009){Wada}, {Tanaka}, {Suyama}, {Kimura}, \&
  {Yamamoto}}]{Wad09}
---. 2009, \apj, 702, 1490

\bibitem[{{Weidenschilling}(1977)}]{Wei77}
{Weidenschilling}, S.~J. 1977, \mnras, 180, 57

\bibitem[{{Weidenschilling} \& {Cuzzi}(1993)}]{WC93}
{Weidenschilling}, S.~J., \& {Cuzzi}, J.~N. 1993, in Protostars and Planets
  III, ed. E.~H. {Levy} \& J.~I. {Lunine}, 1031

\bibitem[{{Windmark} {et~al.}(2012{\natexlab{a}}){Windmark}, {Birnstiel},
  {G{\"u}ttler}, {Blum}, {Dullemond}, \& {Henning}}]{Win12a}
{Windmark}, F., {Birnstiel}, T., {G{\"u}ttler}, C., {et~al.}
  2012{\natexlab{a}}, \aap, 540, A73

\bibitem[{{Windmark} {et~al.}(2012{\natexlab{b}}){Windmark}, {Birnstiel},
  {Ormel}, \& {Dullemond}}]{Win12b}
{Windmark}, F., {Birnstiel}, T., {Ormel}, C.~W., \& {Dullemond}, C.~P.
  2012{\natexlab{b}}, \aap, 544, L16

\bibitem[{{Windmark} {et~al.}(2012{\natexlab{c}}){Windmark}, {Birnstiel},
  {Ormel}, \& {Dullemond}}]{Win12c}
---. 2012{\natexlab{c}}, \aap, 548, C1

\bibitem[{{Winn} \& {Fabrycky}(2015)}]{WF15}
{Winn}, J.~N., \& {Fabrycky}, D.~C. 2015, \araa, 53, 409

\bibitem[{{Wittenmyer} {et~al.}(2020){Wittenmyer}, {Wang}, {Horner}, {Butler},
  {Tinney}, {Carter}, {Wright}, {Jones}, {Bailey}, {O'Toole}, \&
  {Johns}}]{Wit20}
{Wittenmyer}, R.~A., {Wang}, S., {Horner}, J., {et~al.} 2020, \mnras, 492, 377

\bibitem[{{Woodward} {et~al.}(2021){Woodward}, {Wooden}, {Harker}, {Kelley},
  {Russell}, \& {Kim}}]{Woodwardetal2021}
{Woodward}, C.~E., {Wooden}, D.~H., {Harker}, D.~E., {et~al.} 2021, \psj, 2, 25

\bibitem[{{Wurm} {et~al.}(2005){Wurm}, {Paraskov}, \& {Krauss}}]{Wur05}
{Wurm}, G., {Paraskov}, G., \& {Krauss}, O. 2005, \icarus, 178, 253

\bibitem[{{Yang} \& {Ciesla}(2012)}]{YC12}
{Yang}, L., \& {Ciesla}, F.~J. 2012, Meteoritics and Planetary Science, 47, 99

\bibitem[{{Youdin} \& {Johansen}(2007)}]{YJ07}
{Youdin}, A., \& {Johansen}, A. 2007, \apj, 662, 613

\bibitem[{{Youdin} \& {Goodman}(2005)}]{YG05}
{Youdin}, A.~N., \& {Goodman}, J. 2005, \apj, 620, 459

\bibitem[{{Youdin} \& {Lithwick}(2007)}]{YL07}
{Youdin}, A.~N., \& {Lithwick}, Y. 2007, \icarus, 192, 588

\bibitem[{{Zhu} {et~al.}(2019){Zhu}, {Zhang}, {Jiang}, {Kataoka}, {Birnstiel},
  {Dullemond}, {Andrews}, {Huang}, {P{\'e}rez}, {Carpenter}, {Bai}, {Wilner},
  \& {Ricci}}]{Zhuetal2019}
{Zhu}, Z., {Zhang}, S., {Jiang}, Y.-F., {et~al.} 2019, \apjl, 877, L18

\bibitem[{{Zsom} {et~al.}(2010){Zsom}, {Ormel}, {G{\"u}ttler}, {Blum}, \&
  {Dullemond}}]{Zso10}
{Zsom}, A., {Ormel}, C.~W., {G{\"u}ttler}, C., {Blum}, J., \& {Dullemond},
  C.~P. 2010, \aap, 513, A57

\end{thebibliography}



\end{document}